\documentclass[preprint,secnumarabic,preprintnumbers,aps,superscriptaddress,floatfix]{revtex4}
\usepackage[pdftex]{graphicx}
\usepackage{amssymb}
\usepackage{mathptmx}
\newcommand{\be}{\begin{equation}}
\newcommand{\ee}{\end{equation}}
\newcommand{\bea}{\begin{eqnarray}}
\newcommand{\eea}{\end{eqnarray}}

\newcommand{\lsim}{\mathrel{\lower4pt\hbox{$\sim$}}
\hskip-11.5pt\raise1.6pt\hbox{$<$}\;}

\newcommand{\gsim}{\mathrel{\lower4pt\hbox{$\sim$}}
\hskip-11.5pt\raise1.6pt\hbox{$>$}\;}

\def\lsim{\mathrel{\raise.3ex\hbox{$<$\kern-.75em\lower1ex\hbox{$\sim$}}}}
\def\gsim{\mathrel{\raise.3ex\hbox{$>$\kern-.75em\lower1ex\hbox{$\sim$}}}}

\begin{document}

\pagenumbering{roman}

\preprint{Fermilab-TM-2424-E}
\preprint{BNL-81896-2008-IR} 
\preprint{LBNL-1348E} 

\title{ ~~~ \\ ~~~ \\ 
Report on the Depth Requirements for a Massive Detector at Homestake\\
}

\author{Adam Bernstein} 
\affiliation{Lawrence Livermore National Laboratory, Livermore, CA 94550}

\author{Mary Bishai}
\affiliation{Department of Physics,
Brookhaven National Laboratory, Upton, NY 11973}

\author{Edward Blucher}
\affiliation{Department of Physics, University of Chicago,  Chicago, IL 60637}

\author{David B. Cline} 
\affiliation{Department of Physics, University of California, Los Angeles, CA 90095}

\author{Milind V. Diwan} 
\affiliation{Department of Physics,
Brookhaven National Laboratory, Upton, NY 11973}

\author{Bonnie Fleming}
\affiliation{Department of Physics, Yale University, New Haven, CT 06520}

\author{Maury Goodman}
\affiliation{Argonne National Laboratory, Argonne, IL 60439} 

\author{Zbigniew J. Hladysz}
\affiliation{Mining Engineering and Management Department, South Dakota
 School of Mines and Tehnology, Rapid City, SD 57701}

\author{Richard Kadel}
\affiliation{Physics Division, Lawrence Berkeley National Laboratory, Berkeley, 
CA 94720, USA}

\author{Edward Kearns} 
\affiliation{Physics Department, Boston University, Boston, MA 02215}

\author{Joshua Klein} 
\affiliation{Department of Physics and Astronomy, 
University of Pennsylvania, Philadelphia, PA 19104}

\author{Kenneth Lande} 
\affiliation{Department of Physics and Astronomy, 
University of Pennsylvania, Philadelphia, PA 19104}

\author{Francesco Lanni} 
\affiliation{Department of Physics,
Brookhaven National Laboratory, Upton, NY 11973}

\author{David Lissauer} 
\affiliation{Department of Physics,
Brookhaven National Laboratory, Upton, NY 11973}

\author{Steve Marks}
\affiliation{Physics Division, Lawrence Berkeley National Laboratory, Berkeley, 
CA 94720, USA}

\author{Robert McKeown}
\affiliation{Department of Physics, California Institute of Technology,
 Pasadena, CA 91125}

\author{William Morse} 
\affiliation{Department of Physics,
Brookhaven National Laboratory, Upton, NY 11973}

\author{Regina Rameika}
\affiliation{Particle Physics Division, MS 220, Fermilab, Batavia, IL 60510}

\author{William M. Roggenthen }
\affiliation{Department of Geology and Geological Engineering, South Dakota
 School of Mines and Tehnology, Rapid City, SD 57701}

\author{Kate Scholberg} 
\affiliation{Department of Physics,
Duke University, Durham, NC 27708}

\author{Michael Smy}
\affiliation{Physics Department, University of California, Irvine, CA 92697}

\author{Henry Sobel}
\affiliation{Physics Department, University of California, Irvine, CA 92697} 

\author{James Stewart} 
\affiliation{Department of Physics,
Brookhaven National Laboratory, Upton, NY 11973} 

\author{Gregory Sullivan} 
\affiliation{Physics Department,
University of Maryland, College Park, MD 20742} 

\author{Robert Svoboda}
\affiliation{Physics Department, University of California, Davis, CA 95616}

\author{Mark Vagins} 
\affiliation{Institute for the Physics and Mathematics of the Universe,
University of Tokyo,  Kashiwa, 277-8568, Japan}

\author{Brett Viren} 
\affiliation{Department of Physics,
Brookhaven National Laboratory, Upton, NY 11973} 

\author{Christopher Walter} 
\affiliation{Department of Physics, Duke University, Durham, NC 27708}

\author{Robert Zwaska} 
\affiliation{Fermi National  Accelerator Laboratory, Batavia, IL 60510}

\date{23 July  2009}

\newpage

\begin{abstract}

This report provides the technical justification for locating a large
detector underground in a US based Deep Underground Science and
Engineering Laboratory. A large detector with a fiducial mass
in the mega-ton scale will most likely be a multipurpose facility. The
main physics justification for such a device is detection of
accelerator generated neutrinos, nucleon decay, and natural sources of
neutrinos such as solar, atmospheric and supernova neutrinos.  The
requirement on the depth of this detector will be guided by the rate of
signals from these sources and the rate of backgrounds from cosmic rays
over a  very wide range of energies (from solar neutrino
energies of ~5 MeV to high energies in the range of hundreds of GeV).

For the present report, we have examined the depth requirement for a
large water Cherenkov detector and a liquid argon time projection
chamber. There has been extensive previous experience with
underground water Cherenkov detectors such as IMB, Kamioka, and most
recently, Super-Kamiokande which has a fiducial mass of 22~kTon and a 
total mass of 50~kTon at a 
depth of 2700 meters-water-equivalent in a mountain. Projections for signal and
background capability for a larger and deeper (or shallower)
detectors of this type can be scaled from these previous detectors.
The liquid argon time projection chamber has the advantage of being
a very fine-grained tracking detector, which should provide enhanced
capability for background rejection.
We have based
background rejection on reasonable estimates of track and energy
resolution, and in some cases scaled background rates from measurements in
water.

In the current work we have taken the approach that the depth should
be sufficient to suppress the cosmogenic background below predicted
signal rates for either of the above two technologies. Nevertheless,
it is also clear that the underground facility that we are examining
must have a long life and will most likely be used either for future
novel uses of the currently planned detectors or new technologies.
Therefore the depth requirement also needs to be made on the basis of sound
judgment regarding possible future use of the planned investment.  
In particular, the depth
should be sufficient for any possible future use of these cavities or
the level which will be developed for these large structures.

 Along with these physics justifications there are practical issues
 regarding the existing infrastructure at Homestake and also the
 stress characteristics of the Homestake rock formations.
 In this
 report we have examined the various depth choices at Homestake from
 the point of view of the particle and nuclear physics signatures of
 interest. We also have sufficient information about the existing
 infrastructure and the rock characteristics to narrow the choice of
 levels for the development of large cavities with long lifetimes. We
 make general remarks on desirable ground conditions for such large cavities
 and then make  recommendations on how to start examining
 these levels to make a final choice. 
In the appendix 
we have outlined the initial requirements for the detectors. These requirements 
will undergo refinement during the course of the design.

The depth requirements associated with the various physics processes
considered in this report are listed in
table~\ref{tab:depth_summary} for water Cherenkov and liquid
argon detector technologies.
While some of these physics processes can be adequately studied at
shallower depths, none of them require a depth greater than 4300~mwe
which corresponds to the 4850~ft level at Homestake. It is very important to 
note that the scale of the planned detector is such that 
even for accelerator neutrino detection (which allows one to use 
the accelerator duty factor to eliminate cosmics) a minimum depth is needed 
to reduce risk of contamination from cosmic rays. 
After consideration of both the science  
and the practical issues regarding the Homestake 
site, 
we strongly recommend that
 the geotechnical studies be commenced at the 4850ft level, 
which we find to be the most suitable, in a timely manner.

\end{abstract}

\maketitle

This document contains figures in color.

This work was performed under the auspices of the U.S. Department of
Energy, Contract No. DE-ACO2-98CH10886 and Contract
No. DE-AC02-05CH11231 and No. DE-AC02-07CH11359.

This report was prepared as an account of work sponsored by an agency
of the United States Government. Neither the United States Government
nor any agency thereof, nor any of their employees, makes any
warranty, expressed or implied, or assumes any legal liability or
responsibility for the accuracy, completeness, or usefulness of any
information, apparatus, product, or process disclosed, or represents
that its use would not infringe privately owned rights. Reference
herein to any specific commercial product, process, or service by
trade name, trademark, manufacturer, or otherwise, does not
necessarily constitute or imply its endorsement, recommendation, or
favoring by the United States Government or any agency thereof. The
views and opinions of authors expressed herein do not necessarily
state or reflect those of the United States Government or any agency
thereof.

\bigskip 

First Edition 23 December 2008.  

\bigskip 

Second Edition 23 July 2009. 

For the second edition, the major change was a 
changed and clarified description of levels in 
Table \ref{levels1}. The author list was expanded to 
correspond to additional contributions to this document. 
There were a few other minor changes to 
 plots and captions. 

\newpage

\newpage  

\tableofcontents

\newpage

\pagenumbering{arabic}

\section{Cosmic ray muon rate in Homestake DUSEL  }

\label{sec:cosmics} 

The most important reason for locating sensitive detectors deep
underground is to eliminate the background events caused by cosmic ray
muons that originate in the atmosphere of the Earth.  We follow the
PDG \cite{pdg} to briefly summarize the rate of cosmic ray muons as a
function of depth.  Muons are the most numerous cosmic ray charged
particles at the surface of the Earth. They are produced in the upper
atmosphere by the collision of cosmic ray primaries (protons, and
nuclei); and they lose about 2 GeV in the atmosphere before reaching
the surface.  The integral intensity of vertical muons above 1 GeV/c
at sea level is $\sim 70 m^{-2} s^{-1} sr^{-1}$. The energy spectrum
is flat below 1 GeV; it steepens gradually from 10 to 100 GeV, and
then it steepens further beyond 100 GeV. The muon spectrum structure
reflects the energy spectrum of the primaries as well as the energy
dependence of the pion interaction cross section in the atmosphere.
The energy-averaged angular distribution of muons at ground level is
$\sim cos^2 \theta$ where $\theta$ is the angle with respect to the
vertical.  Low energy muons have a steeper angular dependence, whereas
high energy ones have a flatter dependence.

\begin{figure}
\centering\leavevmode
\includegraphics[angle=0,width=0.9\textwidth]{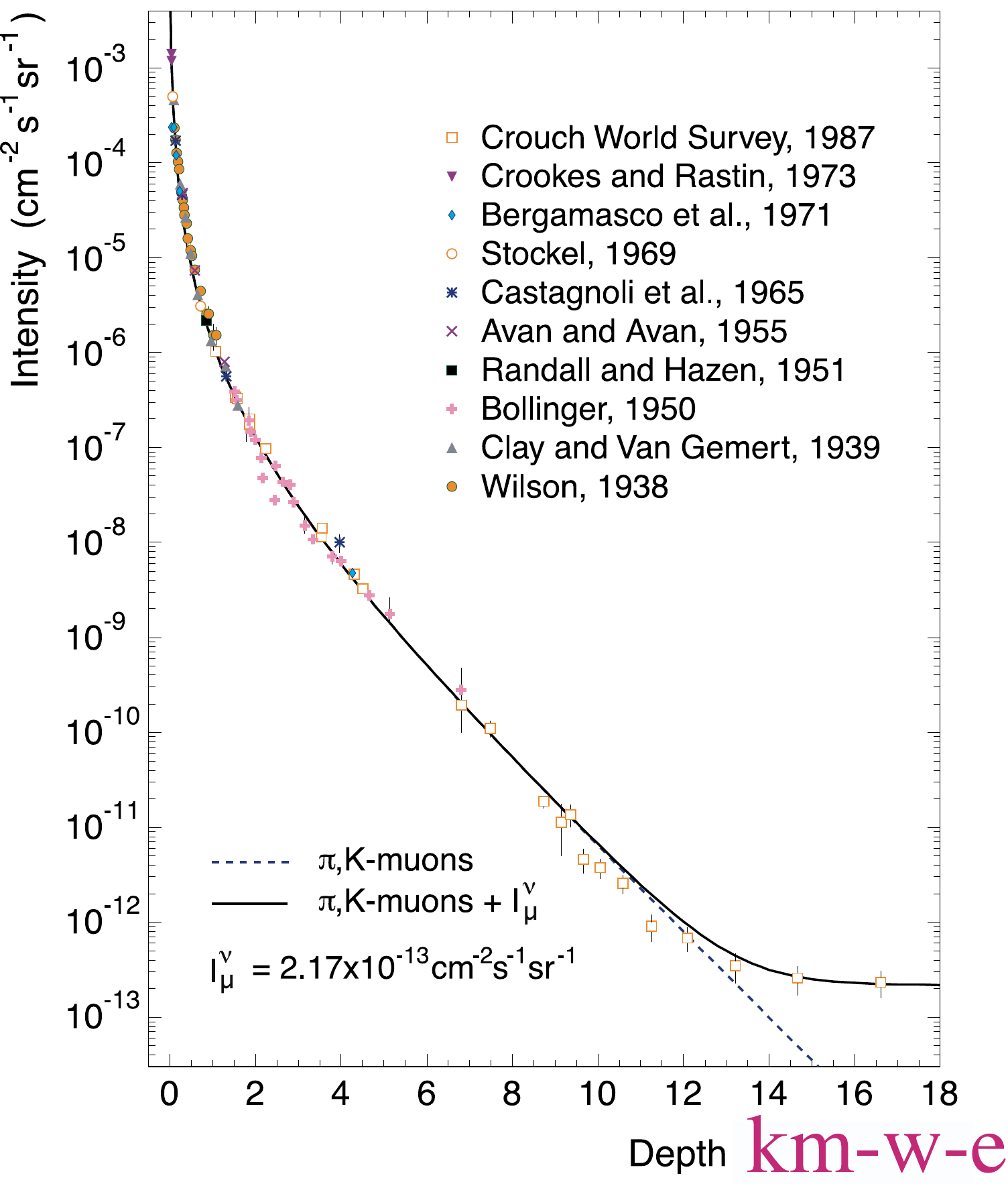}
 \caption{(in color)
The Crouch world survey \cite{crouch} 
 of vertical muon intensity  versus depth.
The unit of depth is km-w-e or kilometers water equivalent. 
 The quantity $I_\mu^\nu$ is the 
flux of muons from atmospheric neutrino interactions in the rock. 
It becomes the dominant contribution beyond 13 km-w-e. 
  \label{crouch} }
\end{figure}

\begin{figure}
\centering\leavevmode
\includegraphics[angle=0,width=0.75\textwidth]{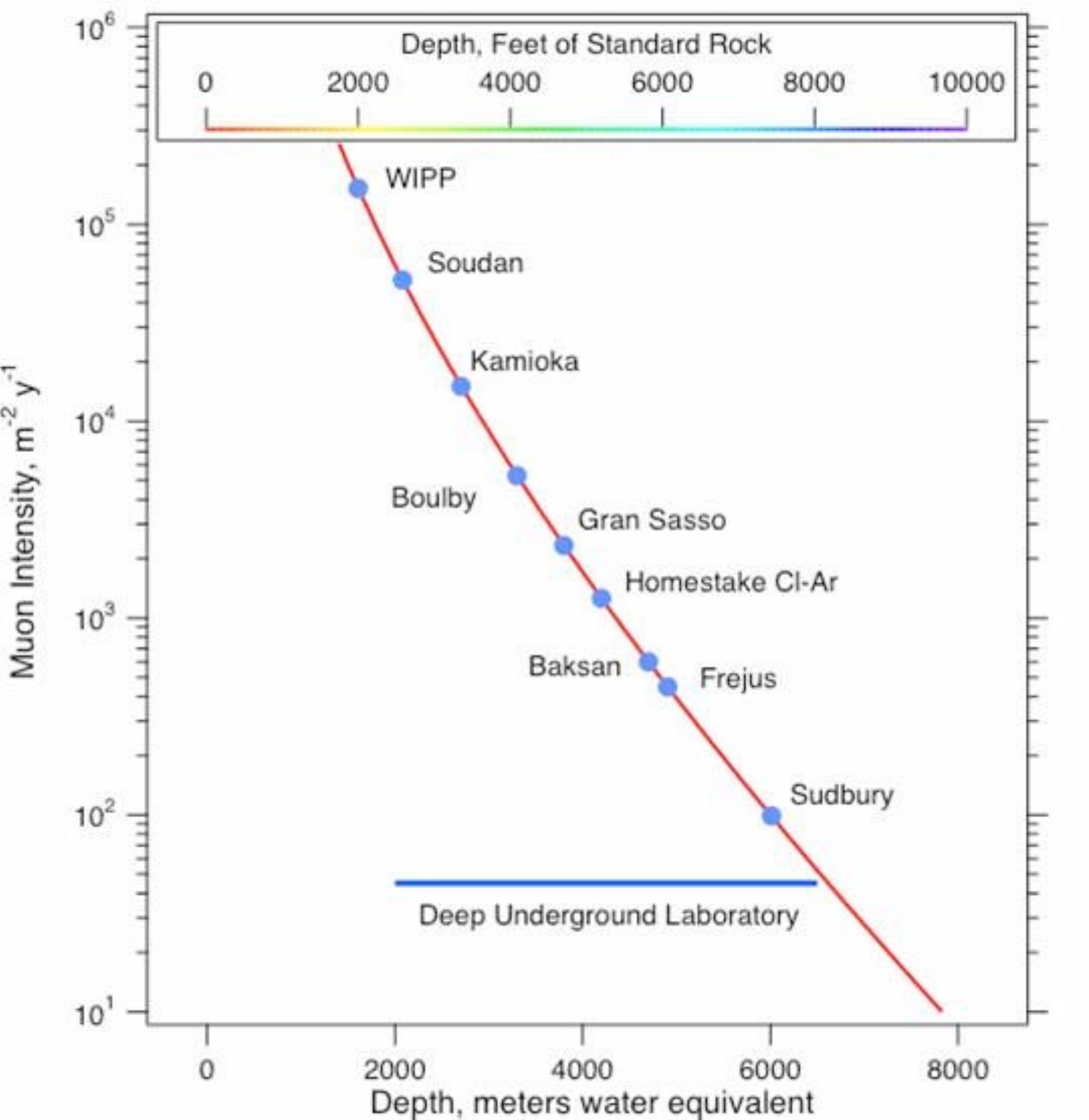}
 \caption{(in color)
Comparison of various sites in terms of the muon flux. The various
 interesting levels
in the Homestake site are indicated\cite{nrcrep}. 
  \label{crouch2} }
\end{figure}

Only muons and neutrinos penetrate to significant depths
underground. The muons produce tertiary fluxes of photons, electrons,
and hadrons.  The goal of the underground laboratory is to reduce all
such sources of backgrounds by shielding the detectors under rock.  The
shielding is commonly expressed as either ft of standard rock (with
density of 2.65 gm/cc) or in meters-water-equivalent (mwe).  As muons
penetrate underground they lose energy by ionization and by radiative
processes. One can calculate the rate of muons underground by using a
model for the surface flux and a simulation of muon traversal in the
rock.  A number of reviews exist that have details of such
calculations \cite{pdg, gaisser, bugaev}.  A detailed compilation of
muon rate data as a function of depth exists from \cite{crouch} and is
shown in figure \ref{crouch}.  The shielding at various underground
laboratory locations is shown in figure \ref{crouch2}\cite{nrcrep}.

An accurate calculation of the muon rate and the energy spectrum at
any location within the Homestake mine is possible, but it will
require careful modeling of the surface features above the chosen
location.  For example, the Davis chamber (of the Chlorine experiment)
was determined to be at an effective shielding depth of 4200
meters-water-equivalent by examining the density of rock above the
site ($2.9 gm/cc$) and the depth of rock along several angular paths
\cite{lande}.  Such detailed modeling is underway, but will not be the
subject of this report. For the purposes of this report, we have
assumed a flat overburden equal to the depth of rock above a given
level in the mine with rock density of $2.9 gm/cc$. Because of surface
features at Homestake the overburden could have an error of as much as
200 mwe corresponding to an error of $\pm$30\% in muon rate at the
4850 ft level (see figure
\ref{crouch2}\cite{nrcrep}). 
This is sufficient accuracy to determine the depth required 
for the physics goals given here. 

Table \ref{murate} shows the calculation of muon flux  as a
function of depth assuming a flat overburden. 
The depth levels chosen in the table
correspond to the levels that are discussed in Section
\ref{geotech}. The average muon energy also needs to be considered for
some calculations that involve muon interactions with rocks; it
increases with depth from $\sim50$~GeV at shallow depths to $\sim300$~GeV
for depths greater than 3000 mwe.

\begin{table}
\begin{tabular}{|l|l|l|l|}
\hline
Homestake &     Depth &     Rate      \\
depth (ft) &  (m.w.e.)    &   $m^{-2}s^{-1}$  \\
\hline
300 &    265   &    0.75  \\
1000 &  880    &  0.10  \\
\hline
2600 &  2300   &  $1.3\times 10^{-3}$   \\
3350   &  2960  & $3.3\times 10^{-4}$  \\
\hline
3950 &  3490   & $1.5\times 10^{-4}$     \\
4100  & 3620     & $6.7\times 10^{-5}$   \\
\hline
4850  & 4290    & $2.3\times 10^{-5}$      \\
\hline
\end{tabular}
\caption{Muon rate 
as a function of depth assuming a flat overburden.
 The actual effective shielding
depth of the various Homestake levels
depends on the rock density and surface topography.
For the 4850 ft level,
we expect a variation in rate of $\sim 30\%$. \label{murate}}
\end{table}

\newpage

\section{Detector Technologies}

In this section we briefly describe the technique of a 
water Cherenkov detector and a liquid argon time 
projection chamber. These are the two technologies under consideration 
for building a very large detector in DUSEL.  We will outline 
how these detectors work and nature of the cosmic ray 
and neutrino  signals from these devices.

\subsection{Water Cherenkov Detector  }

Large volume water Cherenkov detectors have been 
operated very productively in particle physics for over 25 years.
 The first large scale water Cherenkov detector was the IMB detector,
 constructed in a  salt mine in the United States, which began
 operation in the early 1980's. Following closely on the IMB, the
 Kamiokande detector, built in a zinc mine of Japan, began operations.
 Both detectors' original purpose was primarily a search for nucleon
 decay. However, these detectors went on to make important
 contributions to particle physics with measurements of the
 atmospheric neutrino flux in the GeV energy range. At Kamiokande, the
 detector's energy threshold was successfully lowered far enough to
 enable ground breaking measurements of the lower energy solar
 neutrinos in the 10 MeV range using Kamiokande-II.  The more recently
 constructed big brother of Kamiokande, Super-Kamiokande, has gone on
 to make important contributions in nucleon decay searches and
 neutrino oscillation physics. Using atmospheric neutrinos,
 Super-Kamiokande published the first definitive evidence of neutrino
 flavor oscillations, and therefore non-zero neutrino mass and lepton
 flavor violations, in 1998.

Some of the virtues of water Cherenkov as technology for massive
detectors is the low cost, relative simplicity of design and ease of
operation. The active target medium is water, which provides a very
abundant, very cheap and easy to handle source for the target material
with which to build the massive detectors required for the physics
being explored.  The wall of the water container is instrumented with
photomultiplier tubes (PMTs) whose signals are readout  with
well understood electronics, which includes charge to digital converters
and time to digital converters. The PMT readouts are then
used to analyze the arrival time and the number of photons produced by the
Cherenkov radiation of charged particle tracks in the water and
detected by the PMTs to reconstruct vertex, direction and energy of
the track.

Cherenkov photons are generated in water when a charged particle has
velocity greater then the speed of light in water $c/n$, where $c$ and
$n$ are the vacuum speed of light and index of refraction of water
respectively. These Cherenkov photons are emitted in a cone around the
direction of the charged particle (with charge z) track with a half angle,
$\theta_{c}$ given by:
\[
\cos \theta_{c} = \frac{1}{\beta n(\lambda)}
\]
where $\beta$ is the particle's velocity with respect to $c$, and
$\lambda$ is the wavelength of the Cherenkov light. For highly
relativistic particles ($\beta \sim 1$) and for the nearly pure water
in these detectors $n \approx 1.33$ in the wavelengths of sensitivity
for the PMTs resulting in a Cherenkov angle of $\theta_{c} =
42^{\circ}$.
The number of Cherenkov photons emitted per unit length (x) traveled per
unit photon energy is given by: \[ \frac{d^2 N}{dEdx} = \frac{\alpha
z^2}{\hbar c} \sin^{2} \theta_{c}
 \; \approx \; 370 \: z^2 \, \sin^{2} \theta_{c} \; eV^{-1} cm^{-1} \]
 For a highly relativistic particle of unit elementary charge
 traveling in water, several hundred Cherenkov photons will be
 generated in the wavelength range of PMT sensitivity per centimeter of
 travel.

\begin{figure}[htbp] 
   \centering \includegraphics[width=2.5in]{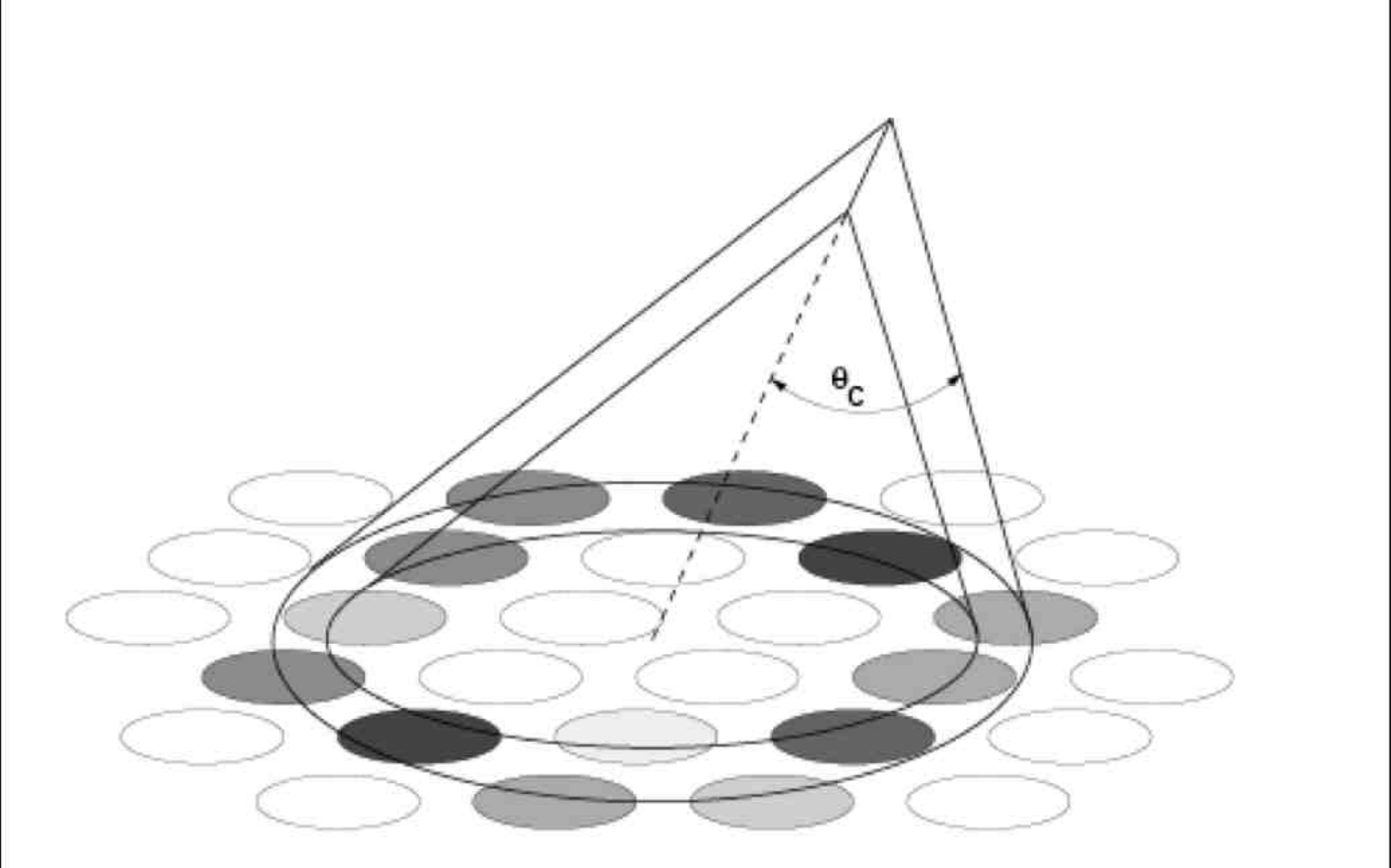}
   \caption{The Cherenkov light generated by the charged particle's
   path inside the detector arrives at the detector wall, where it
   produces a ring pattern that can be used to reconstruct the track's
   geometry and energy.}  \label{fig:cone}
\end{figure}

Water Cherenkov detectors use the nature of the Cherenkov light
emission described above in a technique called Cherenkov ring
imaging. The cone of Cherenkov light produced by the particle's path
inside the water volume of the detector travels through the clear
water volume and arrives at the detector wall, where it produces a
ring pattern. The PMTs lining the walls of the detector detect this
light pattern. The pattern is used 
 to uniquely reconstruct the geometry (vertex,
direction and ending point) of the particle's path as well as estimate
the energy and identify the type of the particle
(Figure~\ref{fig:cone}).

\begin{figure}[htbp] 
   \centering
   \includegraphics[width=5in]{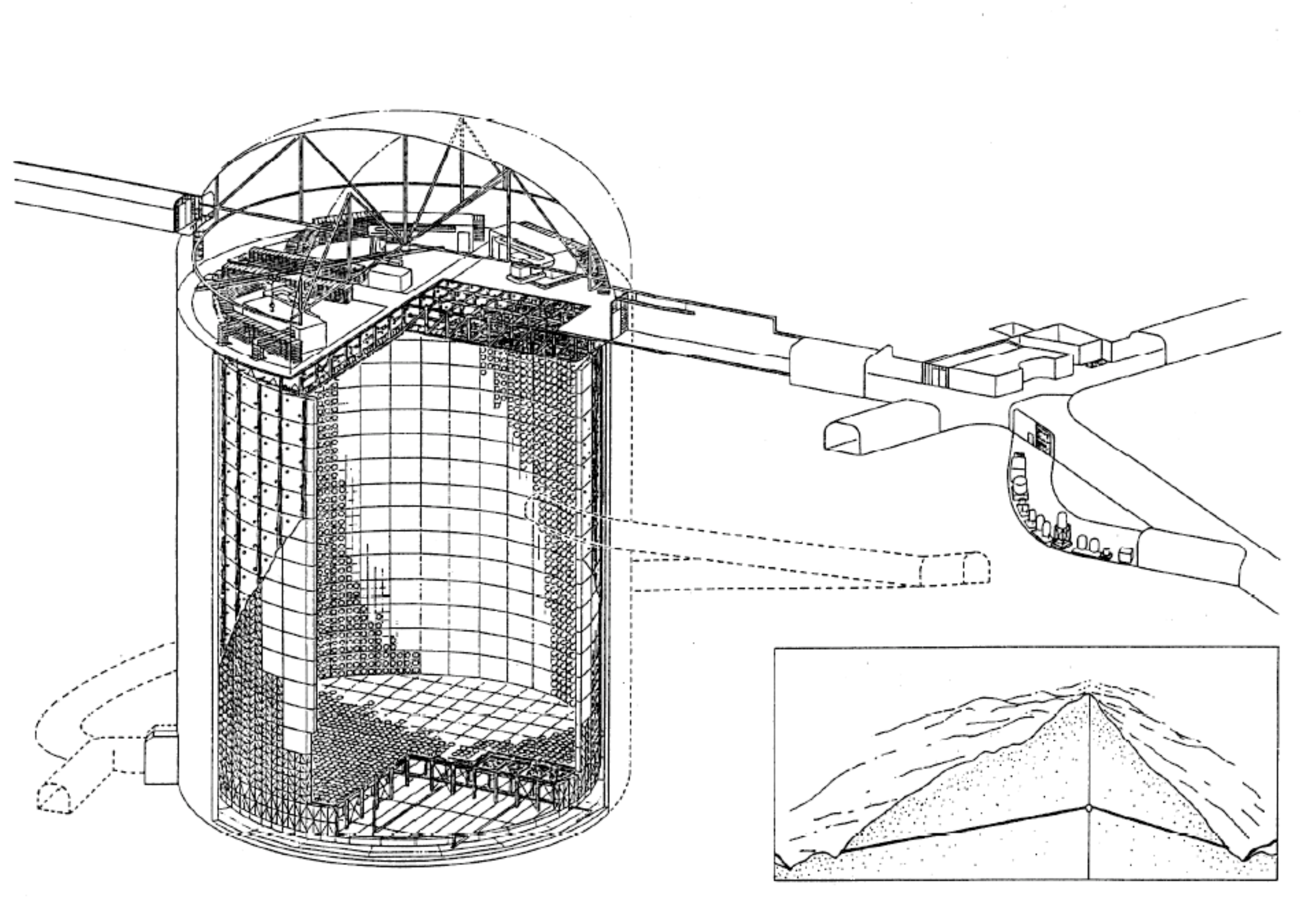} 
   \caption{Schematic view of the Super-Kamiokande 50 kiloton water Cherenkov detector in Japan. The detector is accessed by vehicle through a 2 kilometer long tunnel \cite{sknim}.}
   \label{fig:SK}
\end{figure}

The largest operating water Cherenkov detector, with a completely
man-made detector volume, is the Super-Kamiokande detector in
Japan. The Super-Kamiokande detector is located in a zinc mine
approximately 1 km deep inside a mountain (2700 mwe). 
The detector volume is a
cylinder approximately 41m high and 39m in diameter holding 50~kTon of 
highly pure water. The walls of active inner region of the
detector are lined with more then 11,000 PMTs (each with 50 cm diameter),
making about 40\% of the wall surface sensitive to Cherenkov
photons. The detector is illustrated in Figure~\ref{fig:SK}.

\begin{figure}[htbp] 
   \centering \includegraphics[width=2.75in]{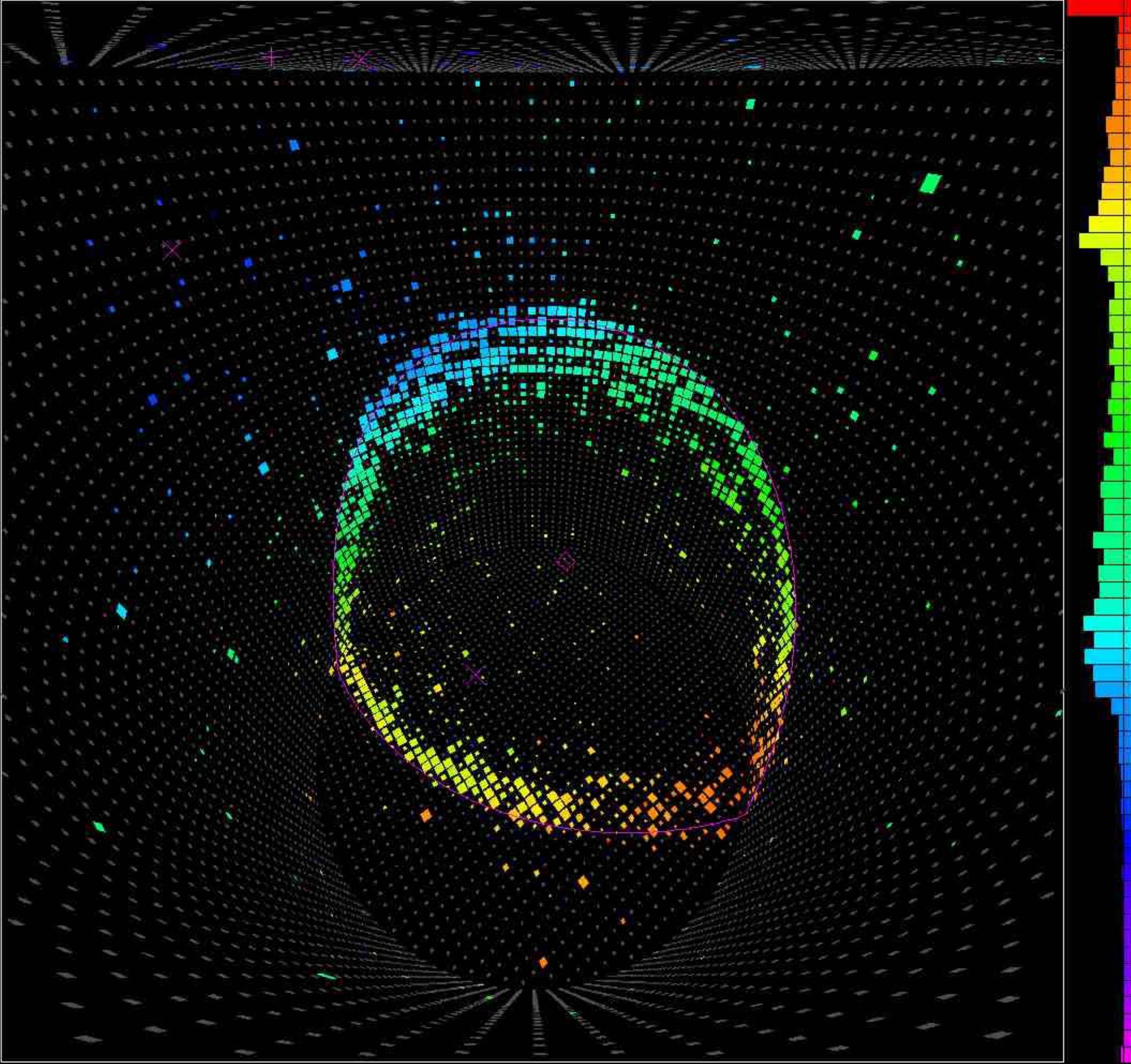}
   \includegraphics[width=2.75in]{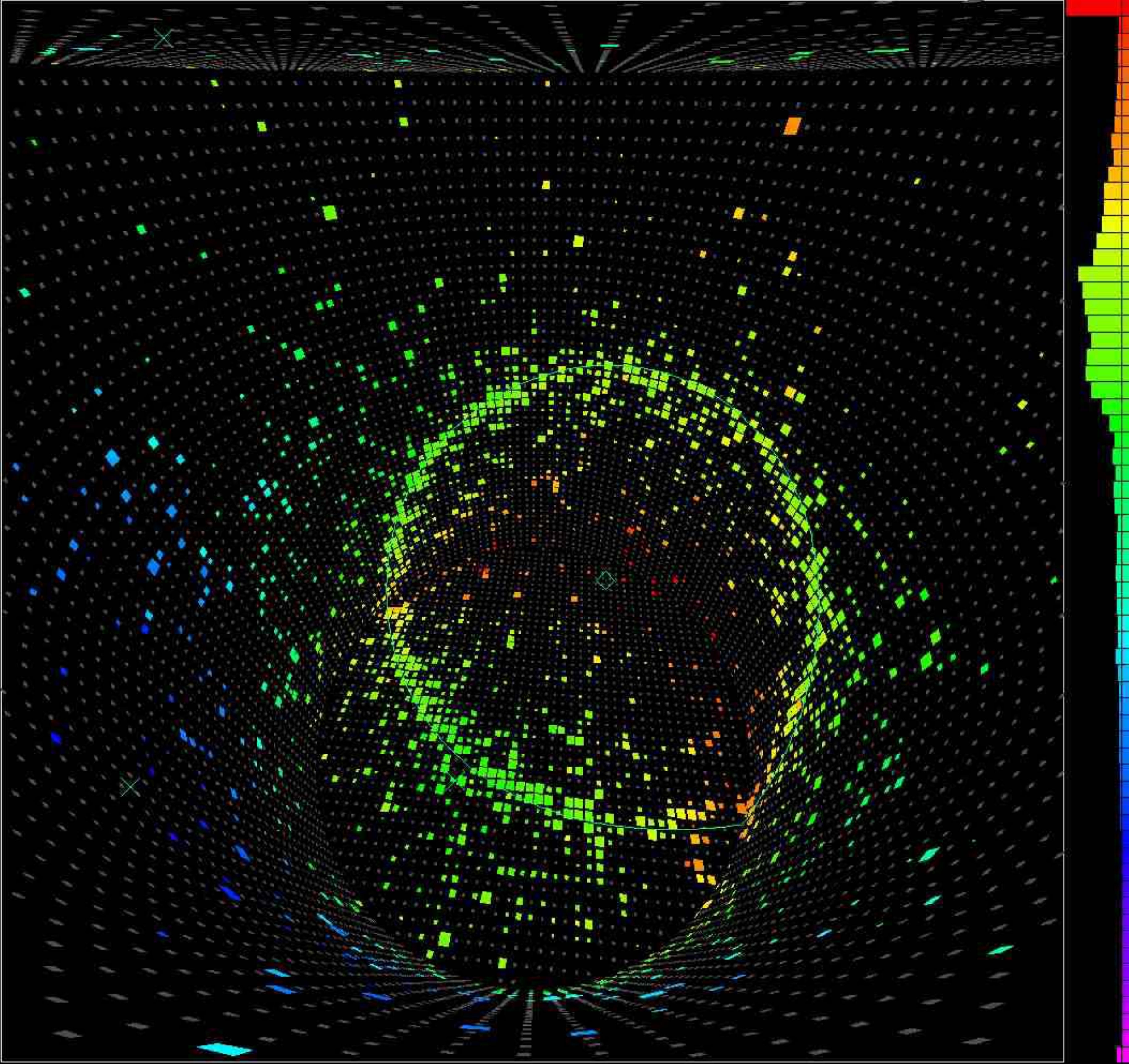} \caption{Two
   simulated events displayed for the Super-Kamiokande detector. Left: a muon
   event. Notice the cleaner outer ring of the Cherenkov cone. Right:
   an electron event. Notice that the ring is much more ragged due to
   the many particles of the electromagnetic shower and multiple
   scattering of the shower particles.\cite{skcollab}  \label{fig:partID}}
\end{figure}

Water Cherenkov detectors use the Cherenkov ring imaging technique in
order to search for and measure various physics processes that can
occur within the detector volume. For example, a classical mode of
proton decay that would be searched for is:
\[
p \rightarrow e^+ \pi ^0 \rightarrow e^+ \gamma \gamma
\]
where the gammas are of sufficient energy that they interact within a
radiation length or so and produce an electromagnetic shower similar
to an electron. The signature in the detector would therefore be three
electron-like tracks, with two of the tracks reconstructing to the
$\pi ^0$ mass. Neutrino events would be detected by measuring the
particle tracks resulting from neutrino interactions within the
detector volume, such as the charged current processes:
\[
\nu_{l}  + N \rightarrow l^{-} + X
\]
\[
\overline{\nu_{l}} + N \rightarrow l^{+} + X
\]
The direction, energy and flavor of the incoming neutrino ($\nu$) is
indicated by measuring the direction, energy and flavor of the lepton
($l$) produced by the interacting neutrino. Muons can be distinguished
from electromagnetic showering particles, such as electrons and
gammas, with high efficiency using the morphology of their respective
Cherenkov cones. Muons undergo very little multiple scattering and
therefore travel straight and produce a neat outer edge to the ring
projected onto the detector walls. In contrast, a particle such as an
electron or gamma produces an electromagnetic shower of multiple
particles, many of which undergo some multiple scattering as they
travel through the water, thus causing a very ragged Cherenkov light
cone on the detector walls. Figure~\ref{fig:partID} illustrates this
difference for muons and electrons with event displays for both in the
Super-Kamiokande detector.

A cosmic ray muon in a water Cherenkov detector will leave a distinctive 
signature. Because of the high energies of cosmic ray muons at the depths 
of interest, a large fraction of the muons will penetrate completely 
through the detector leaving very large deposits of energy or light. 
Generally, on the average, half of the photo-multiplier tubes will detect 
some light from such events. The ring pattern from these muons will be 
completely filled with large deposits of light at photo-tubes near the 
exit points of these muons. Muons that stop in the detector will either 
be absorbed by the oxygen nuclei or decay. The decay will create a
low energy electron signature sometime later after the muon stops (with 
lifetimes of 2.2 microseconds). Muons can also undergo inelastic  
interactions in the rock surrounding the detector or in the detector.
Such events can create neutrons that have delayed hits in the detector.
In addition, muon interactions can create light radioactive nuclei that will 
decay (with wide ranging livetimes) mainly by beta-decay. These spallation 
product beta decays can cause backgrounds to low energy ($\le 10$~MeV) 
neutrino events. 
Depth will reduce the rate of muons as well as 
the rate of all events associated with the muons.  A complete review is in 
\cite{formagio}.  

After traversal of a 
 cosmic ray muon the photo-multiplier tubes and the electronic readout 
chain will require some time to recover (generally in the range of 
$\sim$100 ns). This will cause of loss of data for more interesting 
events such a nucleon decay or neutrinos. The muon, if not properly 
reconstructed could also cause background. The quantification of 
this data loss and backgrounds will be in section \ref{physics}.

\subsection{Liquid argon TPC  }

Liquid argon time projection chambers (LArTPCs) record 3 dimensional
``photo-like'' images of passing particle tracks along with the energy
deposited by those tracks.  The few-millimeter-scale spatial
granularity of a LArTPC combined with energy at each step make it a
very powerful detection technique. This technique, pioneered by Carlo
Rubbia \cite{rubbia} and the ICARUS collaboration\cite{t600}
 in Europe, has been tested at the
300 ton scale with successful operation above ground of one module of
the ICARUS T600 detector.  Modifications to the T600 design to scale to
larger sizes that can be build underground are under study in Europe and 
the US where a staged program of LArTPC detectors is underway.

\begin{figure}[htb]
  \centering \includegraphics[width=3.8in]{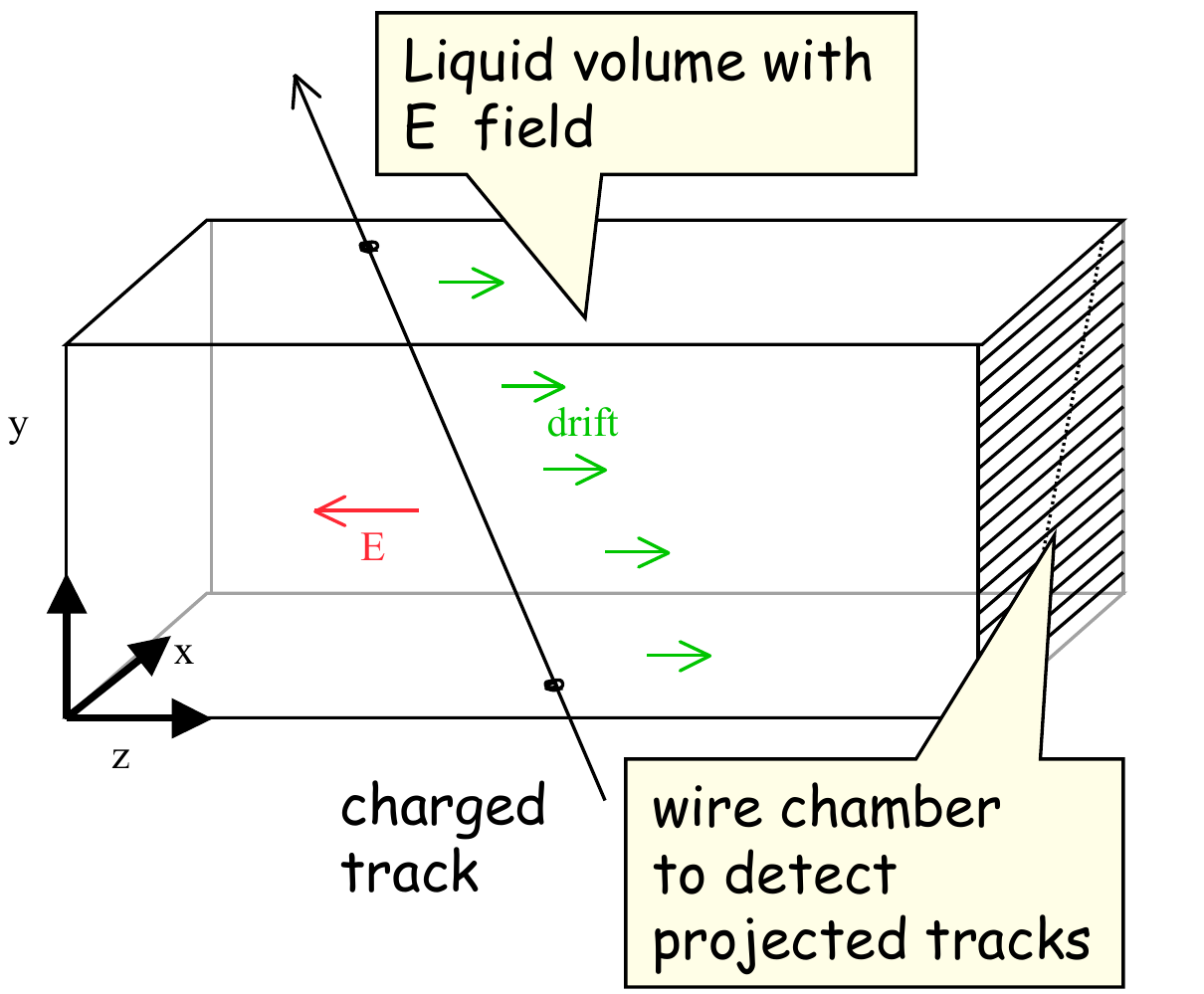}
  \caption{
Schematic of the functioning of a liquid argon time projection chamber. 
The multi wire proportional chamber (MWPC) reads out the x/y position 
of drifted ionization while the time of the hit allows the 
determination of the 
z distance from the plane from the knowledge of the drift velocity.}  
\label{fig:illustration}
\end{figure}

In a time projection chamber ionization, electrons from passing charged
particles are drifted by a strong electric field in ultra pure liquid
argon to the edge of the detector.  A series of wire chamber readout
planes then record the passing charge. The time of the charge at the
wire plane location is also recorded. From the knowledge of the time
and the position on the flat wire plane a 3 dimensional picture of the
event can be reconstructed.  The technique to read out the ``shadow''
of the event is illustrated in Figure~\ref{fig:illustration}.

\begin{figure}[htb]
\vskip 0.cm
  \centering
  \includegraphics[width=3.8in]{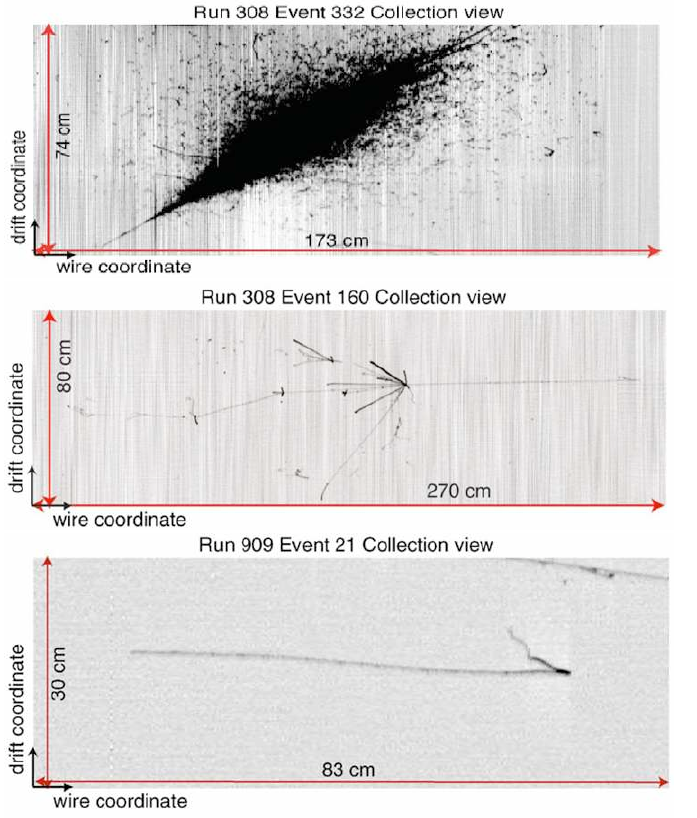}
  \caption{Events collected by the ICARUS test run. An electromagnetic 
shower (top), hadronic interaction (middle), and a muon decay (bottom). 
\cite{t600}
}
  \label{fig:tracks}
\end{figure}

The granularity of track sampling depends upon the distance between
readout electrodes on the wire chamber planes which is typically
3-5mm. The final of the typically three readout planes collects the
passing charge to record the deposited energy at each step.
Figure~\ref{fig:tracks} shows a few examples of events 
 in the ICARUS test detector.  The granularity of the detector
allows for these detailed  images, and the differing intensity of
the tracks shows  the energy deposition measurement.

The topology of the events and the $\frac{dE}{dx}$
measurement can be used  to differentiate signal from background for
neutrino physics measurements and proton decay.
For example, a cosmic ray muon will be seen as a clear incoming track,
whereas a neutrino event will be a track that originates inside the 
chambers. As a more complex example,  
 single electrons from charged current interactions of electron type neutrinos 
can be differentiated from single gamma rays  from mis-identified
interactions of the muon type neutrino 
by using  the overall event topology and the energy
deposition in the first few centimeters of the events.  This is
difficult to achieve in detectors with limited spatial resolution 
near the vertex.  
Single high energy gamma rays  will produce electromagnetic 
showers that are almost indistinguishable from electron induced 
gamma rays except in detectors with very fine granularity. 


We expect that the fine resolution of the LArTPC will allow very high
efficiency for electron neutrino selection compared to a water
Cherenkov detector. The combination of signal efficiency and
background rejection 
makes  the LArTPC 
 more sensitive to long baseline oscillation physics than a water
Cherenkov detector, so that 
the detector 
mass needed for liquid argon to reach the same 
sensitivity is less by a factor of 
3 to 6, than for a water Cherenkov detector. 
Similarly, in the case of proton
decay, LArTPCs are sensitive to $p \rightarrow
\bar\nu K$ by detecting and identifying the final state 
kaon by its  $\frac{dE}{dx}$ signature.  
The nuclear  re-interactions of the kaon in the Argon nucleus are expected to 
broaden the final state momentum distribution, but the loss in detetion 
 efficiency is 
expected to be small $\sim4\%$ \cite{bueno}. 
 The kaon is invisible in a water
Cherenkov detector because it is below Cherenkov threshold.  It is
expected that the LArTPC's have high efficiency to this decay.  The
water Cherenkov detector is likely to have much more mass than
LArTPCs; nevertheless the high efficiency will allow a LArTPC to have
equal or better sensitive to this particular decay mode.

A water Cherenkov detector  can be scaled up for large mass, and  
it has demonstrated high  dynamic range in energy, and extensive 
experience in construction. The liquid argon TPC needs extensive R\&D
to demonstrate how to scale it up to the needed 50~kTon scale. 
Nevertheless, 
it could have unique capability because of the expected 
high efficiency for important physics goals. Therefore, the two technologies 
are considered complementary.

With a drift speed of about a meter per millisecond, a LArTPC with a
3-5m drift, as envisioned for these detectors, will have 3-5ms of data
to read out, per event.  Coincident with the event of interest will be
passing cosmic ray background events.  While these can in principle be
rejected as background via their topology -- that they enter from
outside the detector, they are a background to consider if there are
many to reconstruct and if they overlap an event of interest.  Both of
these factors are mitigated by overburden to reduce the overall cosmic
background rate.  How much overburden is needed depends upon the
signal process, as described in the sections below.

For the long baseline experiment, very large detectors are
needed to obtain sufficient events for the physics of 
CP violation.  The key issues for construction,
installation, and operation of the very large LArTPCs envisioned are
\begin{itemize}
\item Achieving and maintaining the required purity in the large,
  non-evacuable cryostats housing the TPCs.
\item Development of cold, 
low-noise electronics with multiplexed readout in the
  detectors.
\item Underground construction of caverns for the LArTPC modules and
  safety features required for the large volumes of cryogenics needed.
\end{itemize}

A program of LArTPC development to address these questions is
underway in the US. What is learned from the R\&D components of this
program will guide the design, construction, and installation of an
initial 5~kTon and later additional 25~kTon detectors at DUSEL.

Details on the physics sensitivity of the liquid argon detector for nucleon
decay, neutrino physics, and astrophysics at a given depth is
presented below along with some  issues related to construction and
operation of the detector underground.

\newpage

\section{Depth requirements for physics }
\label{physics} 

In each of the following subsections we examine important physics
signatures in the two types of detectors and how they are affected by
the depth of the detector.  We will generally rely on previously
published reports and other material and will not attempt a complete
review. Our intention is to arrive at criteria that do not overly
depend on detailed software analysis or reconstruction of events for
setting the depth requirement. Detailed event reconstruction
capability will depend on detector technology and the large number of
decisions regarding the design of the detector and electronics. A
conservative approach to evaluating cosmogenic background is to rely
mainly on measurements such as total energy, time, and position in
fiducial volume to distinguish background from signal. If the
background rate is satisfactory with such considerations, then a more
detailed analysis using improved methods is likely to allow additional
reduction of background rates.

\subsection{Accelerator neutrinos }

In this section we briefly discuss the overburden issue in the
context of accelerator  neutrinos. 
The event rate from a Fermilab based broad band neutrino beam has been 
extensively studied \cite{study}. There are still many beam optimization 
issues 
to be resolved, nevertheless the charged current muon neutrino event rate is
summarized for two possible beam choices in table \ref{accrate}. 
The total rate for a 1 MW beam operation is  $\ge$ 20000 events 
per 100 kTon (fiducial mass)  per year with very large effects due to
oscillations. 
\footnote{There will be $5.2\times 10^{20}$ protons on target if the 
beam power is 1 MW, proton energy 120 GeV, and running period of $10^7$ sec. 
We assume a running period of $2\times 10^7$ sec per year for our accelerator 
rate calculations. }
The fraction of muon neutrinos that convert to electron 
neutrinos will be small and depends on $\sin^2 2 \theta_{13}$, the CP angle,
and the mass hierarchy.  The measurement of these effects is one of the 
central 
goals of   this project. We do not address the sensitivity issues here. 
They are addressed in  detail in many reports\cite{study}. 
Once the cosmic ray background is made negligible for selection of 
neutrino events, cosmic rays will have no effect on the sensitivity.
It is, therefore,  
very important that the choice of depth be made in such a way as to 
completely eliminate the possibility of cosmic ray contamination of 
beam neutrino data. 

\begin{table} 
\begin{tabular}{|c|c|c|}
\hline 
Event type & 100 kTon & 100 kTon \\
Proton Beam Energy  &   120 GeV &  60 GeV \\
Angle &   0.5$^o$ &  0$^o$ \\
\hline 
CC $\nu_\mu$  &  27000 & 45000 \\
No Oscillations &    &  \\
\hline  
CC $\nu_\mu$ & 11400 &  21000 \\ 
With Oscillations &  &  \\
\hline  
\end{tabular} 
\caption{ Rate of accelerator muon neutrino beam events in a 
100kTon detector per year at Homestake with a beam from Fermilab. The details 
of the beam spectrum and running conditions can be found in \cite{bishai}. 
} 
\label{accrate} 
\end{table} 

The background rates in a large
detector due to cosmic rays have been calculated for both surface
and underground locations for a generic detector in the shape of a
cylinder.  The reduction of cosmic background can generally be
facilitated by: increasing the depth of the detector, event timing
with the beam pulse, and an active veto in conjunction with pattern
recognition software to remove incoming muon events. The
detector-related issues relevant to cosmic ray background are:
\begin{itemize}
\item the ability to handle the raw (depth-dependent) background
 event rate, and
\item the ability to reject background events efficiently.
\end{itemize}

A preliminary evaluation of both data acquisition rates and
background rejection capability without overburden leads to the
following conclusions:
\begin{enumerate}
\item  It is not possible to operate a large water Cherenkov detector
($> 50$~kT) on the surface.
\item A liquid argon TPC could be operated on the surface
during a short ($\sim 10\mu$sec) beam spill\cite{cosmicswriteup} if
high background rejection factors of $\sim 10^8$ ($\sim 10^3-10^4$)
for cosmic muons (photons) can be achieved.
\end{enumerate}
In general, the exceptional performance of a fine-grained tracking
detector such as a liquid argon TPC will enable a higher degree of
cosmic background rejection at any given depth of overburden.
Therefore, we expect that the water Cherenkov detector will require a
depth that is greater or equal to that of a liquid argon TPC.

\noindent
 \underline{\bf Water Cherenkov detector}

For a cylindrical tank of size 50 m height/diameter (approximately
100kT of water) the rate of cosmic muons (with momentum $>0.5$~
GeV/c) at the surface will be 250 kHz from the top plus 250 kHz from
the sides. This implies that during a 10 $\mu s$ beam spill there
will be an average of 5 muon tracks in the detector per spill. For a single
volume water Cherenkov detector in which the photo-multipliers are
mounted on the walls looking inwards, each muon on the average will
produce a hit in more than 50\% of the PMTs. Therefore, each cosmic
ray will produce enough light over a period of 
the crossing time through  the detector (200 ns for a 40 m length) that 
it will render  the entire detector ineffective for up to $\sim 1~\mu$sec. 
With a rate of
$0.5$~MHz at the surface the dead-time fraction is unacceptable. For
example, for a detector similar in technology to Super-Kamiokande,
the dead-time from the above event rates will exceed 50\%
\cite{sknim}. It is considered impractical  to mitigate this problem using 
fast pulse digitizers coupled with significant software and hardware
R\&D to resolve overlapping pulses to reconstruct multiple
simultaneous events with contained vertices. However, the
consequences of such electronics and analysis 
 for background rejection and resolution are at present
unknown.

Therefore, we will conservatively assume that sufficient overburden
is necessary to reduce the cosmic background to a manageable level.
The depth required to reduce the number of cosmic events during a
$10 \mu$sec beam spill to various levels is given in Table
\ref{intime}. A depth of at least $\sim 1000$ meters water
equivalent is needed to reduce the muon rate to a level comparable
to the rate of events from the neutrino beam so that minimal
dependence on pattern recognition (and a modest active veto
capability) is needed to separate beam related events.

\begin{table}
\begin{center}
\begin{tabular}{rrr}
Rate(Hz) &  In-time cosmics/yr & ~~~Depth (mwe) \\
\hline
500 kHz  &  $5\times 10^7$ &  0 \\
3 kHz  & 300,000	& 265 \\
400 Hz & 40,000	& 880  \\
5 Hz   & 500     & 2300 \\
1.3 Hz  &  130	& 2960  \\
0.60 Hz &  60	& 3490 \\
0.26 Hz  & 26	& 3620 \\
0.09 Hz &  9	& 4290 \\ 
\hline
\end{tabular}
\end{center}
\caption{The rate  of  cosmic ray muons in a 50 m height/diameter detector
 assuming a $\cos^2 \theta$ distribution (there will be a small correction at the 
deepest levels). 
The second column is the number 
in  10~$\mu s$ long pulses for $ 10^7$ pulses, corresponding
 to approximately
1 year of running, versus depth in meters water equivalent. In comparison,
1 year of running time with 1 MW of beam from FNAL will produce $\ge$20000
muon charged current 
beam neutrino events in a 100~kTon detector in the absence of 
oscillations depending on the detailed choices
of the beam \cite{study}.   Oscillations will reduce this number 
by a factor of $\sim$2. 
}
\label{intime}
\end{table}

\noindent
\underline{\bf Liquid argon TPC}

A 50 kT liquid argon TPC can be contained in a cylindrical tank of
size 35.5 m height/diameter; such a detector on the surface will
have a cosmic ray muon rate of 125 kHz from the top and 125 kHz from
the sides.
An examination of cosmic rays \cite{cosmicswriteup} in a liquid
argon TPC has considered their effects on data acquisition and event
reconstruction, and as a source of background.  The rate of cosmic
rays was shown to be tolerable  with the proposed drift-time ($\le 10$ ms) and
data acquisition system for cycles up to 5 Hz. In this scheme the
detector takes data in a short time interval (currently proposed to
be 3 drift times, or about 30 msec) near the beam time.
The high granularity of the detector should allow removal of cosmic
muons from the data introducing a small ($<0.1\%$) inefficiency to
the active detector volume, so that most of the accelerator-induced
events are unobscured. If a cosmic ray muon (photon) event mimics a
contained in-time neutrino event it must be rejected based on
pattern recognition. The rejection required is estimated to be $\sim
10^8$ for muon cosmics and $\sim 10^3-10^4$ for photon cosmics;
given the fine grained nature of the detector this rejection is
likely achievable using 
the incoming angle of the photons and 
by sacrificing fiducial volume at the edges, but still needs 
to be demonstrated by detailed simulations.

\subsection{ Improved Search for Nucleon Decay  }
\label{sec:pdk} 
The depth requirement for proton decay experiments is dominated by the
practical effect of livetime loss due to event overlap with cosmic ray
muons. This is particularly serious for water Cherenkov detectors,
where there is no current instrumentation or analysis that can
untangle illumination of the detector on timescales of order the time
it takes light to cross the detector, i.e. $\sim$220 ns for a 50-m diameter
detector. If we assume that the deadtime for each crossing muon, 
after inclusion of 
reflections and electronic effects,  is 1~$\mu$sec, then to achieve $\le$ 1\%
deadtime requires a rate of less than 10 kHz. 
Fortunately, even a modest overburden of order 1000 mwe (370
meters of rock) is sufficient to keep the deadtime due to cosmic ray
muon crossing well below 1\% (see Table \ref{murate} and Table \ref{intime}).
 The IMB experiment was successful with an
overburden of 1600 mwe. A liquid argon detector 
is very likely to have much less deadtime loss 
 at shallow depths, in this regard, as the fine segmentation
in space and drift time might allow one to exclude regions of the
detector around each passing muon. Bueno et al.\cite{bueno}
 estimate an effective
loss of detector mass of less than 4\% for a 100~kT liquid argon
detector with
mountainous overburden of only 200 m. Thus, based only on livetime
arguments we find that a proton decay detector must be underground,
although a depth of $<1000$ mwe is sufficient.

Further considerations regarding depth relate to specific signatures
associated with particular nucleon decay modes. For water Cherenkov
and liquid argon detectors, the mode $p \rightarrow e^+ \pi^0$ would
be fairly easy to distinguish, with similar efficiencies, at any
depth due to the significant visible energy and event topology. This
leaves atmospheric neutrino interactions of energy 1 GeV as the most
serious background for proton decay. Depth cannot  reduce
background due to atmospheric neutrinos.

The mode $p \rightarrow K^+ \bar\nu$ is detected in water Cherenkov
detectors using a more sophisticated analysis that combines detection 
of the kaon decay with  coincident
tagging of $6.3$ MeV gamma rays from the de-exitation of the 
remanent nucleus ($p_{3/2}$ excited state of $^{15}N$). This tag could 
 suffer at shallow depths.
Cosmic ray induced spallation events can mimic these gamma rays, and therefore 
all candidate events near in time with
 a muon need to be rejected.  The time window 
for the gamma ray near a candidate event is $\sim 30 ns$. To keep the 
inefficiency due to spallation $\le$1\%, the rate from spallation should be 
$< 300$ kHz. Even if one assumes 1 to 5  spallations per muon, such a rate 
can be easily achieved with modest overburden. 

\begin{figure}[h]
\includegraphics[width=0.49\textwidth]{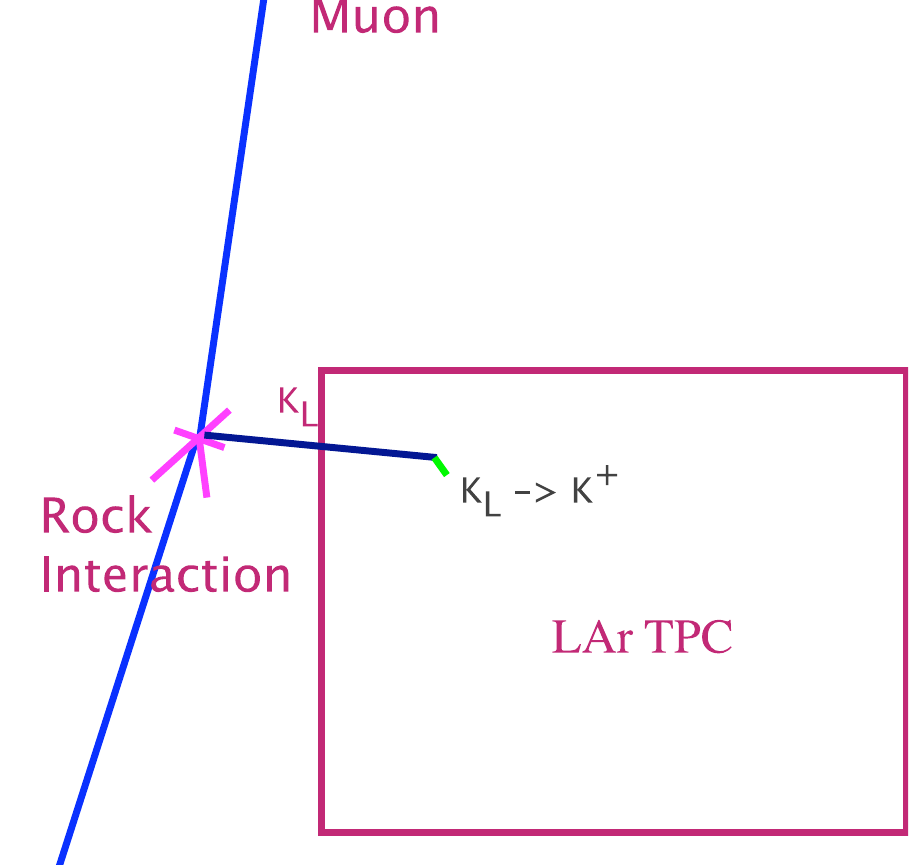}
\includegraphics[width=0.47\textwidth]{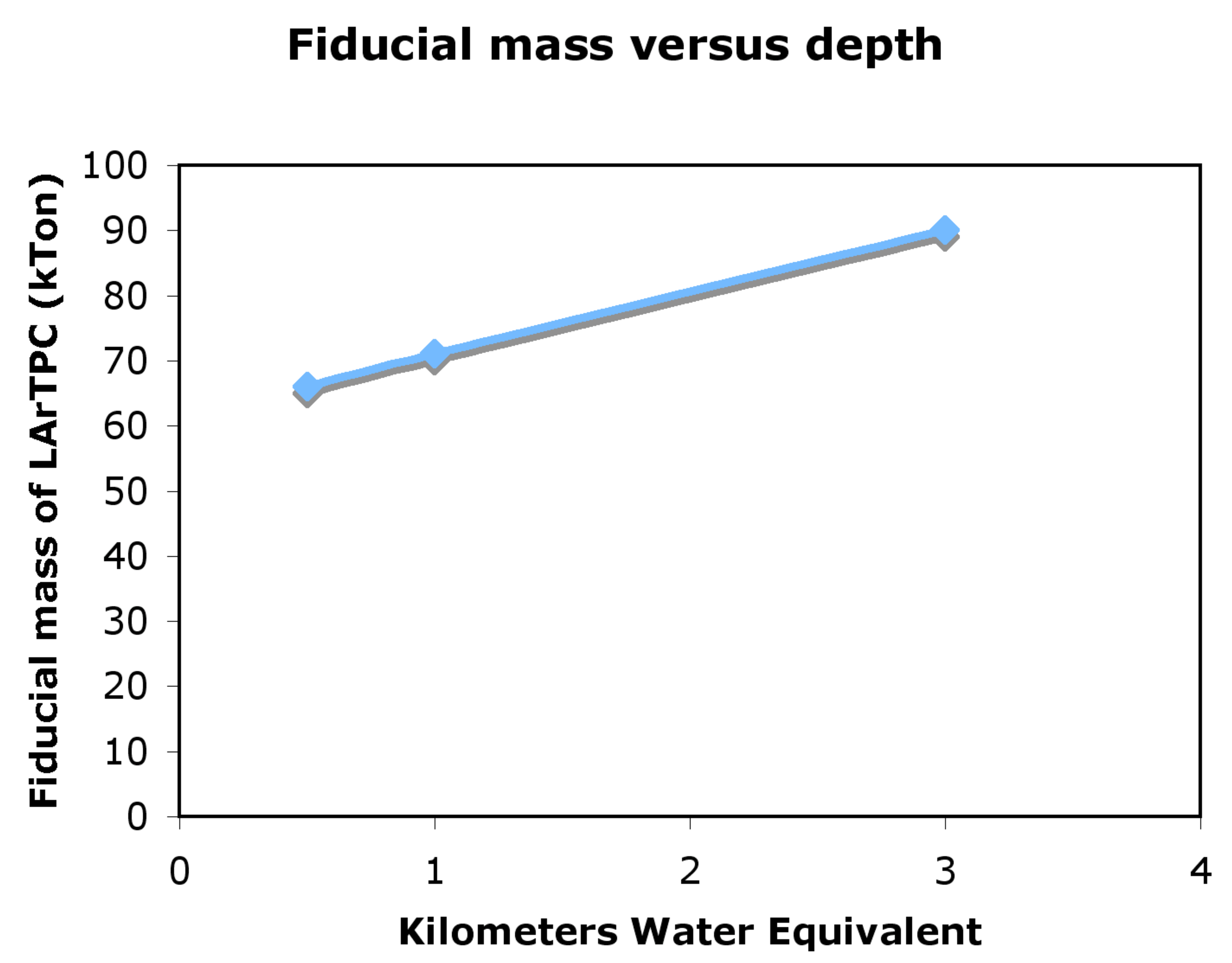}
  \caption{Schematic illustrating a possible background for the 
$p\to K^+ \bar\nu$ mode in which a neutral 
kaon is generated by muon interaction 
in rock (left). Right hand side shows the fiducial volume 
that can be retained to reject this cosmogenic background 
down to  0.1 events/year 
for a liquid argon TPC  with a total mass of 100 kTon in a single module
\cite{bueno}.  }
\label{pdkbck}
\end{figure}

However, for both water Cherenkov 
and LAr TPC detectors, for the $\bar\nu K^+$
mode,  a
potentially indistinguishable background proportional to the cosmic
ray rate appears. Nearby energetic cosmic rays may have photonuclear
interactions with the rock surrounding the detector and produce
hadrons including neutrons and $K^0_L$ that enter the detector. These
neutral particles evade any surrounding active veto and may interact
in the fiducial volume creating a contained vertex interaction that
can mimic proton decay. However, sacrificing fiducial mass
effectively shields against these interactions, which do not
penetrate to the center of the detector. The most troubling is a
charge exchange interaction of a $K^0_L$ producing a $K^+$. Bueno et
al.\cite{bueno} estimate a background of 0.1 events per year
background to $p \rightarrow K^+ \bar\nu$ at a depth of 3000 mwe, after
reducing the LAr fiducial volume from 100 kton to 90 kton. This
estimate is in agreement with an independent check by W. Morse
\cite{morse}.
Morse also considered the nucleon decay mode $n \to K_s \bar\nu$, 
which requires 10\% greater depth to reduce the background to an acceptable level. 
 Shallower depths decrease the effectiveness of LAr
mass, for example, 500 mwe (570 ft at Homestake) 
 would reduce the effective mass of a 100 kTon detector by
33\% compared to 3000 mwe (see figure \ref{pdkbck}). 
 This reduction in effective mass could be mitigated by an
active veto surrounding the detector\cite{bueno}.
Of course, passive background suppression is always preferable to 
active. If the liquid argon detector is built in 
modules smaller than 100 kTon, the loss in fiducial volume 
will be much greater.

In summary, proton decay, with signatures in the 0.1-1 GeV scale,
require some overburden but not the great depth needed for other
experiments such as dark matter and double beta decay that work at
much lower energies. From considerations of 
data-taking capabilities alone 
water Cherenkov detectors should be sited at a
depth of at least 1000 mwe. However, when considering 
potential backgrounds to the proton decay mode, $p\to K^+ \bar\nu $,
the optimum depth appears to be greater than 3000 mwe to maintain
background level of $<0.1$ event per year. This calculation is applicable 
to either technology.   
LAr detectors may be sited at shallower
depths, but with significant  loss in effective  mass. This loss is 
greater if the liquid argon detector must be built in modules smaller than
100 kTon.

\subsection{Observation of Solar Neutrinos  }
\label{sec:solar}

    Neutrinos from $^8$B decay within the Sun have been studied in
    great detail over the past decade  by the Sudbury Neutrino
    Observatory (SNO) and the Super-Kamiokande Collaborations.  With
    the additional reactor antineutrino disppearance measurements by
    the KamLAND collaboration, it has become clear that at energies
    above 1~MeV, solar neutrino flavor transformation 
is dominated by 
     the Mikheyev-Smirnov-Wolfenstein (MSW) mechanism
    or `matter effect'.  Nevertheless, some of the most interesting
    predictions of the MSW mechanism have remained elusive, because
    the mixing parameters are in a region that makes much of the
    phenomenology unobservable by existing detectors.

    The most direct and convincing demonstration of the matter effect
    would be the observation of a change in the flavor content of a
    neutrino beam with and without intervening matter. Such a measurement will 
also result in an independent and precise  measurement of the mixing angle 
$\theta_{12}$. 
     The solar
    $^8$B neutrino beam provides us with just such a possibility:
    neutrinos from the Sun pass through the dense core of the Earth at
    night, and the difference between the forward scattering amplitude
    of $\nu_e$s and the other flavors leads to a flavor transformation
    similar to that which occurs within the Sun.  As the beam from the
    Sun arrives at the Earth, it is nearly a pure $\nu_2$ state and
    therefore its flavor content is only $\sim$1/3 $\nu_e$.  The
    flavor transformation within the Earth thus leads to a net gain in
    $\nu_e$ content -- the Sun `shines brighter' in $\nu_e$s at night
    than during the day.   

\begin{figure}[h]
\includegraphics[width=0.5\textwidth]{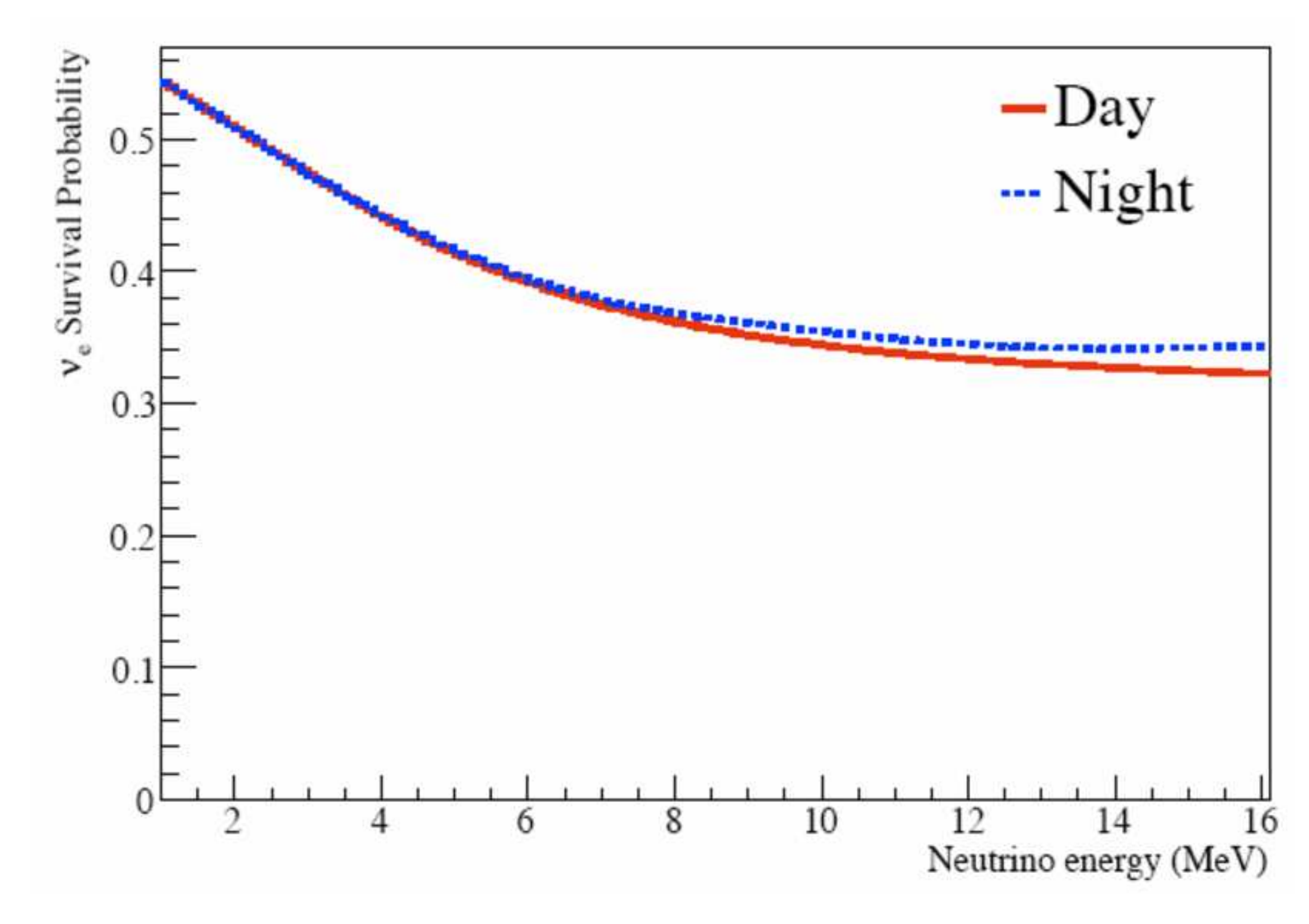}
  \caption{Electron neutrino survival probability as a function of energy, for
day and night~\cite{seibert}.\label{fig:psurvdn}}
\end{figure}

    Fortunately, for the best fit values of the mixing parameters, the
    Day-Night $\nu_e$ flux asymmetry is largest at 
    energies higher than 5 MeV. These energies are 
    accessible by a large detector with reasonable 
light collection ($\sim$ 30\% coverage with photocathode of 
20\% quantum efficiency) 
and no special 
requirements on the purity of detector materials. 
Figure~\ref{fig:psurvdn} shows the
    solar $\nu_e$ survival probability as a function of energy, for
    both `day' and `night' neutrinos, for the central LMA region.
    For the discussion here, we will assume that there will be an
analysis cut at 7~MeV, above which radioactive backround
becomes unimportant and only spallation events remain as important backgrounds. 

    A measurement of the day-night asymmetry can take several
    forms. At its simplest, an integral asymmetry measurement can be
    made:
\begin{equation}
A = \frac{2(\phi_{\nu_e}^{night} - \phi_{\nu_e}^{day})}{\phi_{\nu_e}^{night}+\phi_{\nu_e}^{day}}
\end{equation}
Currently, the measurements by the Super-Kamiokande and SNO
Collaborations on this integral asymmetry have found
$A=0.021\pm0.02^{+0.013}_{-0.012}$~\cite{smy} and
$A=-0.037\pm0.063\pm0.032$~\cite{nsp}, respectively, each within
1$\sigma$ of $A=0$ when both statistics and systematics are included.
For a 300~kTon water Cherenkov detector, the event rate in the detector
is roughly 130/day, and consequently the statistical precision on this
asymmetry after a year should be significant, $\sim 0.005$,
depending on the achievable analysis energy threshold.  For the
current best fit LMA parameters, the integral asymmetry is expected to
be near 0.02.  More sophisticated analyses, involving fits to the
energy and zenith-angle dependent survival probabilities, have
already provided noticeably better measurements of the asymmetries in
both Super-Kamiokande and SNO, and could be applied in a larger
detector as well.

\begin{figure}[h]
\includegraphics[width=0.5\textwidth]{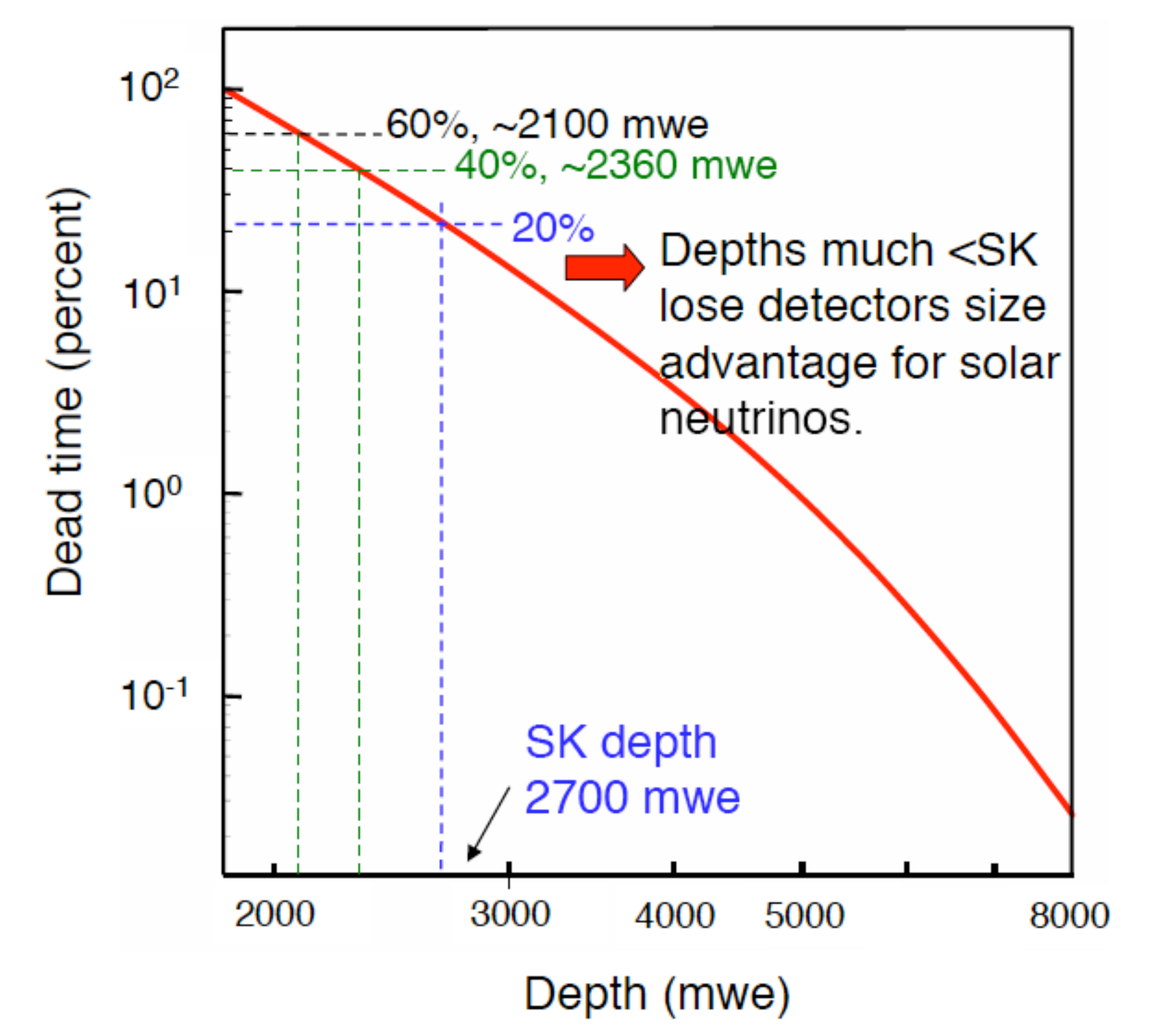}
  \caption{
Spallation induced deadtime versus depth in solar neutrino
measurements. The deadtime for Super-Kamiokande is 20\%. 
The deadtime fraction is to first order, independent of 
detector volume.  
\cite{hsobeldeep}
\label{spalldead}}
\end{figure}

Depth affects the solar neutrino measurement in two ways:
by introducing deadtime and by introducing unwanted asymmetries 
in the background that remains after analysis cuts.  
The signal in the very large water Cherenkov detector 
under consideration here is due to elastic scattering of solar neutrinos 
on the electrons in the detector.  The distribution of electrons from 
this signal points back to the Sun.
For a liquid argon detector absorption of neutrinos on argon nuclei is expected 
to be the dominant detection mechanism ($\nu_e + ^{40}Ar\to ^{40}K^* + e^-$). 
Backgrounds, in both detector types, associated with cosmic rays are mainly
decays of  radioactive spallation nuclei. For each cosmic ray 
muon traversing the detector, events from a tubular region around the 
muon must be rejected for as long as 100 milliseconds.  
This will create deadtime for collection of these events. This deadtime 
fraction 
is approximately independent of the volume of the detector. 
In figure \ref{spalldead}
we show the spallation related deadtime in a large water Cherenkov detector
versus depth in mwe.  
The deadtime fraction is
 approximately the same 
in a liquid argon TPC since the spallation mechanisms 
and time scales are similar.  
To keep the deadtime fraction below 20\%, a minimum depth of 
2700 mwe, or equivalent to 
Super-Kamiokande depth is recommended.

	The second way the spallation backgrounds could affect the 
day/night measurement is by inducing fake asymmetries in the 
event rates.  
    The day-night asymmetry measurement is very robust to backgrounds,
    as long as the backgrounds are reasonably small and symmetric day
    and night.  For example, the spallation related events could be 
affected because  the number of cosmic rays
    may differ day and night because of atmospheric conditions. 
The best way to eliminate such systematic effects is to reduce the 
rate of background to be negligible.  
For a depth of 4300 mwe the backgrounds become small, and the
    asymmetry in the backgrounds even smaller.  
The number of muons passing through a
    single 100~kTon module at this depth is 
  roughly 0.1 Hz.  At Super-Kamiokande,
    $\sim$1.7\% of the throughgoing muons created detectable
    spallation events\cite{blaufuss}, and while this fraction may be higher at
    greater depth (the average energy of muons is higher at greater depths) 
we take this as a baseline estimate.  These numbers
    therefore imply a rate for the creation of spallation nuclei of
    about 150/day, before any cuts are applied.
Similarly roughly 0.05~Hz of through going cosmics
are expected in a 50~kton liquid argon module at 4300 mwe.  
Here we assume a similar
fraction of spallation events as is seen at Super-Kamiokande.  Therefore, we
would expect the creation of about 75 spallation nuclei per day.

    Very few spallation nuclei have decay energies above 7~MeV with
    lifetimes longer than 0.5 seconds~\cite{admarino}.  Two
    exceptions are $^{11}_{4}$Be, with a 13.81 second half-life and a
    $\beta$ endpoint energy of 11.5~MeV, and $^{16}_{7}$N with an
    endpoint of 10.42~MeV and a half-life of 7.13 seconds.  A very
    simple analysis then, which just removes all events within 0.5
    seconds of a throughgoing muon and with energies reconstructing
    below 7~MeV, removes a majority of these nuclei:
    $^{11}_{4}$Be, for example, made up just 5$\times 10^{-5}$ of the
    observed spallation products/day in
    Super-Kamiokande~\cite{koshio}, while $^{16}_{7}$N is a larger
    fraction at 1.4$\times 10^{-3}$.  
If we assume that the $^{16}_{7}$N is the remaining background then  
we are left with roughly 75 spallation background events/year
    in each 100~kTon detector module.  After  additional removal by
    reconstruction cuts and the fitting of the elastic scattering
    directional peak we expect to  have a negligible
    background to the day-night asymmetry measurement at a depth of
    4300 mwe.

   After elimination of backgrounds, the day-night asymmetry
   measurement is more likely to be limited by systematic
   uncertainties associated with understanding the signal detection
   asymmetries (like top versus  bottom) within the detector. The
   consideration of these backgrounds for a liquid argon TPC are
   similar   if the low energy threshold ($\le$ 7 MeV) can
   be achieved. There are important differences in the signal detection
technique: the water Cherenkov signal detection is  through elastic 
scattering of neutrinos off electrons whereas in liquid argon there is expectation that
absorption of neutrinos  on $^{40}$Ar will be dominant. The event rates from 
elastic scattering and  absorption  on $^{40}$Ar are expected to be
in the ratio of 1:$\sim$3  in a liquid argon detector, but the 
exact ratio depends on the energy threshold \cite{larsolar}. 
Nevertheless, it is clear that the 
depth requirements for a water detector are applicable to a liquid argon 
detector as well.

    In addition to a measurement of the day-night asymmetry, a
    measurement of the solar {\it hep} flux (the highest energy expected solar
    neutrino flux component has rate about 1/2000 of the $^8$B flux) 
could be made, if the detector's energy resolution
    is good enough.  Limits on the flux of solar antineutrinos, and
    the neutrino magnetic moment, might also be made if backgrounds
    are small enough.  While these measurements are not as high
    priority as the day-night measurement, they are noticeably less
    robust to spallation backgrounds, and therefore a shallower depth
    than 4300 mwe would make them more difficult.

In summary, the signal for solar neutrinos in a very large water Cherenkov 
 is  elastic scattering of 
neutrinos on electrons. The background at energies of interest (above 
5 MeV) mainly comes from products of spallation interactions of  cosmic
ray muons.  Rejection of such  background causes loss of signal due to 
deadtime. To limit this deadtime to a reasonable level ($<$20\%) requires 
a minimum depth similar to the depth of Super-Kamiokande. To reduce this 
background so that the day/night asymmetry does not have significant
 contribution from asymmetries in the background requires $\ge 4300$mwe.  
The background contributions to a solar signal in a liquid argon detector 
are less well known, nevertheless since the signal event rates per unit mass 
are similar for the two technologies (within a factor of few), 
the depth requirements for liquid argon 
should be similar to the  water detector requirements.

\subsection{Observation of Supernova Burst Neutrinos  }

A nearby core collapse supernova will provide a wealth of information
via its neutrino signal (see\cite{Scholberg:2007nu} for a review).
In 1987, much was learned from about twenty detected
neutrino
 interactions resulting from the explosion of a supernova in the Large
 Magellanic Cloud (SN1987a).  
The neutrinos are emitted in a burst of a few tens of seconds
duration, with about half in the first second. Energies are in the few
tens of MeV range, and luminosity is divided roughly equally between
flavors.  The observed neutrino signal will shed light on several 
topics of current interest. 

\begin{itemize}
\item \textbf{Astrophysics:} The time, energy and flavor distribution of the detected
neutrinos will give valuable information on the astrophysics of core
collapse:  the explosion mechanism,
accretion, neutron star cooling, possible transitions to quark matter or
to a black hole. 

\item \textbf{Particle physics:} As a copious source of neutrinos, we will also
  learn about the properties of neutrinos.  In particular,
  oscillations in the core can provide information on oscillation
  parameters, mass hierarchy and $\theta_{13}$, possibly down to
  very small values of $\theta_{13}$, inaccessible to conventional accelerator 
experiments,  
if the systematics of the supernova models are well understood
\cite{Dighe:2008dq, Mirizzi:2006xx, Raffelt:1997ac, Hannestad:2001jv}.  

\item \textbf{Early alert:} Because the neutrinos emerge promptly after core
  collapse, in contrast to the electromagnetic radiation which must
  beat its way out of the stellar envelope, an observed neutrino
  signal can provide a prompt supernova alert \cite{Antonioli:2004zb,
    Scholberg:2008fa}.  This could allow astronomers to find the
  supernova in early light turn-on stages, which may yield information
  about the progenitor. 

\end{itemize}

The better one understands the astrophysics, the better the quality of
information about neutrino physics, and vice versa.
Hence it is essential to gather as much high-quality 
information as possible.  Ability
to tag the different neutrino flavor components of the flux
will be especially valuable.\\

\subsubsection{The Supernova Neutrino Signal}
In water, the dominant neutrino interaction is 
$\bar{\nu}_e+p\rightarrow e^++n$.
Gd added to the water will result in improved tagging 
of $\bar{\nu}_e$
via $\gamma$-rays resulting from neutron capture on Gd.
Other interactions of interest are shown in table \ref{tab:rates}
\cite{Kolbe:2002gk}.  
Elastic scattering, 
$\nu +e^{-}\rightarrow\nu +e^{-}$,
while representing only
a few percent of the total signal, will allow pointing to the supernova
in a water Cherenkov detector, thanks to its directional nature.

In liquid argon, a  tagged $\nu_e$ channel is
available, $\nu_e+^{40}{\rm Ar}\rightarrow e^{-}+^{40}{\rm K}^{*}$, in which 
 the $^{40}{\rm K}^{*}$ de-excitation $\gamma$-rays are observable and
 provide a tag\cite{Bueno:2003ei,Cline:2006st}.  The $\nu_e$
 sensitivity of liquid argon should be contrasted with the $\bar\nu_e$
 sensitivity of a water Cherenkov detector. With similar event rates
 for supernova, the two detector technologies provide important
 independent measurements and therefore should be considered
 complementary.  A very strong argument for this complementarity can
 be seen in table \ref{tab:rates}, which has the number of
 interactions from a supernova at 10~kpc for a 100~kTon water and a 
50~kTon liquid argon detector.  
At 10~kpc (the center of our galaxy), a
 supernova produces a few hundred interactions per kton in both water
 and LAr. The numbers in the table assume no effects of
 oscillations.  Strong enhancements for $\nu_e$ are expected from
 oscillation effects. Detection of the $\nu_e$ enhanced rate  can be made in a 
LAR detector and compared to the rate of   $\bar\nu_e$ in a water Cherenkov
 detector \cite{autiero, botella}.  Comparison of data from a water Cherenkov
 and a liquid argon detector would be remarkable. 

\begin{table}
\begin{center}
\begin{tabular}{|l|l|}
\hline  
\textbf{100 kt water} & No. of interactions\\ \hline
Inverse beta decay $\bar{\nu}_e+p\rightarrow e^++n$ & 23000 \\ 
CC $\nu_e + ^{16,18}\rm{O} \rightarrow ^{16,18}{\rm F} +e^-$ & 1000 \\ 
NC $\nu + ^{16}{\rm O} \rightarrow \nu + ^{12}{\rm O}^{*}$ & 1100  \\ 
ES  $\nu + e^{-}\rightarrow\nu + e^{-}$ & 1000 \\ 

\textbf{50 kt LAr} & \\ \hline
CC $\nu_e+^{40}{\rm Ar}\rightarrow e^{-}+^{40}{\rm K}^{*}$ & 3100 \\ 
CC $\bar{\nu}_e+^{40}{\rm Ar}\rightarrow e^{+}+^{40}{\rm Cl}^{*}$ & 260  \\ 
NC $\nu +^{40}{\rm Ar}\rightarrow \nu +^{40}{\rm Ar}^{*}$ & 15000 \\ 
ES   $\nu +e^{-}\rightarrow\nu + e^{-}$ & 500\\ 
\hline  
\end{tabular}
\end{center}
\caption{\label{tab:rates}Summary of expected core collapse signal at 10~kpc.
The expected number of events  scale by distance as $1/D^2$, where 
$D$ is the distance to the supernova. 
These numbers are for no oscillation effects. Oscillation effects 
will very likely  create large effects on the charged current $\nu_e$ and 
$\bar \nu_e$ event rates.  }
\end{table}

\subsubsection{Depth Considerations}
Depth affects the level of background seen during a supernova burst,
via background related to cosmic ray muons, including imperfectly
vetoed muons themselves, radioactive decay of spallation products, and
Michel electrons from unvetoed entering muons.
A supernova within our own galaxy (out to $\sim$20 kpc)
will yield a signal bright enough within a short  period of time
that fairly high levels of background can be tolerated, especially
since background can be well characterized outside of the burst time window.
Cosmic rays can be
vetoed; spallation products can also be removed, at some cost
in deadtime.  Some simple scaling calculations serve to estimate the
severity of background as a function of depth. 

\begin{figure}[!ht]
\begin{centering}
\includegraphics[height=2.8in]{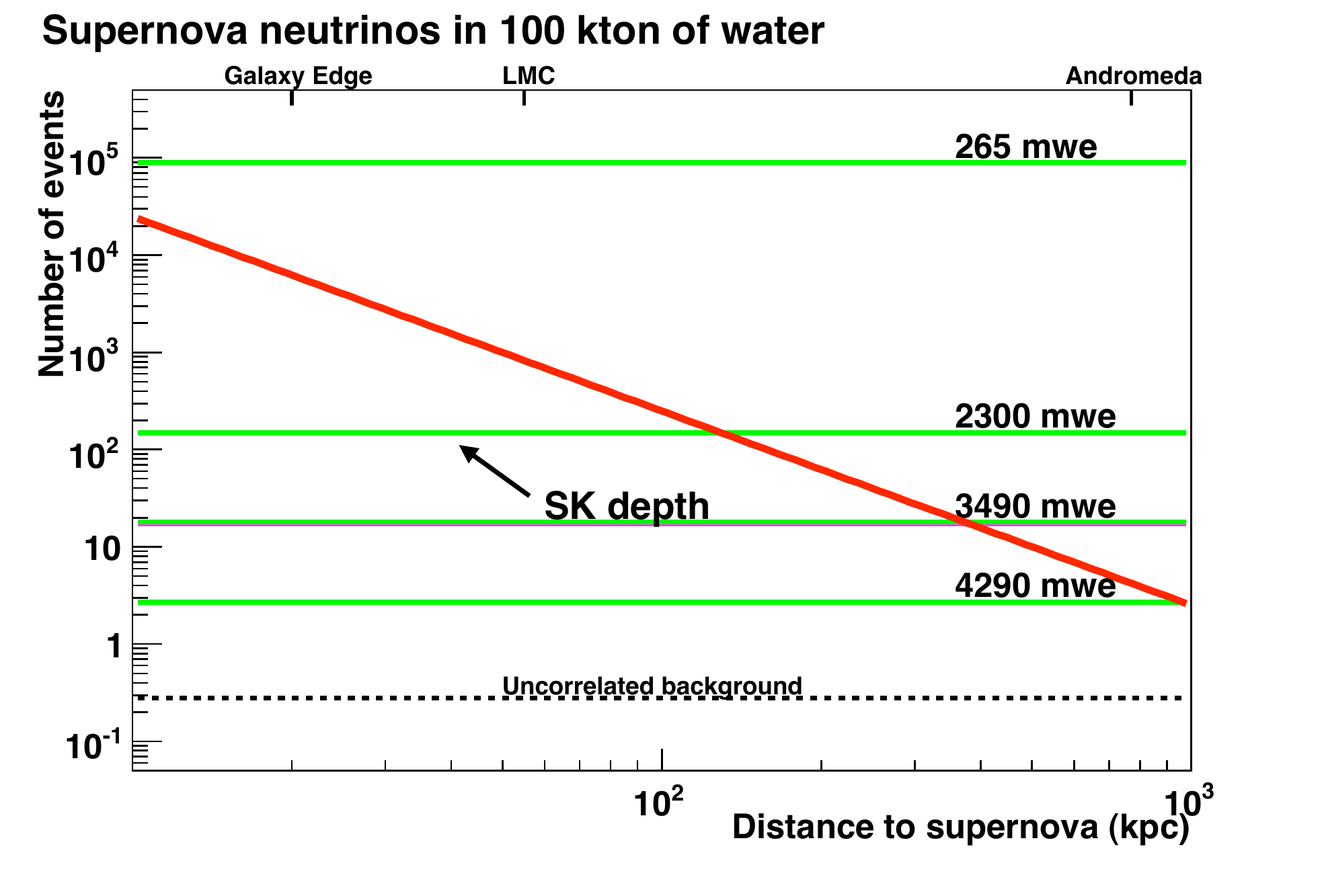}
\caption{Supernova neutrino interactions in a 30 second time window
 as a function of
distance to the core collapse (red); horizontal lines represent numbers of
expected background events (see text).  
}
\label{fig:burst1}
\end{centering}
\end{figure}

Figure~\ref{fig:burst1} shows  the expected total signal events as a
function of distance to the supernova in 100 kTon of water.
The assumed energy threshold is about 7 MeV and
duration of the burst is assumed to be 30 seconds.  Also shown as
green solid lines are expected numbers of cosmic ray muons in the 30
second burst time window for different depths. 
Shown as a black solid line is the uncorrelated background,
estimated by scaling the background rate from Super-Kamiokande offline
supernova burst analysis \cite{Ikeda:2007sa} (180 events/day) 
by mass. 
The uncorrelated background  may consist of
\textit{e.g.} radioactivity, flashing photo-tubes, unvetoed spallation events,
``invisible muons'' from atmospheric neutrinos,
and solar neutrinos.  The simple scaling to a DUSEL 100~kt detector 
may or may not hold depending on the nature of typical
detector noise and rejection efficiencies.  
It should be relatively independent of depth, although
any spallation component will be depth-dependent.

From this plot can be read off the muon rejection factor required
for a reasonable signal to noise for burst supernova neutrinos at
a given distance and at a given depth.  In Super-Kamiokande 
 the muon-related background can be further reduced, 
using a muon veto that surrounds 
the inner detector, by a factor of
 $>10^{3}$.  If we assume that the~100 kton detector
configuration is such that  a rejection factor of $10^{3}$ is
possible, we can see that for all depths beyond 300 ft the signal to
noise for bursts from within the Galaxy can be made reasonably high.
Nevertheless,   considering a supernova in Andromeda, for which
one expects a handful of signal events,
 the signal window
will suffer very little contamination at 4850~ft even without a muon veto.  
However, at 300~ft, the Andromeda 
supernova neutrinos must be extracted from among
 several thousand muons. Although this may not be impossible,
the final sample will most likely be contaminated by muon related 
backgrounds.  
Furthermore, the greater the background, the worse the ability to
separate components of the flux, and any long tail features (perhaps
illuminating neutron star cooling
processes \cite{Pons:1998mm,Pons:2001ar}) will  be obscured.

\begin{figure}[!ht]
\begin{centering}
\includegraphics[height=2.8in]{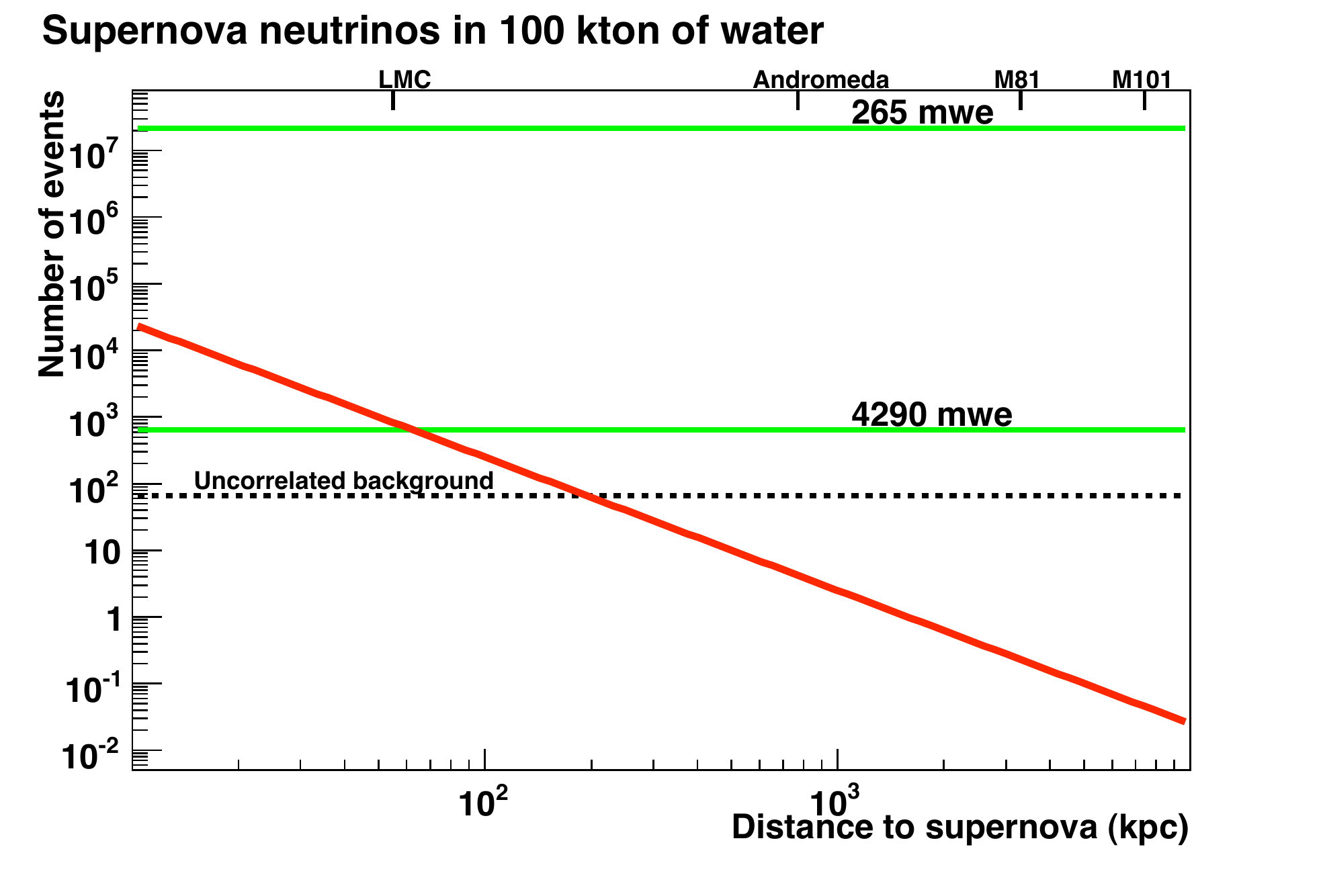}
\caption{Same as Figure~\ref{fig:burst1}, with plot extended to farther 
distances; here the time
window is two hours.}
\label{fig:burst2}
\end{centering}
\end{figure}


Although we could  learn something about Galactic supernovae
even at the shallow depths, farther-reaching supernova neutrino
searches require quieter environments.  It has recently been
proposed\cite{Ando:2005ka} to collect neutrinos one by one in
coincidence with optically-observed supernovae, over a long time frame.
To estimate the effect of depth on such a search,
Figure~\ref{fig:burst2} extends the scale of the previous plot to the
distance of nearby galaxies beyond the Local Group.  This plot
optimistically assumes that one could estimate core collapse time to
within a two hour window based on the optical observation.  Here, one
can see that at the 4850 ft level, with good muon rejection one may
achieve a reasonably clean sample. However the limiting factor could be 
the uncorrelated background for which simple scaling from Super-Kamiokande
may not hold.   This background will have to be rejected 
further by  analysis cuts. 
  At 300 ft, however, it is clearly 
 a daunting task to pick the single supernova neutrino events  from the
haystack of muon-related background.

Depth is also important for the early alert.
It is reasonable to assume that at Super-Kamiokande  depth or
deeper, one could  reproduce the Super-Kamiokande early alert distance
sensitivity of $\sim$ 100~kpc.  
Greater fiducial mass and lower muon flux from a deeper location will 
improve this early alert capability. 
But  early alert rates
from muon-correlated (\textit{e.g.} spallation burst) and
detector-noise-related (\textit{e.g.} flasher, calibration-related)
are highly tunable by threshold selection and online background
reduction algorithms; therefore an estimate of the early alert reach 
will need to be studied as part of the detector optimization.  
Note that coincidence with other
experiments will only help if other detectors of extra-galactic sensitivity
are online.

In summary, we have made estimates to show that for a Galactic core collapse, 
a shallow depth ($<$1000 mwe)  is sufficient for detection of the supernova 
neutrino events. 
However, 
the quality of the information becomes degraded the shallower one
goes, and muon rejection using an active veto system 
must be employed  to compensate. 
 To extend the supernova reach beyond the edge of Milky Way, we recommend
a depth of 3500 ft or greater, combined with rejection of muon
background by  $\sim 10^{3}$ which can be easily achieved by an active veto. 
A location at 4850~ft will reduce the background to a level where an active 
veto may not be needed; 
further rejection of both correlated and uncorrelated backgrounds can be achieved by refined 
analysis, but will need to be studied.  
Lastly,  most of this  study has been done for a water Cherenkov detector because 
a lot of information is known about backgrounds from previous 
experiments.   Since the 
number of charged current signal events per unit mass is smaller for liquid argon,
one needs more care in background rejection for a liquid argon detector.  Due to finer
granularity of the detection mechanism, we expect better muon
rejection in a liquid argon detector, but it is likely that the depth requirements 
are similar to a water Cherenkov detector. 
It should be emphasized again that the water Cherenkov and liquid argon technologies are
highly complementary for supernova detection:  water Cherenkov is mainly sensitive to 
$\bar\nu_e$ events, while liquid argon is sensitive to $\nu_e$ events. If we were to obtain 
supernova spectra of equal statistics from both detectors simultaneously, 
the scientific outcome 
will be extraordinary.

\subsection{Observation of Diffuse Supernova  Neutrinos}

The explosion of a core collapse supernova releases about 99\%
 of its energy in form of neutrinos in a time period on the order
of ten seconds. Unfortunately, due to the small neutrino cross section,
 even such massive neutrino bursts can only be detected in our own galaxy
or nearby. 
The combined supernova explosions throughout the universe left behind
a diffuse (sometimes called relic) 
 background of neutrinos that may be detected on Earth. The flux
and spectrum of this astrophysical source of neutrinos 
 contains information about the rate of
supernova explosions (and consequently the star formation rate) in the past
and also enhances our understanding of the universe to redshifts of 
$z\sim 1$. It is also affected by neutrino properties such as mixings and mass 
ordering.

\begin{figure}[b]
\centering\leavevmode
\includegraphics[angle=0,width=0.7\textwidth]{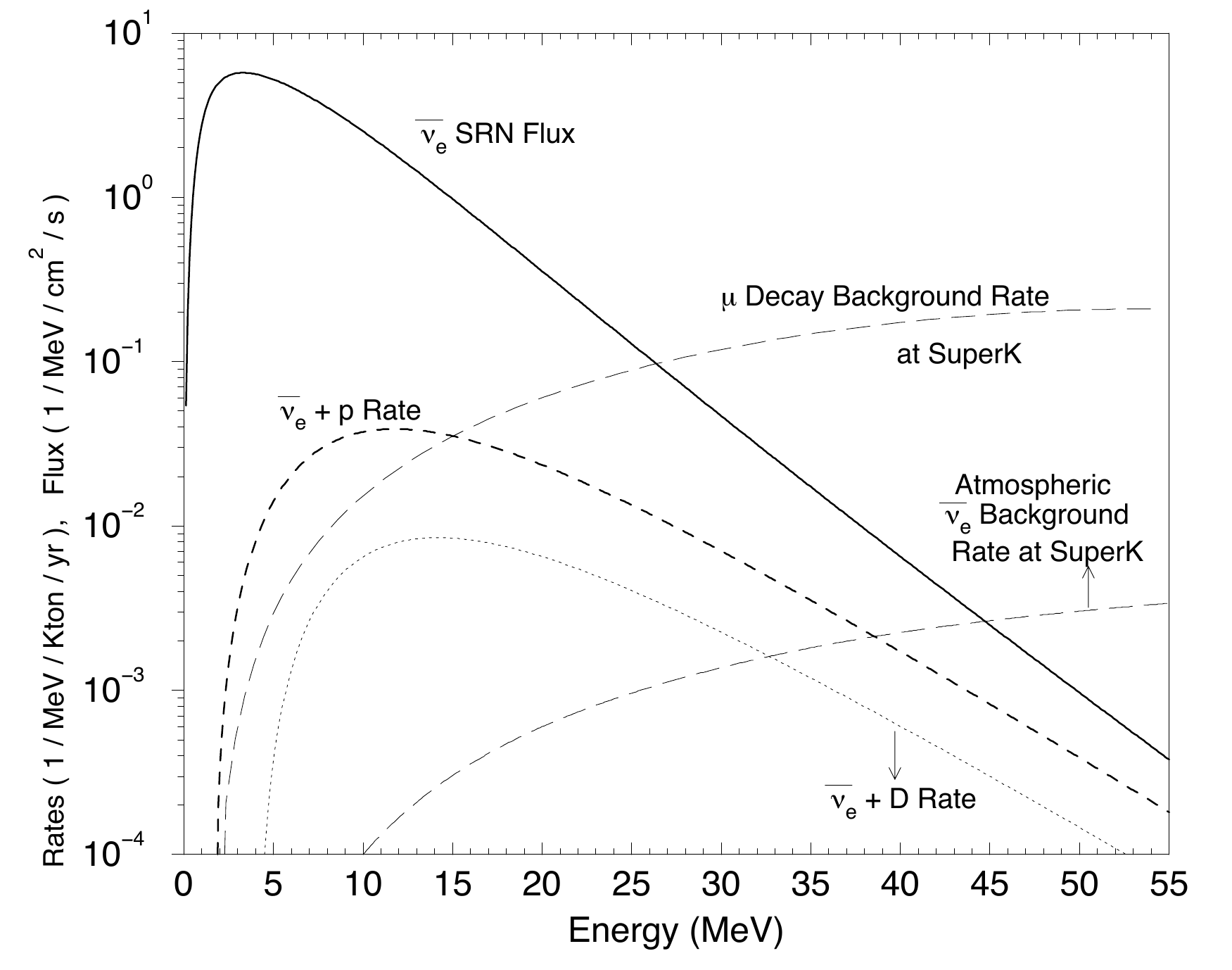}
 \caption{(in color)
Flux of diffuse supernova neutrinos (solid $\bar\nu_e$).
The rate of $\bar\nu_e + p$ events (dashed).  
Rate of  muon decay and
atmospheric neutrino backgrounds (long dashes).
And the rate of $\bar\nu_e$ absorption on deuterium for comparison (dotted).  
The energy window for detection of diffuse supernova neutrinos is approximately 
10 to 30 MeV. 
\cite{ksw}
  \label{relicfig1} }
\end{figure}

\underline{\bf Signal in  water Cherenkov detectors}
The best signal for diffuse supernova  neutrinos in water Cherenkov
detectors is the positrons resulting from the inverse $\beta$
reaction with electron antineutrinos. The predicted spectrum 
and event rate of the
diffuse antineutrinos is shown in Fig.~\ref{relicfig1} along with 
the muon decay and atmospheric neutrino backgrounds.  While the
maximum flux is at lower energies ($< 5$~MeV), there is significant
background below 10~MeV due to antineutrinos from nuclear power
reactors. The reactor background is extremely 
site dependent, and therefore is not 
shown in Fig. \ref{relicfig1}. For example, the rate of reactor events 
at the Kamioka site due to the high concentration of Japanese reactors is 
about 533 events/kTon/year. At the Homestake site  the rate is considerably 
lower  at about 37 events/kTon/year. 
 Therefore, 10~MeV is considered the practical lower limit for
detection of positrons from the very rare diffuse ${\bar \nu}_e$'s.

 The main background in the region 10-25~MeV is
from cosmic ray muon spallation which is depth dependent. Even though such radioactive
background can be tagged by the detection and reconstruction of the
preceding muon, surviving spallation events in Super-Kamiokande-I
still overwhelm the expected supernova diffuse neutrino interaction
rate below $\sim 18$~MeV. Super-Kamiokande-I therefore limited the
search to above 18~MeV positron energy (or 19.3~MeV diffuse neutrino
energy) and placed a 90\% C.L. limit on the flux above that of
1.25/cm$^2$-sec with a data set taken in about five years (1496 live
days). In the Super-Kamiokande-I analysis, the remaining irreducible
backgrounds in the region above 18~MeV were due to atmospheric
$\nu_\mu$ producing invisible muons ($T_\mu<50$~MeV, below Cherenkov
threshold) that subsequently decay and atmospheric $\nu_e$ and
${\bar \nu}_e$ interactions. We note that a same-style
analysis for a 300~kTon (fiducial) detector would improve the exposure
(for 5 years) by about 13 and the sensitivity by a factor of 3.6,
so the 90\% limit would reach 0.34/cm$^2$-sec. Strigari,
Kaplinghat, Steigman and Walker \cite{steigman} have estimated the
lower limit of the diffuse neutrino flux above 18~MeV positron energy
to be approximately 0.3/cm$^2$-sec. Thus a large water Cherenkov detector 
 should be able to detect the diffuse supernova neutrinos if the
backgrounds were reduced below the Super-Kamiokande rates. There are two
methods that can be utilized to reduce the backgrounds: coincident
neutron detection and increased overburden (to reduce the spallation
background).

\begin{figure}[h]
\centering\leavevmode
\includegraphics[angle=0,width=0.65\textwidth]{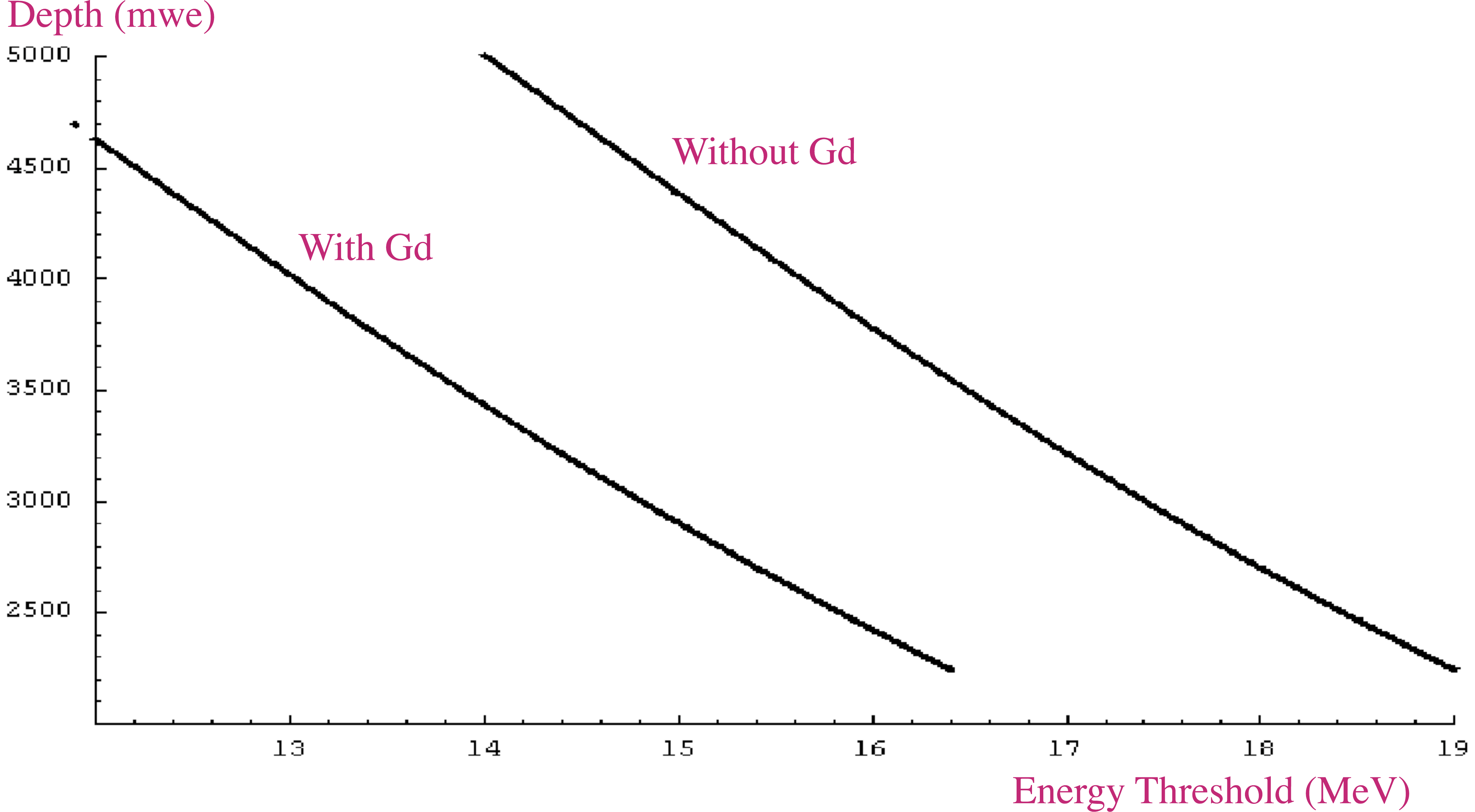}
\includegraphics[width=0.34\textwidth]{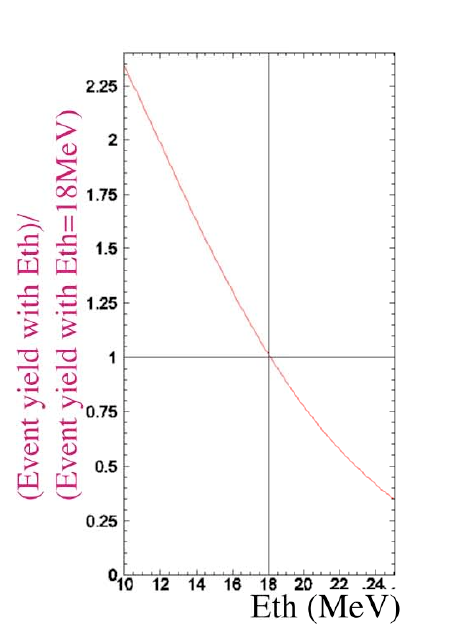}
 \caption{(in color)
Left plot is Energy threshold versus depth for a water Cherenkov
detector with Gd loading and without Gd loading. The right plot is diffuse
supernova neutrino rate relative to energy threshold of 18 MeV.
  \label{relicfig2} }
\end{figure}

The residual backgrounds above 10~MeV could be substantially reduced
by detection of the neutron produced by the inverse $\beta$
interaction in delayed coincidence with the positron. The
atmospheric $\nu_\mu$ and $\nu_e$ events are not generally
accompanied by a neutron, and so the expected background from these
sources will be much lower using the neutron tag. A neutron capture
in delayed coincidence will also reduce the spallation background,
since only isotopes with accompanying neutrons can be confused with
diffuse neutrinos. It is difficult to estimate the remaining
spallation event rate as the spallation production rates of
particular isotopes are not well-known. In general, however, there
are fewer isotopes and the energy of the decay  $\beta/\gamma$
is less, if a neutron has to be produced. We conservatively assume
that the spallation rate would be reduced by an order of magnitude.

The capture of neutrons on hydrogen produces a 2.2 MeV $\gamma$-ray
which is virtually undetectable in present-day water Cherenkov
detectors; only about seven photo-electrons would be detected in
Super-Kamiokande-I. However, doping the water with Gd salt
\cite{vagins} (on the order of 0.1\%) results in the capture of most
of the neutrons ($>90\%$) on Gd and produces a cascade with a total
energy of 8 MeV. Super-Kamiokande-I would see about 30
photo-electrons and could therefore detect these captures.

Assuming a delayed coincidence neutron tag would reduce the
background due to decay electrons from sub-threshold muons by at
least a factor of four,  a 300 kTon detector with 18~MeV threshold could
reach a flux sensitivity of $<0.2$ cm$^{-2}$sec$^{-1}$.  A lower energy
threshold of 10~MeV increases the predicted flux by a factor of 2.3,
so using a neutron tag and lowering the threshold could provide
excellent sensitivity even at depths shallower than Super-Kamiokande.

 To estimate the impact of overburden, we use the antineutrino spectrum
 from the Kaplinghat, Steigman and Walker model \cite{steigman}
 and parameterize the
 spallation spectrum with a simple fit to Super-Kamiokande-I data ($s(E_n)=e^{18.6-0.9 E_n/MeV})$
where $E_n$ is the visible energy. 
We approximate the total spallation rate as a function
 of detector depth $h$ and energy threshold $E$ as the integral of the
Super-Kamiokande-I spallation spectrum up to 25~MeV scaled by the muon
intensity from figure \ref{crouch}. 
\begin{equation}
S(E,h) = I(h) \times \int_E^{25MeV} s(E_n) dE
\end{equation}
where
\begin{equation}
I(h) = 2.18\times 10^{-13} +
e^{(-11.24-2.64h/km)}+e^{(-13.98-1.2227h/km)}
\end{equation}
$I(h)$ is a parametrization of the 
muon intensity data in figure \ref{crouch}.  
From the energy threshold of the Super-Kamiokande-I analysis
(18~MeV) and the Super-Kamiokande depth (2.7~km water equivalent),
 $S(E,h)$ and the integrated Supernova diffuse neutrino interaction
rate R(E), the energy threshold for a detector at depth $h$ can be
 estimated using the equation
\begin{equation}
{S(E,h)\over S(18MeV, 2.7km)} = {R(E)\over R(18MeV)}
\end{equation}
The estimated achievable energy threshold for a Gd-doped detector is
then 2.6~MeV less than for a detector without Gd doping.
Table~\ref{tab:SN} compares various depths with and without Gd
doping, and Figure~\ref{relicfig2} shows the same information in a plot.

\begin{table}[h]
\centering
\begin{tabular}{|l|l|l|l|l|l|}
\hline
rock depth & water equiv & Energy thres. & Energy thres.  & Signal rate without Gd  & Signal rate with Gd \\
 ft.   & km-w-e   &   without Gd (MeV)    &  with Gd (MeV)  & relative to 18 MeV   &  relative to 18 MeV  \\
\hline
4850  & 4.3 &  15.5  & 12.0 &  1.4   &   2.0    \\
3500  & 3.1  &  17.5 & 15.0  &  1.1    &   1.5   \\
3300  & 2.9  &  18.0  & 15.5 &   1.0    &   1.4   \\
2000  & 1.8  &  20.5  & 18.0   &  0.73 &    1.0  \\
300  &  0.27 &  25.0  &  22.5 &  0.36  &  0.55  \\
\hline
\end{tabular}
\caption{Expected energy threshold with and without Gd doping as a
function of depth for detection of diffuse supernova neutrinos. }
\label{tab:SN}
\end{table}

\underline{\bf Signal in Liquid Argon Detectors}
Supernova diffuse neutrinos and anti-neutrinos will interact in liquid argon
detectors via both charged and neutral current interactions as shown in 
table \ref{tab:rates}. 
These interactions are detected  in all but the case of the neutral
current channel, by the  electron or positron in the final
state.  A 5~MeV electron travels 2~cm in LAr, long enough to identify
the event.  Below 5 MeV, the electron could be identified by the
energy deposited, but for the purposes of this document, we 
assume a 5 MeV threshold for electrons for detection of these
interaction channels.  


\begin{figure}[t!]
\centering\leavevmode
\includegraphics[width=4in]{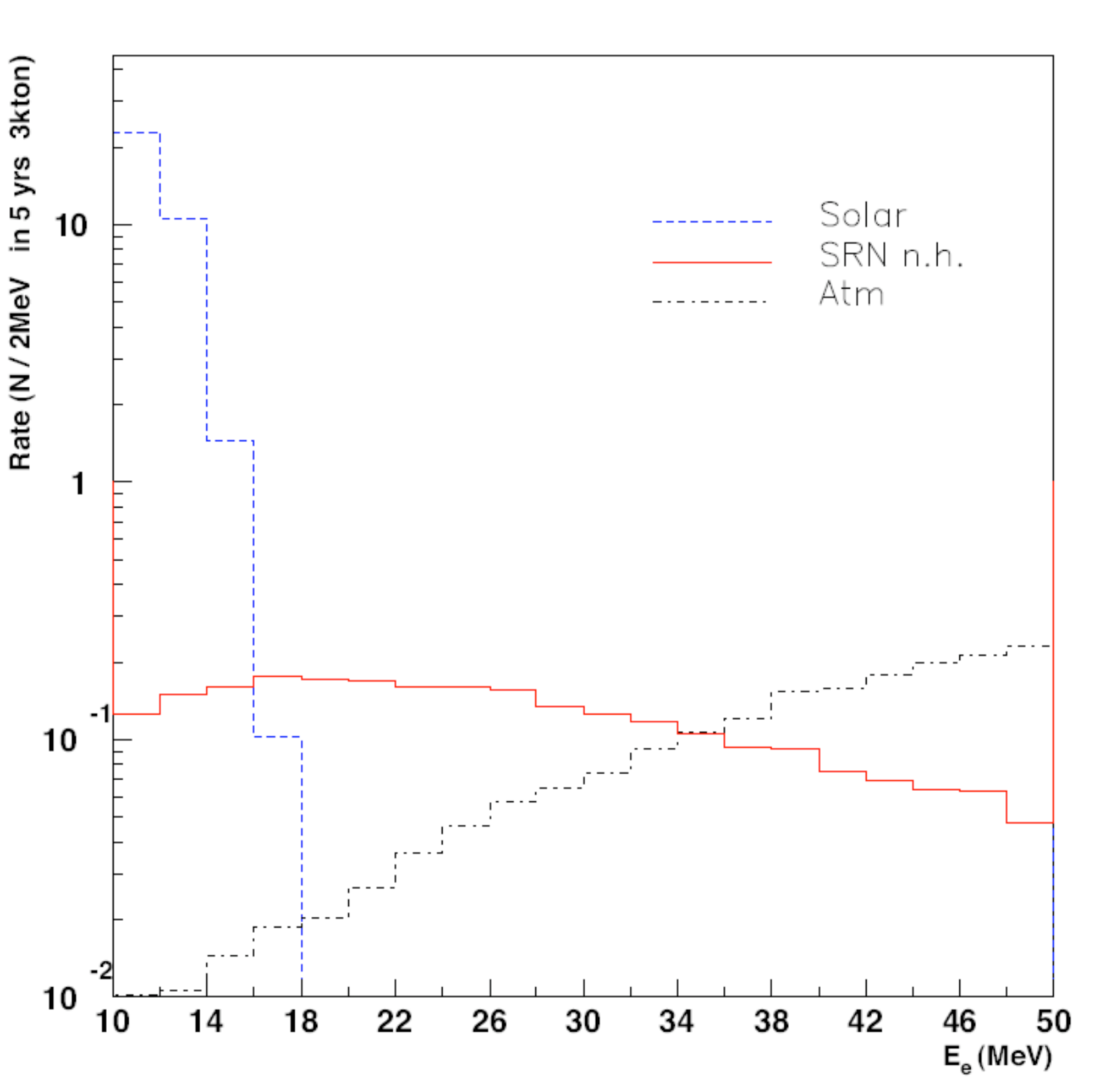}
 \caption{ Number of expected events for diffuse supernova neutrinos
 and backgrounds,   assuming 100\% detection efficiency for electrons with energy
   greater than 5 MeV, per year, for a 3kton detector~\cite{Cocco:2004ac}.
}
 \label{fig:SRNLAr}
\end{figure}

The expected diffuse supernova neutrino signal 
including effects of energy resolution ($\sim 4$\% at 10 MeV) 
\cite{Amoruso:2003sw}
 is shown in
Figure~\ref{fig:SRNLAr}~\cite{Cocco:2004ac} with backgrounds from
solar neutrinos  at low energies and from atmospheric neutrinos
at higher energies. The rate shown in the figure also includes effects of oscillations that 
are expected to enhance the rate for $\nu_e$.  
From the figure, it is clear that the window in which the signal
is best identified above background is from 16 to 40 MeV. 
Integrating this signal window for 5 years of data taking with a
100~kTon liquid argon detector yields 
$ 57 \pm  12$ events in the signal window
16-40 MeV~\cite{Cocco:2004ac}.  The error is a combination of
statistical and systematic due to the atmospheric flux
normalization.  The sensitivity depends 
strongly on the choice of signal window.  
The predictions for the solar neutrino
backgrounds are well known, but the backgrounds from
atmospheric neutrinos suffer uncertainties in the absolute flux of
primary cosmic rays and cross sections for hadronic interactions.
 Other backgrounds due to decay electrons, spallation, and nuclear recoils
are considered negligible within this signal window~\cite{Cocco:2004ac}. 
The very fine granularity of liquid argon allows rejection of 
the cosmogenic backgrounds by  detection of low energy muons
and nuclei from atmospheric neutrino interactions, in addition to 
incoming cosmic ray muons. Nevertheless, it  would be prudent to 
have sufficient overburden to keep the rate of spallation events 
manageable for data analysis purposes; e.g. a good understanding of 
the spectrum of events from 10 to 50 MeV.   The rate and nature of 
spallation in a liquid argon detector is not well known at present. 
If one assumes the same rate as a water Cherenkov detector 
(see section \ref{sec:solar}), and assumes  a rejection factor of 
$\sim 10^3$ by muon tagging, a depth of 2500 mwe is needed to 
keep the spallation rate approximately the same as the solar neutrino event rate
above 10 MeV. Of course, a deeper location for a liquid argon detector would result 
in easier separation of the components of neutrino events in figure \ref{fig:SRNLAr}. 


In summary, for a 300~kTon water Cherenkov  detector, we estimate a positron
energy threshold of 15.5~MeV would be possible, and the achievable
$90\%$~CL sensitivity at the 4850~ft. depth would be 
$<0.3$ cm$^{-2}$sec$^{-1}$, which is below 
the predictions in ~\cite{steigman}. At shallower depths, the
sensitivity would not quite reach the predicted flux. However, a
Gd-doped detector would have sensitivity below the predicted flux
for depths greater than 3300~ft, and significantly enhanced
sensitivity relative to an undoped detector at any depth.
For a liquid argon detector, the  detection mechanism is different,
and the backgrounds could be lower because of the fine granularity of the 
detector. However, there is less information at present on spallation 
events in liquid argon. We have made a rough estimate that a location deeper 
than 2500 mwe is preferrable to keep spallation  backgrounds low so that 
an analysis of the spectrum can be performed.  
We want to emphasize again that the water Cherenkov and liquid argon detectors 
are highly complementary for the detection of diffuse supernova neutrinos. 
Water Cherenkov detector detects $\bar\nu_e$ while liquid argon detector
detects mainly $\nu_e$ events. A combination of the two would allow important 
checks on our understanding of astrophysics as well as neutrino properties.

\subsection{Observation of Atmospheric Neutrinos  }

Atmospheric neutrinos are generated in the upper atmosphere in a
uniform spherical thin shell around the Earth.
For any detector located  at a modest depth within
 the Earth, the flux of neutrinos is
isotropic to a good approximation.
The detection and study of atmospheric neutrinos has resulted in
the remarkable discovery  that the
muon type neutrinos that come from below the detector over a
long distance are suppressed by a factor of $\sim$2 compared to
muon neutrinos that come from above the detector. The increase in
statistics with larger detectors has been matched by the greater
accuracy of simulations of the atmospheric flux.  For a future large
detector, the same study will continue; but will require better
control of systematics that could arise from detector geometry,
calibrations, backgrounds,  and electronics. 
Better statistical and systematic precision could allow 
higher precision measurements of neutrino mixing parameters, especially
$\sin^2 2 \theta_{23}$. Comparison of such a measurement with accelerator 
based measurement is important to detect any contributions from 
new physics.    The cosmic ray muon
background is important for all of these issues.

Atmospheric neutrinos typically have energies of 1~GeV and above.
Because of this, low energy backgrounds like the decay of radon
isotopes are not relevant. The most important source of backgrounds
for atmospheric neutrinos are cosmic-ray muons and particles created
with their passage through matter.
The atmospheric neutrino event rate measured by 
the Super-Kamiokande detector, including effects of oscillations,
and all event identification cuts is $140$~kT$^{-1}$yr$^{-1}$. 
The Kamioka mine is located in a mountain with an effective 
shielding of about 2400 mwe. This corresponds to an approximately 3~Hz rate of
muons passing through the 40~m high by 40~m diameter
Super-Kamiokande detector.
Figure~\ref{crouch2} shows the rate of downward going muons at the
4850~ft level in Homestake is approximately 900~$m^{-2}/yr$. For a
100 kTon fiducial volume detector at 4850 ~ft level we have estimated the 
rate to be ~0.1 Hz (see table \ref{intime}). 
Therefore for each 100 kTon (fiducial) detector module
at 4850 ft  we expect 14000 atmospheric
neutrino events per year and $\sim 3 \times 10^6$ cosmic ray muons per year. 


An important issue arising from the passage of cosmic-ray muons
through the detector is deadtime.  If the detector is completely dead
after a muon passes through than the livetime is reduced.  
As explained in section \ref{sec:pdk}  this is a
more serious issue for water Cherenkov detectors than liquid argon
since the argon detectors are fine grained. Even at a modest depth of 
1000 mwe this should not be a limiting factor for atmospheric neutrino analysis, 
 since the detector does not need
to be deadtimed  for more than $1 \mu$sec after the muon passes to avoid
cosmic muon related backgrounds. 


The Super-Kamiokande detector operates  well at effective 2400~mwe with
negligible background for atmospheric neutrino events. This detector  
makes use of an outer detector veto with
approximately 1/10th the number of 8~inch phototubes as the 11,146
20~inch PMTs in the inner detector. Therefore,  it is clear that a similarly
configured detector could perform well at 4300 mwe (4850 ft level). However we 
need to examine if the deeper location can allow us to operate the 
Homestake based detector without the veto shield. 
The IMB experiment~\cite{BeckerSzendy:1992hr} successfully ran without a veto.
IMB was located at 1570~mwe with a muon rate almost an order of
magnitude higher than that of Kamioka but with almost the same muon
rate since the detector (3.3~kTon) was much smaller with correspondingly
 lower atmospheric neutrino rate.  

As described in~\cite{Ashie:2005ik} the background contamination in the
atmospheric muon-like neutrino sample for both the fully contained 
and partially contained topologies was approximately 0.1\%, and 0.2\%,
respectively, which translates into  $4.0\times 10^{-2}$ events per kTon-year
for fully contained and  
$2\times 10^{-2}$ events per kTon-year for partially contained. 


Assuming a similar configuration to Super-Kamiokande it is possible to roughly
estimate the contamination due to entering background at any depth by
multiplying the background rate by the ratio of cosmic ray muon rates.
For example for a 100~kTon detector at the 4850 ft level, 
the backgrond rate would be  reduced
by $0.1 Hz/3 Hz$. For a 100~kTon (fiducial) detector the atmospheric signal 
is about 14000 events per year  with a background of 
roughly 0.2 events per year.  
If the Homestake water Cherenkov detector is planned without
a veto shield, then some of the background suppression will be 
lost or might have to be performed by other analysis techniques. 
 A detailed examination of this background rejection will be performed
as part of the detector optimization, but we conclude that 
for detection of atmospheric neutrinos, a depth similar to the 
depth of the Super-Kamiokande detector is adequate.  The 
deeper Homestake location at 4850 ft could allow us to remove the veto shield 
and increase the fiducial mass and reduce overall cost.  
The event rates per unit fiducial mass  for a liquid argon TPC 
will be similar to a water Cherenkov detector. The background rates
are at present unknown, but  because of the finer
granularity the cosmic ray muon rejection will be improved allowing
 a shallower location.





\subsection{Summary of Depth Requirements}
The depth requirements associated with the various physics processes
considered in this section are listed in
table~\ref{tab:depth_summary} for both water Cherenkov and liquid
argon detector technologies.
While some of these physics processes can be adequately studied at
shallower depths, none of them require a depth greater than 4300~mwe
which corresponds to the 4850~ft level at Homestake.

For table \ref{tab:depth_summary} we have considered the
detection of each of the physics signatures above background.
The analysis of all physics processes should benefit from reduced backgrounds at 
the ~4850~ft level.
Lower backgrounds  could also allow us to increase the fiducial volume of the 
detector
and reduce the complexity of construction by removing the veto 
shield.   Such detailed optimization studies will need to be 
performed as part of the detector design. Furthermore, 
as the previous sections show, there is considerable room for 
further  analysis and  technology improvements in the coming years, 
but even with considerations of these 
anticipated improvements, we do not see a strong physics justification  for  
depth greater than 4850 ft.

\begin{table}
\begin{tabular}{|c|c|c|}
  \hline
  \textbf{Physics} & \textbf{Water} & \textbf{Argon} \\ \hline
  Long-Baseline Accelerator & 1000~mwe & 0-1000~mwe \\
  $p \rightarrow K^+ \bar\nu$ & $>$3000~mwe & $>$3000~mwe \\
  Day/Night~${}^8$B Solar $\nu$ & $\sim$4300~mwe & $\sim$4300~mwe \\
  Supernova burst & 3500~mwe & 3500~mwe \\
  Relic supernova & 4300~mwe & $>2500$~mwe \\
  Atmospheric $\nu$ & 2400~mwe & 2400~mwe \\
  \hline
\end{tabular}
  \caption{Estimated depth required to study the physics processes
  with either a water Cherenkov or liquid argon detector.}
  \label{tab:depth_summary}
\end{table}

\section{Existing Infrastructure in Homestake and
Siting Considerations }
\label{geotech}

The Homestake underground has many levels spanning a 
spectrum of possible depths. We have developed a set 
of general criteria in order to identify potential sites 
at representative depths based on expected rock conditions
 and existing infrastructure. Subsequently, we have 
estimated the additional cost of developing the
 infrastructure at these levels above that already 
planned for the DUSEL laboratory. The actual cost of 
developing a level cannot be determined until a 
detailed geotechnical analysis is performed, 
so these estimates should be understood as 
differential costs and not representative of final costs.
 This work is summarized below.

\subsection{Summary of criteria for 
siting and candidate levels}

The Homestake mine provides access from the surface 
to 8000 feet underground by a variety of shafts,
 ramps, and adits\footnote{A horizontal mine access 
tunnel from the outside}. DUSEL plans allow primary access
 via the Ross and Yates shafts from the surface to the 4850
 Level and use of No.6 winze from the 4850 Level to lower levels
\footnote{A winze is a shaft that provides access from level
 to level underground,
 but does not go to the surface.}. 
Below the 1100 level, the levels are spaced at 
approximately every 150 feet vertically. The level
 name denotes feet below the top of the shaft (collar)
 of the Yates Shaft, which is at 5310 ft above sea level.
 Therefore, the 4850 Level is 4850 feet below the Yates collar.
 For further reference, the Ross Shaft collar is at 
5355 ft above sea level.
 Current plans for upgrading and
 tailoring the shafts call for conversion of
 the  Yates shaft and hoisting 
infrastructure to provide access for 
scientific personnel and equipment 
and maintaining the Ross for site maintenance, 
excavation, and mining services. The
 elevation of the Yates Shaft collar is 5310 feet 
above sea level, so the 4850 ft level is only 460 feet above sea level.
Further details can be obtained from the Homestake DUSEL CDR \cite{cdr}.

\subsubsection{Level Selection}

The following criteria were used to select candidate levels. 
It is assumed that excavated rock will be removed via the 
Ross Shaft, and that excavation will be done at the same
 time other laboratory experiments are operating or under 
construction.  It also should be noted that construction and
 refurbishment of infrastructure in the underground can be
 done to make nearly any of the levels suitable for 
experiment occupation.  Some levels, however, tend to 
fit better with the overall mine plan to allow them
 to be used more easily and economically.

\underline{Considerations:}

\noindent 	1) Only candidate levels 
at or above 4850 were considered. This is based 
on physics studies, summarized earlier in this document, 
 which show that all scientific 
goals for this detector can be accomplished without
 having to go deeper than the 4850 ft.

\noindent 2) Although additional underground facilities
 could be constructed, in order to maximize the use of 
currently available infrastructure the candidate level 
should provide access to existing shaft stations on both
 the Ross and Yates Shafts. This would provide redundant 
access for personnel, equipment, and safety considerations 
and also reflects the high demands for access for this project.


\noindent 3) Levels with existing waste rock handling 
facilities, e.g. loading (skip) pockets, rail, conveyors, etc.,
 may be preferable.
(A skip pocket is a structure near the shafts
 that allow a skip to be loaded from that level. Without 
a skip pocket, removal of rock from the 
level using the shafts is not possible.)  
 This does not restrict the range of 
available depths, but it serves to focus studies on 
levels which would not require extensive refurbishment.



\noindent 4) In general, construction using the Ross Shaft 
between 3000 and 3600 feet should
 be avoided because the older mine workings 
are close to the shaft in this area and 
additional infrastructure in this area may 
incur difficulties in maintenance, construction, and operation.

\noindent 5) Large excavations should be on a single level and contained 
entirely within the Yates Member, which is thought to be a good
 host material due to its mechanical strength. This is consistent
 with the recommendations of the DUSEL geotechnical committee, 
which were based on preliminary and non-site-specific analysis 
of Yates Member core samples and from Homestake Mine records.

\noindent  
6) Exposure of Yates Member at a candidate level 
should be large enough to contain 3 to 5 excavations
 providing a total excavated volume to support 
 a 500-1000 kTon detector array. Locating the cavities
 at the same level would provide cost and schedule 
saving for multiple cavities, and does not restrict
 potential depths.

\noindent 7) The Yates Member crops out at the surface to 
the north of the laboratory and plunges to the south.  
Therefore, deeper levels encounter the Yates rock closer
 to the Ross Shaft, as well as the Yates Shaft.  In general, 
shallower levels require longer drifting and haulage distances
 than deeper levels.

\noindent 8) Beyond the ES\&H concerns associated in the 
creation of excavations, the site must support life-safety 
and hazard mitigation measures required for the 
detectors (ventilation, water drainage, redundant access, etc.).
 Note: Safety hazards have not been fully assessed, 
which will require more in-depth studies. The hazards are 
different for a water Cherenkov and a liquid argon detector. 

Within the Yates Member at a given level, there are additional 
criteria based on potential schedule and budget risks:

\noindent  
	9) Avoid   areas of high stress in the 
formation.

\noindent  
	10) Avoid   formation contacts.

\noindent 
	11) Avoid   significant geo-structural 
features (e.g. dikes, shear zones, or fracture zones)



\begin{table} 
\begin{tabular}{|l|l|l|l|l|} 
\hline  
Level &  Lacks access & Yates rock  &  Level within       & Comments \\
      & to both shafts & distant from  & Ross pillar area &           \\
      &                &  from Ross shaft & (no development) &         \\
\hline 
300   & $\surd$  & $\surd$  &   &   \\ 
\underline{\bf 800}  &  & $\surd$  &  &   \\
1250 & $\surd$  & $\surd$   &  &   \\ 
\underline{\bf 1700} & $\surd$  & $\surd$  & &   \\
\hline  
1850 & & $\surd$  & &  \\ 
2300 &  & $\surd$  & &  \\ 
 2600 &  & $\surd$  & &  \\
\hline  
2750 & & $\surd$  &  &  \\
2900 & & $\surd$  &   &  \\
3050 & & $\surd$ & $\surd$  &    \\ 
\hline  
3200 & & $\surd$ & $\surd$  &  \\ 
3350 &  &$\surd$   & $\surd$   &     \\  
3500 & & $\surd$  & $\surd$   & \\
\hline  
\underline{\bf 3650} & & $\surd$  & $\surd$ &   \\
3800 & & &  & Yates member closer to the Yates shaft,    \\
	&  & & & but access between shafts is not best \\ 
3950 & $\surd$  & & &  \\
\hline  
\underline{\bf 4100} &  & &  & Possible alternative, but access  \\
              & & & &   between shafts is not as good as  at 4850 \\  
\underline{\bf 4850}  &  &  & &  Preferred Level  \\ 
\hline  
\end{tabular} 
\caption{Possible Levels in the Homestake Mine for 
development of a 
megaton scale detector cavity.
\label{levels1} } 
\end{table}

\subsection{Candidate Levels} 

Table \ref{levels1} is a list of the existing levels 
 in the 
Homestake mine
that were 
considered as initial candidates.
Underlined  levels are 
recommended levels for consideration due 
to meeting some of the listed criteria. These levels 
span a reasonable range of depths and would be
 the least expensive to develop.  The levels in 
the table fall in 3 general categories: the upper shallow levels
down to $\sim$ 2000 ft, the middle levels from 2000 to 3500 ft, and 
the deep levels below 3500 ft.   
The shaft stations in the upper levels are quite far away from the 
preferred rock to be developed (the Yates Member), a long 
access tunnel will be needed to the rock if the shallow sites
are to be used.  In the middle levels, the area closest to the shafts 
have rock with structural features or formation contacts that are 
best to avoid for laboratory development. Many of these levels also lack 
existing skip pockets or other infrastructure.   The most important 
feature of the deep levels below 3500 ft 
is the proximity of the Yates rock 
to the shafts. In fact, 
the Yates shaft penetrates into the preferred rock 
at both the 4100 and  4850 ft levels.

\begin{figure} 
\centering\leavevmode
\includegraphics[angle=0,width=\textwidth]{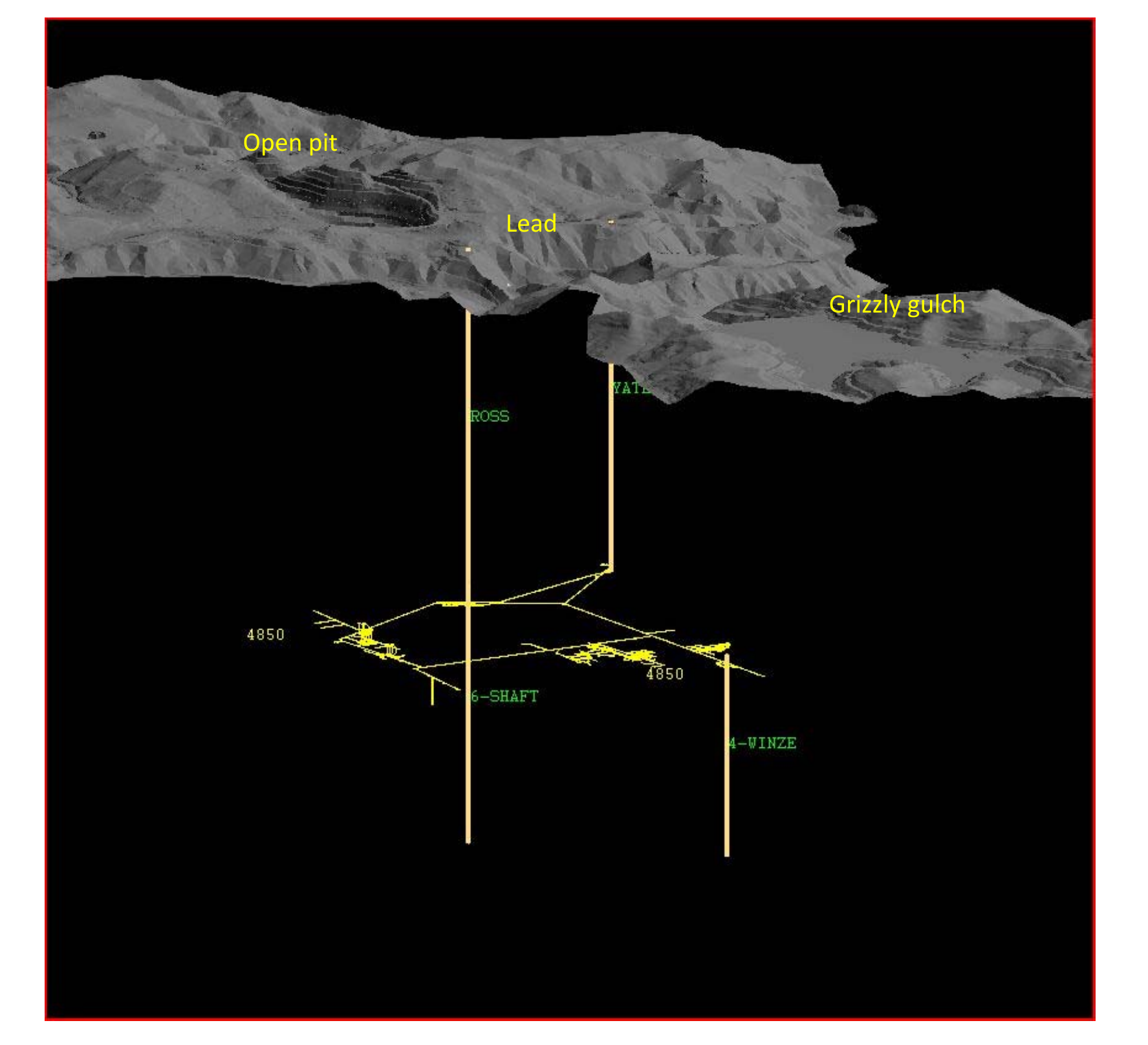}
 \caption{(in color) The approximate 3 dimensional view of the 4850 ft
 level with respect to the surface. A detailed topographical mapping
 and calculation is in progress to determine the effective cosmic ray
 shielding for the laboratory between the Yates and Ross shafts.
 \label{figrock6} }
\end{figure}

\begin{figure} 
\centering\leavevmode
\includegraphics[angle=0,width=\textwidth]{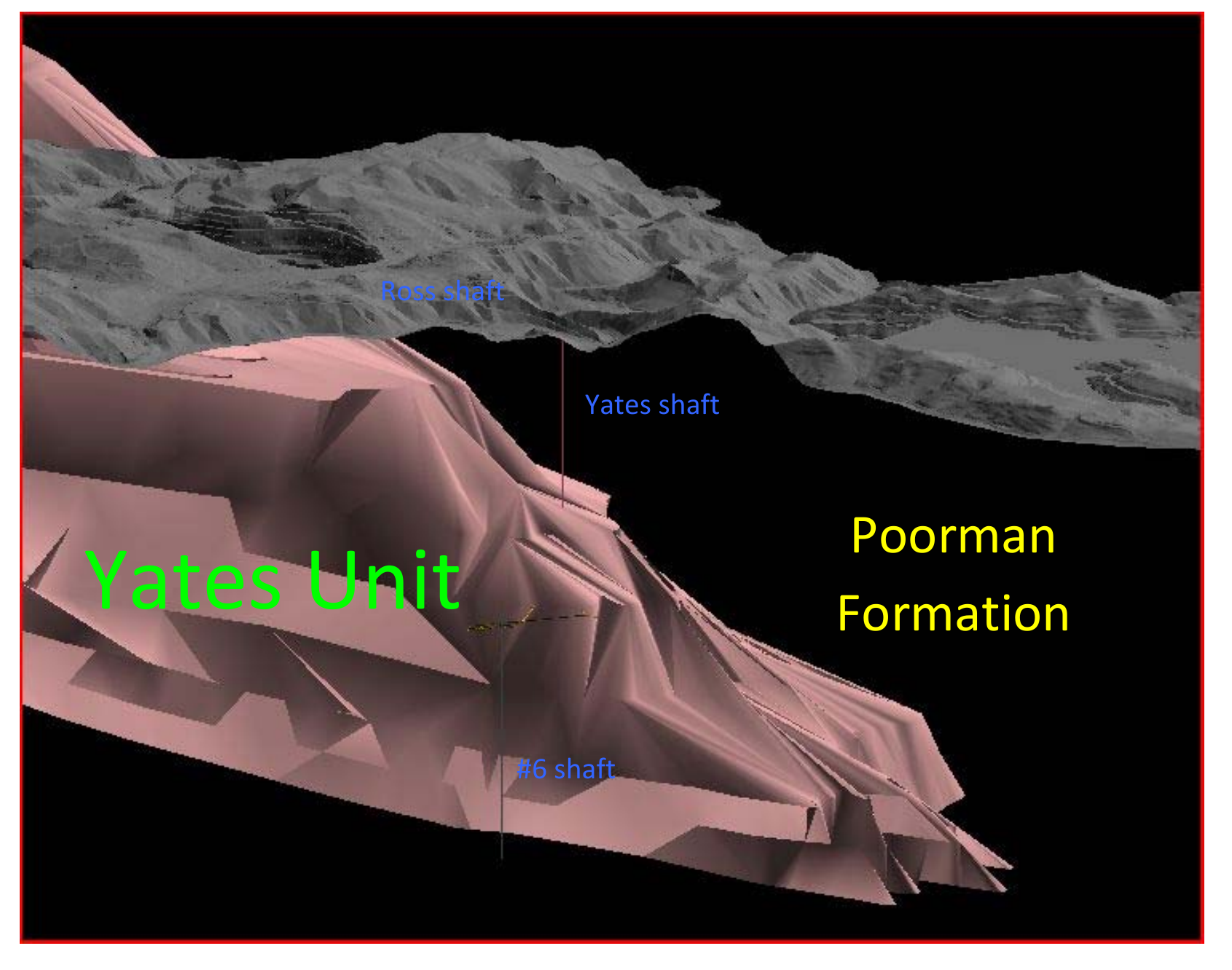}
 \caption{(in color) The approximate 3 dimensional view of the Yates
 unit as it descends to the 4850 ft level using the Vulcan database
 (Maptek inc.) for the Homestake mine.  The Yates unit, underlying the
 Poorman Formation, is considered the strongest rock in the local
 stratigraphic column. The new excavations for the DUSEL laboratory 
 will
 most likely be located in both the Yates unit and the Poorman
 formation. The large cavities, discussed here, are expected to be entirely in
the Yates.   \label{figrock7} }
\end{figure}

\begin{figure}
\centering\leavevmode
\includegraphics[angle=0,width=\textwidth]{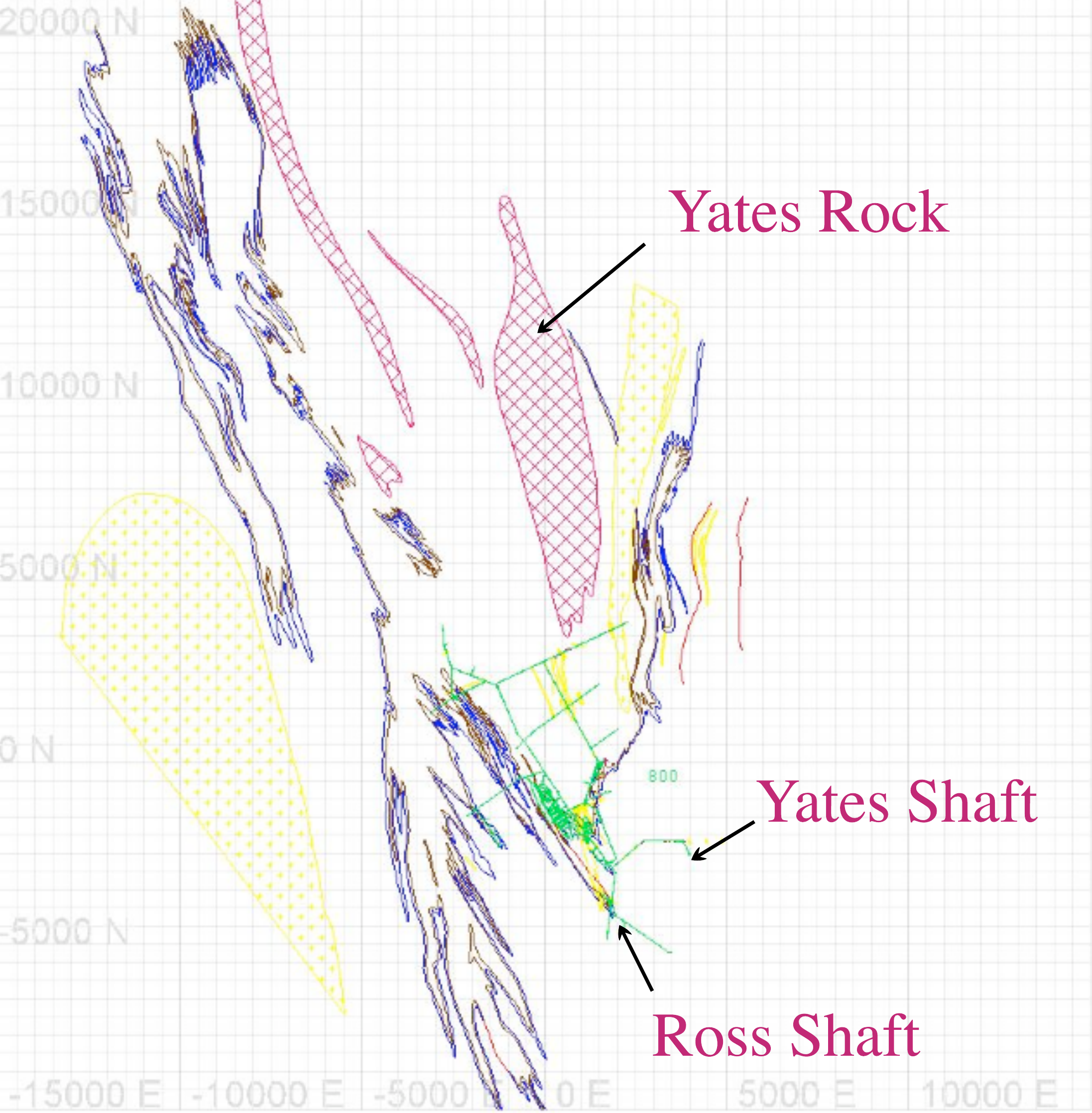}
 \caption{(in color)
The 800 level. Note that the Yates Member is several km from the 
Ross shaft. 
  \label{figrock1} }
\end{figure}

\begin{figure}
\centering\leavevmode
\includegraphics[angle=0,width=\textwidth]{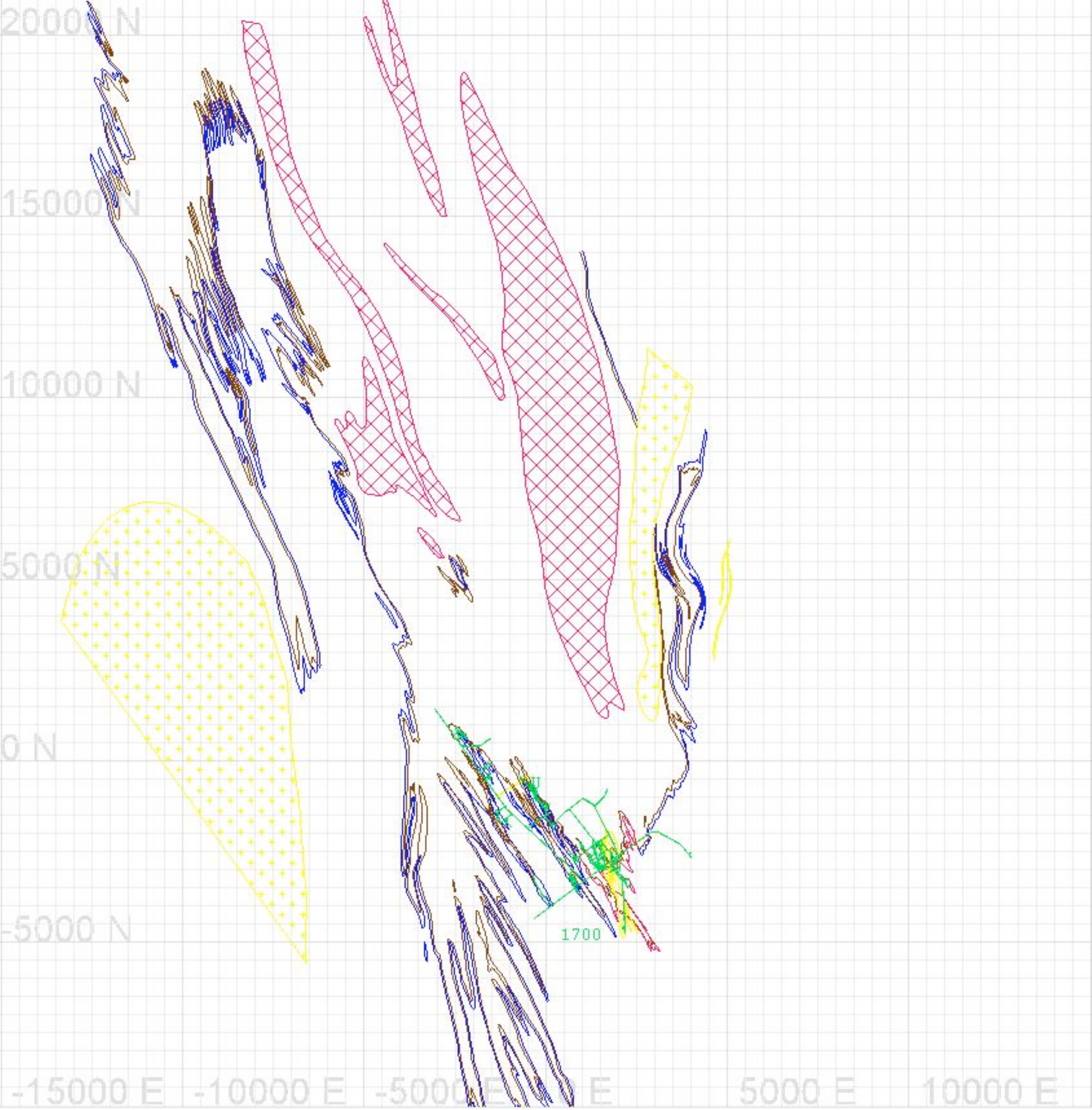}
 \caption{(in color)
The 1700 level. Note that the Yates Member is
 plunging towards the Ross Yates bisector.  
  \label{figrock2} }
\end{figure}

\begin{figure}
\centering\leavevmode
\includegraphics[angle=0,width=\textwidth]{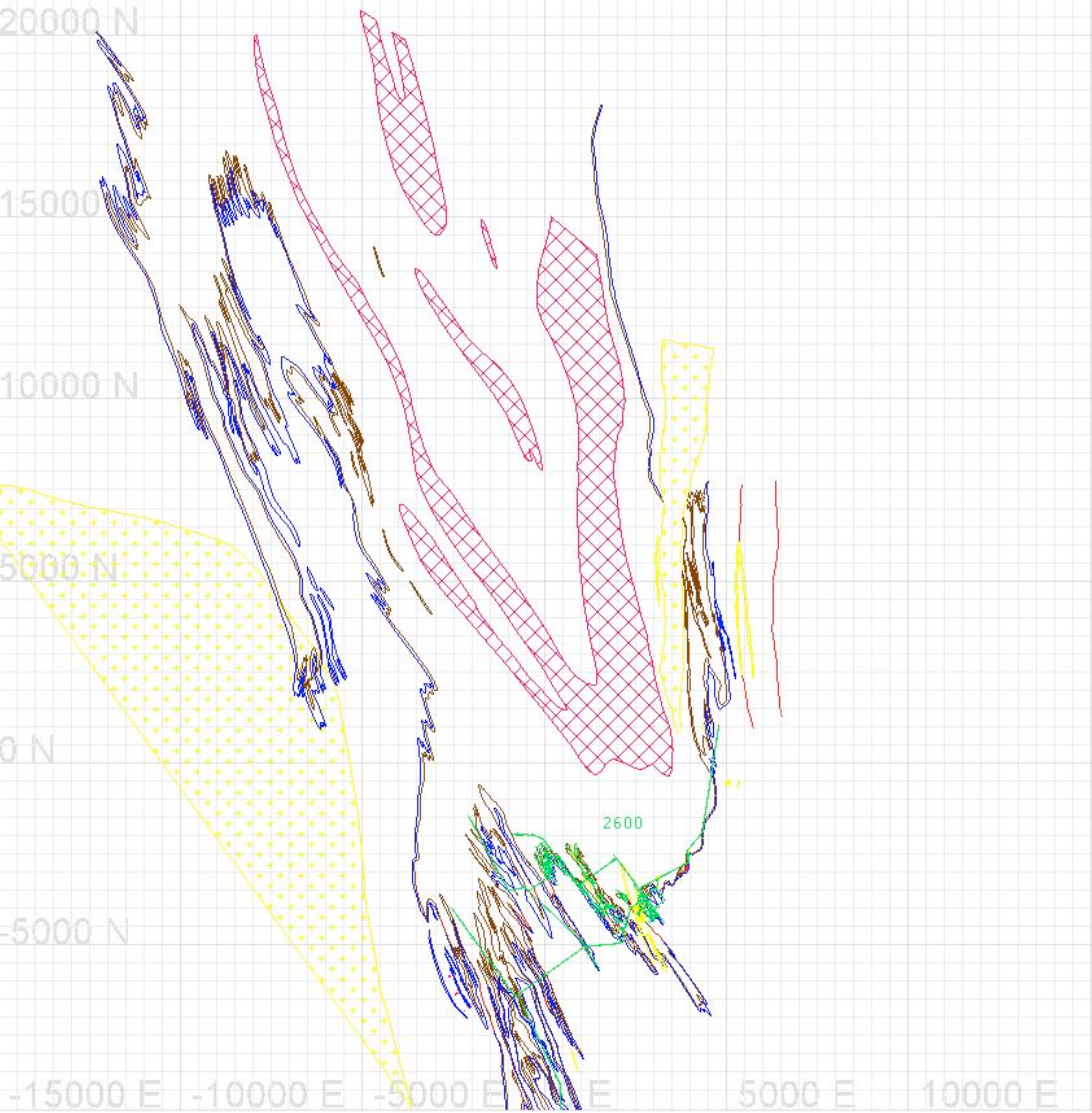}
 \caption{(in color)
The 2600 level. The Yates Member continues to plunge and broaden. 
  \label{figrock3} }
\end{figure}

\begin{figure}
\centering\leavevmode
\includegraphics[angle=0,width=\textwidth]{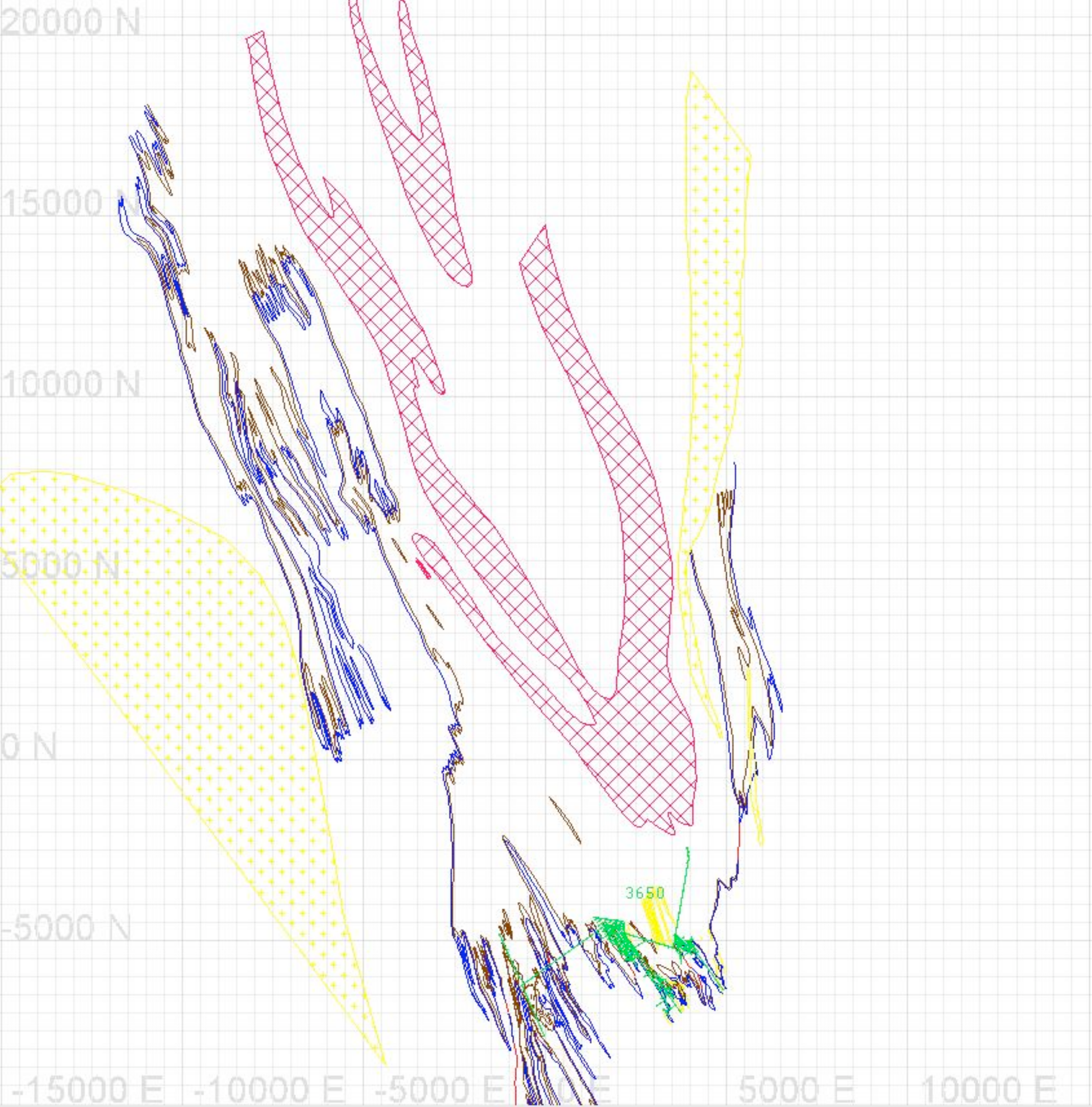}
 \caption{(in color)
The 3650 level.  
  \label{figrock4} }
\end{figure}

\begin{figure}
\centering\leavevmode
\includegraphics[angle=0,width=\textwidth]{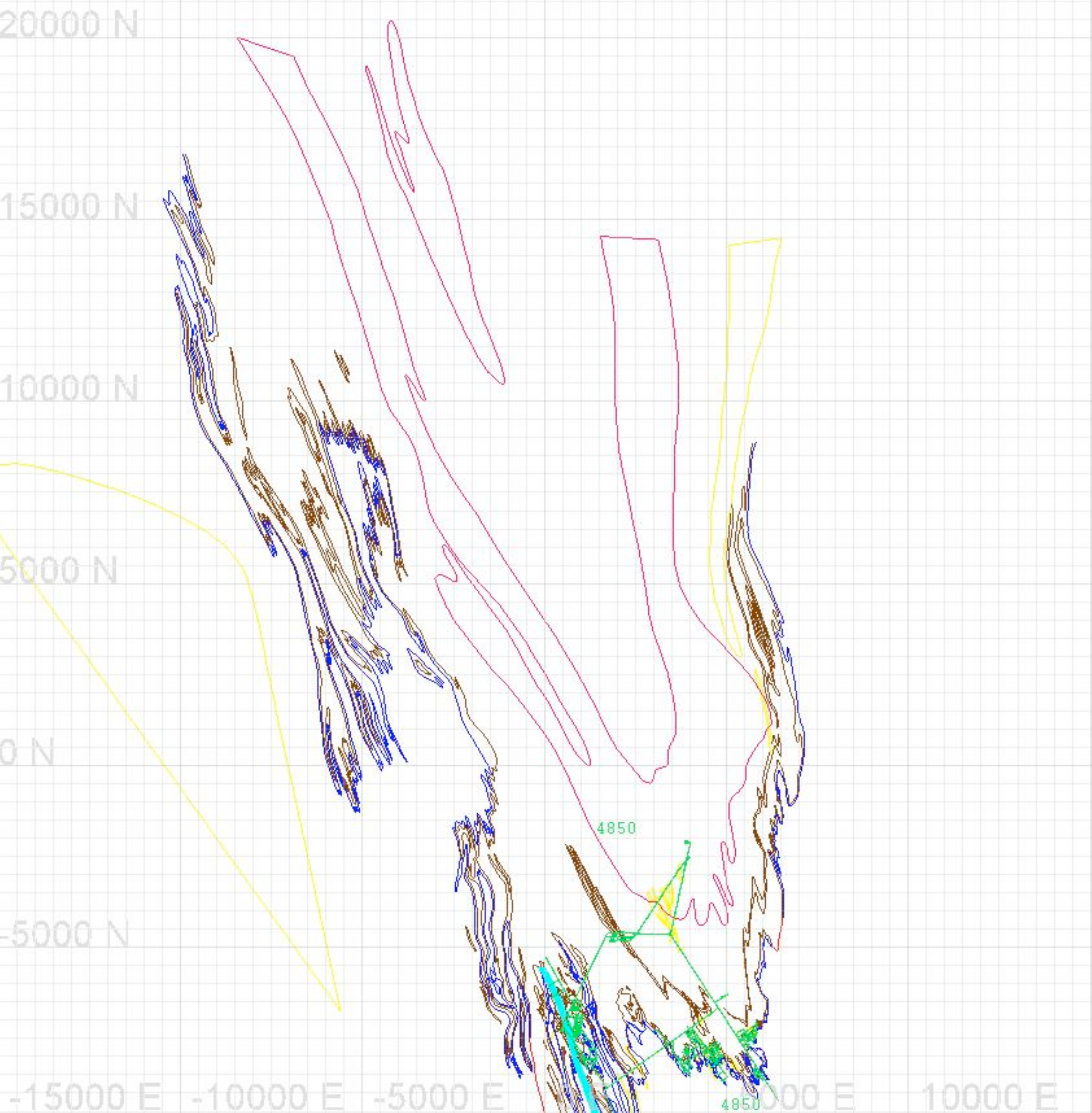}
 \caption{(in color)
The 4850 level. The Yates shaft is now in the Yates unit. 
The green triangle indicates the area of the mine that is to be developed for 
the laboratory.  The Yates shaft is at the upper tip of the triangle.  
  \label{figrock5} }
\end{figure}

Figure \ref{figrock6}   is a 3 dimensional model of the Homestake site.
The Ross and Yates shafts and the 4850 ft level are 
shown with respect to 
various surface features. 
Figure \ref{figrock7} shows the Yates rock unit, 
which is the preferred rock type for 
the cavities under discussion, as it descends from the surface towards the deep levels 
in the mine.  
The geology maps and existing mine workings for
some of the levels from table \ref{levels1}  are shown in 
Figures \ref{figrock1} to \ref{figrock5}. These are cross sections of the mine
at various levels on the same scale.   The principal
 formations are indicated by
 the following color index \cite{hladysz}. 

\noindent 
1. Yates Member: purple or purple-hatched. 

\noindent 
2. Poorman Formation: between brown and purple. 

\noindent  
3. Homestake Formation: between brown and blue. 

\noindent  
4. Ellison Formation: borders blue. 

\noindent  
5. Rhyolite intrusive formations: yellow. 

Mine Workings are shown in green. Not all of these  workings
 are  guaranteed to be habitable at this time.
 Some drifts have been back filled, therefore additional
 information may be required. The locations of the
Ross and Yates shaft stations are shown 
in figure \ref{figrock1}; figures \ref{figrock2} to
 \ref{figrock5} have the same 
scale, and so the shaft can be followed as it descends.

	Table \ref{rock}  is a summary of representative 
rock strength (from Z.J. Hladysz, South Dakota School 
of Mining and Technology) in three directions both in
 compression (C) and tension (T). As can be seen,
 in general the Yates rock is stronger (at least 
in compression) than the rock of the other formations.
 Thus, lacking other information, it is
 best to stay in the Yates Member rock.
The stress field, which is compressive,  in the Homestake mine is known from measurements 
in the 1970's and 1980's between 3050~ft level and the 7400~ft level. 
This stress field is characterized by 
\begin{equation} 
\sigma_v = 28.28 h,  \\
\sigma_{h1} = 14327.8 + 11.99 h, \\
\sigma_{h2} = 834.3 + 12.44 h \\
\end{equation} 
where $\sigma_v$, $\sigma_{h1}$, and $\sigma_{h2}$ are 
the vertical and two horizontal components of the stress in 
kPascal, respectively,  and $h$ is the depth  in meters.  
These parameters have allowed preliminary studies of cavern feasibility
in the Yates formation.
These will be summarized in section \ref{geotech2}.

\begin{table} 
\begin{tabular}{|l|l|l|l|l|l|l|} 
\hline  
Property & \multicolumn{6}{c|}{Formation} \\ 
\hline  
 (psi)  &  Homestake & Ellison &  Poorman & Yates & Yates contact & Rhyolite  \\
        &            &         &          &        &  with Poorman &   \\ 
\hline  
$C_1$ & 20,150  & 13,620 & 11,340 &  (22,000 to & (7,900 to &  (14,000 to  \\  
$C_2$ & 11,550 & 10,000 &  11,410 &   31,000)   & 26,000)   &   34,000)     \\
$C_3$ & 13,270  &  12,270 &  8,150 &            &           &  		    \\ 
\hline   
$T_1$ &  1380 & 2990 & 2350  &    &    & (1800 to  \\
$T_2$ &  1140 &  820  &  590  &   &   &    3300)   \\
$T_3$ &  1920 &  1910  & 1650  &   &   &      \\
\hline
\end{tabular}   
\caption{Homestake  geology: summary of 
rock strength data (Z.J.Hladysz, SDSM\&T). 
The strength in compression (C) and tension  (T) is specified for 
3 directions. Directions 1, and 3 are parallel to schistosity and 2 is 
perpendicular to 
schistosity. 
For the Yates, 
Yates contact with Poorman, and Rhyolite, only a range is known for 
all three directions (The numbers in brackets). 
Schistosity refers to the tendency of some rocks to form visible large 
crystalline structures that are have a planar orientation.  
\label{rock}}
\end{table}

\subsubsection{Incremental costs at levels other than 4850~ft.}

Discussions with engineers familiar with the rock conditions at the
mine, such as they are currently known, lend further support to the
construction of the large cavities at the 4850~ft level.  They include
the following points:

\begin{itemize}

\item  
 There will not be significant cost savings in constructing large
 detectors at shallower levels, 
for example levels in the  3000 ft range, 
 versus the 4850 ft Level. It might
 appear that a shallower location than 4850~ft will allow excavation
 to proceed with somewhat higher speed and with less total power
 consumption. However, these factors are considered minor in the
  cost and schedule of the excavation project at Homestake above the
 4850~ft level.

\item 
 There may be some advantage to excavating the large cavern at a
higher level than the 4850~ft from a ground control perspective, but
there is significant doubt that there would be much difference in how
the ground support is designed or specified. Previous 
 studies \cite{init-stab} 
demonstrate the feasibility of the excavation at 6900~ft, so 4850~ft
level should be better. However, as we point out in the next section, it is 
very important to understand the local rock conditions to make detailed 
estimates. Moreover, it is highly likely that the local conditions at levels 
shallower than 4850~ft are inferior. 

\item Geotechnical studies which might include coring, modelling, etc.
at the 4850 ft level will be  applicable to shallower levels in a 
relatively straight-forward way. Studies at shallower levels will not 
be as applicable to the deeper site. Therefore, it is preferred that the 
initial coring studies start at 4850 ~ft. 

\item 
 Another consideration (and more important
one) is that the Ross Pillar is near the 3000 ft level. The consulting
engineers uniformly agree that this is an undesirable rock formation
for excavating large cavities.  The access at the Ross station on this
level is highly stressed and is currently cable bolted on the 3650
L. Therefore, the middle levels from 3000~ft to 3650~ft are undesirable.  

\item 
It would be costly to drive a new connection between the Ross and
Yates shafts. The distance of such a connection is $\sim$3500~ft. The
cost of such a connection is in the range of \$10~M
(without any considerations of overhead factors). However, such a
construction will have significant impact on the schedule for the
project. It seems unnecessary when levels exist with good connectivity
between Ross and Yates.

\item 
 The Ross Shaft hoisting system will be the
conduit for hoisting 
from any level above the 4850~ft. 
For levels (in the range of 3000~ft)
 that do not have an existing loading pocket, 
it will likely be more expensive to
build a new loading pocket 
than it will be to hoist from the 4850~ft
level by dropping rock down from higher levels.  
 A new loading pocket will  be of order of $\$$3-5M. Hoisting costs
should be roughly $\$$4-$\$$6 per ton so total hoisting from the 4850~ft of
500~kTon of rock (corresponding to 1 water Cherenkov cavity) 
should be $\sim$\$3M. In the past, rock mined from any of the levels above
4850~ft was dumped through an ore pass/waste pass system to the loading
pocket on 5000~ft. It makes more sense to use this system than
installing a new loading pocket.

\item 
 Cost savings on moving the large
cavities to a higher level are only realized  if there is a significant
change in the stress field and significant  reduction in hoisting. 
In the previous studies \cite{init-stab}, the plan was to develop 
these cavities at much deeper levels (6900~ft). The excavation for 
these cavities would have required the use of the No.6 winze system and
transfer of the rock to the Ross lift.  
By moving to the 4850ft Level, the No.6 winze system was
removed from the hoisting conduit, thereby significantly reducing
hoisting time and cost. There is no such gain by moving from 4850~ft 
level to a shallower level.

\end{itemize}

\newpage
\section{Further Geotechnical Considerations Regarding
Deep Placement of Large Caverns  }
\label{geotech2}

In the previous sections we reviewed the depth requirements from the 
point of view of physics and site dependant  issues. In this section we 
will briefly review the existing work on locating large caverns 
for detectors in 
the Homestake mine. The purpose is to inform the reader that 
this preliminary work gives us confidence to proceed further with 
the design of these caverns. Some general  remarks on how such a 
design might proceed follow at the end of this section.  
Needless to say, it is extremely important to proceed with this 
design work to maintain a healthy schedule for the overall project.

\subsection{Review of the Preliminary
Work on Cavern Feasibility}
\label{feasible} 

 Preliminary Plans for the construction of chambers for a multiple
 module megaton Cherenkov detector at the Homestake Mine have been
 described in \cite{prop}. These plans were part of the US long
 baseline neutrino study \cite{study} and were also presented to the
 NuSAG group \cite{nusag} and the P5 panel \cite{p5}.  The intention
 of this work was to establish the feasibility for constructing the
 caverns needed for the large water Cherenkov detector. Clearly, the
 liquid argon caverns which are smaller can also benefit from the same
 considerations, although for liquid argon the safety and handling
 issues of a large amount of cryogenics is a primary concern.

A summary of the various steps, rock strength and stability
evaluation, chamber design and layout, construction planning,
sequencing, and development of a preliminary budget and timetable, are
described below. Although this document is a justification for much
more detailed work that is necessary to evaluate the cost and schedule
of this project, future work will certainly benefit from the existing
 documentation.

The Homestake Mine geology has been extensively studied (see the U.S.
 Geological Survey Bulletin 1857-J (1991) 
and the references cited therein).
 The strength characteristics of the rock have
 been thoroughly studied and
 measured. 
The Homestake Mining Company (HMC) has constructed several large,
 deep underground chambers. Among these are an equipment repair shop at
 the 7400~ft level and an air conditioning plant at the 6950~ft level.
These excavations at great depths provide a strong
 indication that large excavations at depths of 4850~ft to 7000~ft 
can be constructed and will remain stable for  multi-decade periods. 
The region of monolithic rock (Yates formation) being
 considered for the deep underground Cherenkov
 detector has not been mined or explored in detail. 
By extrapolating from above
 and from the west side, this rock region appears homogeneous, and with
 few intrusions. 
The preferred site for the large excavations in \cite{prop} is the 4850 ft level 
close to the Yates shaft and the Chlorine experiment. The rock in this 
region is better known than the rest of the Yates formation, and therefore 
the confidence in the cavern stability and the excavation plan is considered high. 
 The Chlorine Detector chamber was
excavated in 1965 and has been completely stable since then.  In
addition, the Yates and Ross rock dump and rock hoist system are on the same 
level  so that waste rock from the first 
chambers need only be transported within the same level.  

\subsubsection{Determination of Excavation Stability}

A preliminary 2 dimensional large chamber stability evaluation was
carried out in the fall of 2000 by members of the Rock Stability Group
at the Spokane Research Laboratory of NIOSH (National Institute of
Occupational Safety and Health). This evaluation indicated that stable
chambers with dimensions in excess of 50 meters could be constructed
at depths of 7000~ft or more at the Homestake Mine.  In the fall of
2001, rock samples from the Yates formation were taken to the Spokane
Laboratory and strength and stress analyzed.  These measurements
provided more specific input for a three dimensional stability
analysis of large excavations as a function of depth in the Yates rock
formation in the Homestake Mine. This 3D analysis involved a finite
difference evaluation using the FLAC3D program. These results
\cite{init-stab}
 were compared with the empirical prediction charts
 of \cite{charts, barton}.
 The conclusions were that 50~meter diameter by 50~meter high chambers
 could be safely excavated and would be stable for long term occupancy 
at depths up to~2150 meters and probably somewhat deeper.

Using the results of the stability evaluation, a group of former Homestake mining 
engineers, (Mark Laurenti--former Chief Mine Engineer, Mike Stahl--former
 Mine Production Engineer and John Marks--former Chief Ventilation Engineer)
 designed an array of ten 100 kiloton water Cherenkov chambers. The criteria used 
in this design were, a minimum of 50 year safe occupancy of the chambers,
 independent ventilation and access system for each chamber so that completed
 chambers can be used for research while additional chambers are under
 construction, and a structure that will permit a plastic lined water 
tight and radon reducing structural enclosure.  
The original detector construction plan was for a detector
array at the 6950~ft depth.  For \cite{prop} the plan was  adapted  for
the 4850~ft level. 
 The design involves a detailed construction plan, a rock reinforcement 
plan with cable and rock bolts and a cylindrical concrete liner,
 and a coordinated water handling, ventilation and chamber access plan. 
 The top of each chamber is
 connected to the 4850~ft level via a horizontal, radial 
tunnel. A similar tunnel connects the bottom of each chamber to a
 tunnel at the 5000~ft level.
Fresh air will be sent to each chamber via the top tunnel and 
exhaust air removed via the bottom tunnel, thus providing independent air
 supplies to each chamber. 

\begin{figure}[htbp]
  \begin{center}
 \includegraphics*[angle=90.,width=\textwidth]{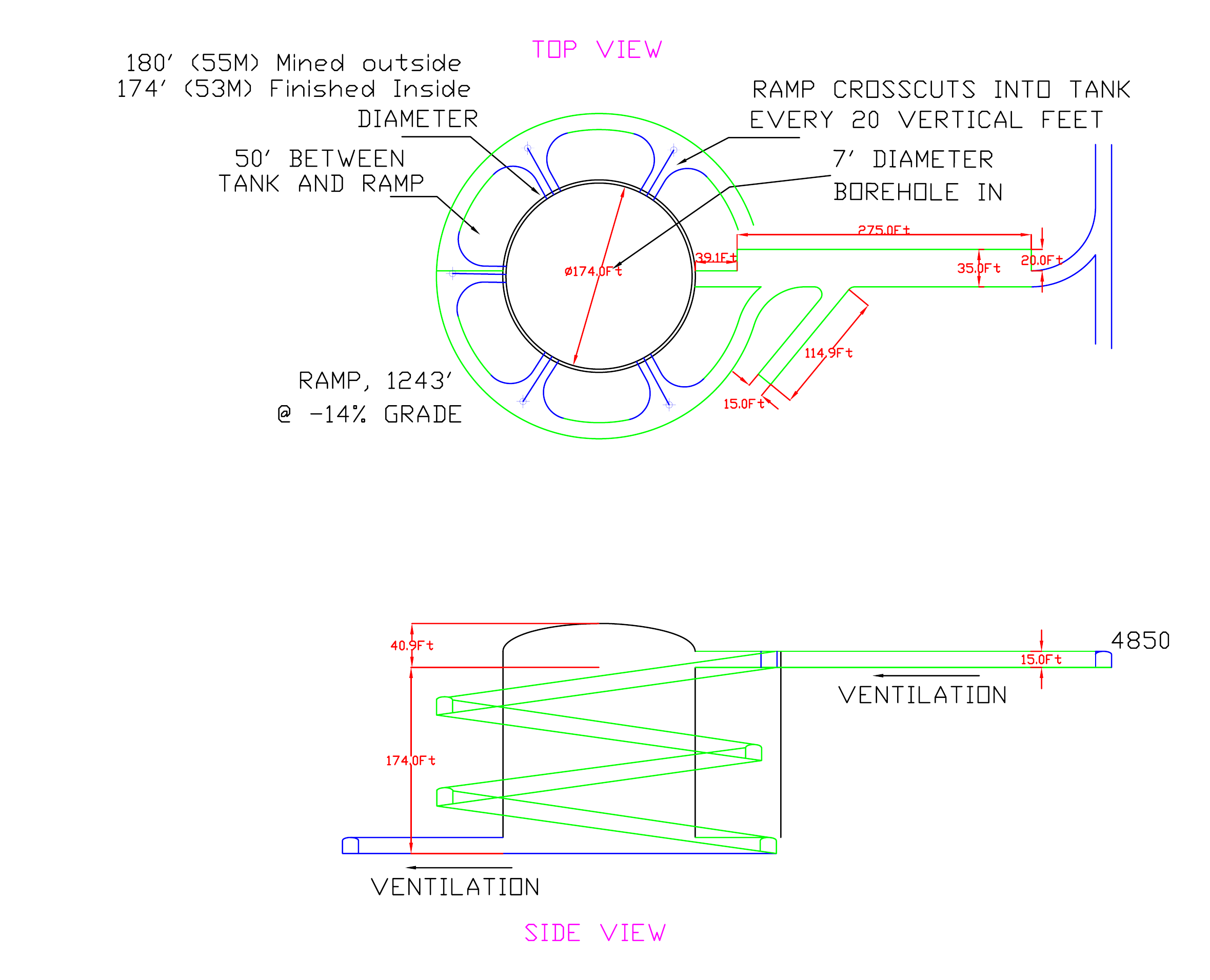}
  \caption[Details of a single cavern]{
Details of the construction of a single cavern. 
 }
   \label{detail1}
  \end{center}
\end{figure}

\begin{figure}[htbp]
  \begin{center}
 \includegraphics*[width=\textwidth]{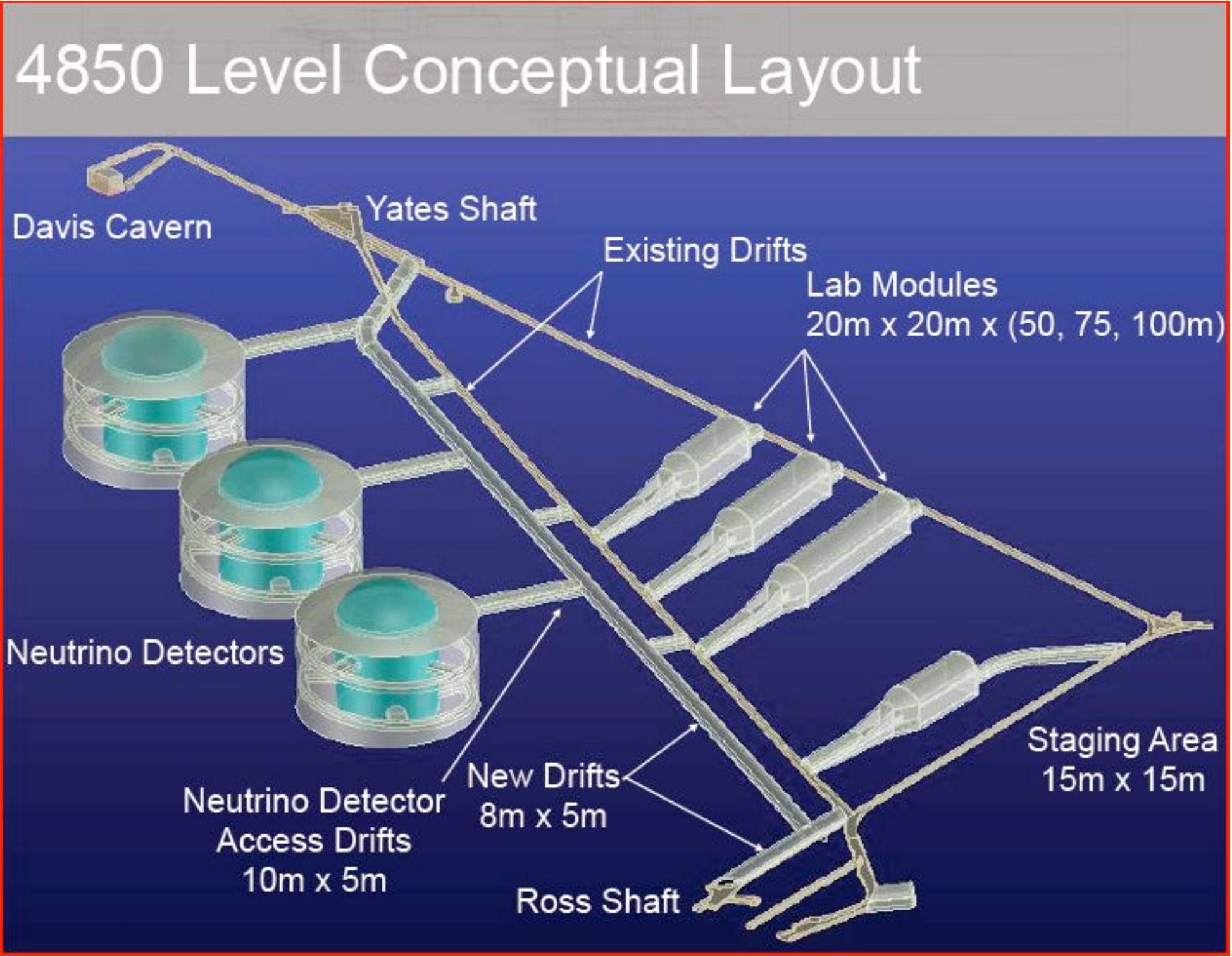}
  \caption[Placement of 3 chambers on 4850 ft]{
Placement of 3 100 kTon water Cherenkov chambers at 4850 ft.  
(Courtesy D. Plate, LBNL)
 }
   \label{detail2}
  \end{center}
\end{figure}

Plan and elevation views of  a single 
cavern
  are shown in Fig. \ref{detail1}. Placement of 3 of these 
chambers is shown in Fig. \ref{detail2}.

As part of the planning process   a detailed timetable and budget was created 
 for the construction of these chambers including initial rock evaluation
coring, construction of both top and bottom access tunnels, removal of
waste rock, maintenance of mining equipment, etc. This was presented in previous 
studies \cite{study, nusag, p5}. We emphasize, however, that the plan must be 
reconsidered once access to the Homestake mine is obtained and extensive site investigation
has been done. 

The excavation process consists of continuous repetition of three
separate tasks (1) drilling and blasting of rock, (2) removal of the
rock rubble, and (3) installation of rock and cable bolts to stabilize
the freshly exposed rock walls. Each excavation cycle is about 10
weeks with 3 weeks for each of the above three steps. There can be  a cost
savings in excavating multiple chambers at the same time, with a three
week phase shift between steps in each module. However, other models of 
excavation are also possible and should be evaluated as part of the ongoing 
studies.

\subsubsection{Rock Removal}
A 100 kiloton chamber ($10^5$~m$^3$) will involve the removal of
about 419,600 tons of rock including access tunnels, domed roof,
etc. Scaling from these numbers for 300~kTons (fiducial mass of water)  this results 
in 1,258,800 tons of rock in $\sim$4
years or 314,700 tons of rock per year. Since the hoisting capacity of
either the Yates or Ross shaft system is 750,000 tons per year, the
simultaneous construction of three chambers utilizes only 40\% of the
capacity of one of the two existing shaft systems. It is important for the 
laboratory to identify   nearby rock disposal sites as part of the planning process. 
In our initial planning generic disposal costs were included.

\subsection{Cavern Engineering Design Plan}

In this section we summarize the necessary next steps in the design and engineering of
the large cavern(s) for the proposed detector. The evolving 
engineering plan is an  essential part of the DUSEL facility plan.  Current plan calls for 
the Geotechnical Advisory Committee (GAC) of the DUSEL facility to carry out this work using 
industry participation and using the best industry practices. The DUSEL facility will be 
aided by the Large Cavity Advisory Board (LCAB). The water Cherenkov and liquid argon 
detector projects  will interact with the facility to define the requirements of 
the detector. See appendices 1 and 2 for preliminary 
requirements. 

There is consensus that the following steps will be needed to arrive at a 
the exact placement of the cavern(s), the size and shape of the cavern, and 
the cost and schedule for building the cavern. We have made this list to be informative 
about how this type of geoengineering can be carried out for the reader. The exact 
scope of this work and the schedule will be determined by the  GAC aided by 
LCAB and the detector collaborations.  

\begin{itemize} 

\item This document, associated science studies, as well as knowledge of the 
Homestake mine  has concluded that Yates rock at 
 the  4850~ft level is the
best location for development of the cavities. Nevertheless, the exact location within 
that rock has not been determined.  The exact location will depend on a detailed site investigation 
coupled with a plan for excavation. 

\item There are other examples of hard rock caverns of sizes that
 are similar to what is proposed here.
There is certainly extensive experience with good documentation for
the excavation of of the Super-Kamiokande cavern \cite{sknim}. There
are also caverns designed and built for high pressure gas storage in a
number of places in the world. A cavern in hard rock was built in
Skallen\cite{skallen} with extensive documentation. There is also a
feasibility study for a deep hard rock cavern in the Pyhasalmi mine in
Finland for the LENA detector \cite{lena}.  The documentation for these
and other examples will be used in the planning for the caverns 
we have described.

\item An initial step in the site investigation is mapping of various types of rocks and rock 
boundaries. This mapping can be done immediately after safe access is obtained to the 4850~ft level. 
Positions of 
geological features that are exposed on the walls of the existing drifts will be recorded. 
Using information from various levels a more detailed 3 dimensional view of the area can be developed. 

\item After the mapping, suitable sites for the cavern(s) can be narrowed using these criteria:  
1) The sites should be outside the zones of influence of
major mapped geo-structures. Where excavation wall, crown or
invert (the bottom of the cavern) rock materials could be
 weakened or local stress or displacement anomalies created.
 2) They should be outside the zones of influence of rock contacts, where
 significant stiffness/stress contrasts could be present across and
 adjacent to boundaries.  3)  Sites should be outside the zones of influence of
 stopes and temporary excavations, where  deterioration and
 overstress/destress may be more frequent and severe than in virgin
 ground.  4) Sites should be  outside the zones of influence of existing or
 planned permanent excavations,  where higher levels of blast vibration
 and stress  superposition could be expected.  5)  Sites should be in close
 proximity to rehabilitated openings and infrastructure,  where
 dedicated development costs to support investigation and construction
 work will be minimized.  
If adverse geotechnical conditions are discovered in these
 Yates rock volumes we can explore other Yates sites, which are more
 remote from existing openings or accessible on other adjacent levels
 (4000 - 5000ft). 

\item There will be initial cavity modeling with existing information.  This will improve on
 the work summarized in section \ref{feasible}. The prior work was carried out 
assuming a cylindrical cavity at 6950~ft level.  Further investigation for 
other shapes and sizes at the 4850~ft level needs to be performed.  This modeling 
will also result in a list of needed  site specific information. 

 \item After identification of suitable site(s) more detailed information about the 
rock is obtained by coring. The cores could be several hundred ft long each. Several 
initial cores are needed for each cavern. Some of the cores will be oriented cores in which 
the orientation of the rock is preserved.  The cores will be analyzed to obtain 
i) strength of the rock in different directions, ii) boundaries or fractures within
the rock mass, iii) the extent of the rock mass. 

\item The stress field in the rock mass can also be obtained from the cores by either 
placement of strain gauges or by pressurizing the drill holes 
 with water. 

\item If there are multiple nearby cores, the pattern of water flow through the rock mass 
can be analyzed by measuring  water flow from the cores or between cores.   

\item The information from cores will be used to refine the modeling of the caverns. Modeling 
will include both analytical and computations methods. The outcome of the modeling will be 
recommendations i) on the rock support needed to make the cavern suitably stable, ii)  on
a possible liner for the cavity,  iii) on the shape and size of the cavity, 
iv) on the motion of the walls over time and the impact on the internal detector structure,  
v) on the excavation 
sequence to minimize the stresses on the cavity wall, and   vi) on a list of additional geotechnical 
investigation needed. 

\item Once the exact location of the caverns is narrowed down, a detailed pattern of 
cores will be needed to map the rock mass at the site(s) with fine granularity. For example,
each cavity may need 7 to 10 cores in a pattern that obtains information about the rock above the roof
as well as on all sides of the cavern.

\item With full information from the detailed pattern of cores, final cavity modeling will be 
performed. The cavity model and the excavation sequence
 will be integrated with the plan for detector installation.  

\item Sensors could be placed in the bore holes to obtain information on
long term monitoring of strains in the rock.  

\item Based on past experience in mapping and feasibility studies from 
other underground construction projects, it is likely that there are 
a number of acceptable sites in the Yates formation, but no site will be 
perfect. Rock strength, structure, and in situ stresses need to be 
measured as early as possible so that the cavity can be engineered with 
the best possible information, and greater confidence can be assigned to the 
cost and schedule. Site investigation should
 allow for types, locations and extents of potential adverse behavior
 to be pinpointed.

\item The above sequence was specific for the very large caverns needed 
for a water Cherenkov detector. The liquid argon chamber will be smaller in size, and 
the cavern engineering can proceed in a similar way, but is not as critical as issues related to 
cryogenic liquids.  Liquid argon technology is also going through extensive R\&D phase, and the size and
shape of the cavern will depend on the length of the drift region and ideas concerning the assembly 
and filling of the detector. Once there is a concept for the liquid argon detector, a site within
the mine must be chosen based on the best way to contain any spill of the liquid argon or vent it 
to the surface.  The safety infrastucture needed for liquid argon will be more important for 
site selection than geoengineering issues.

\end{itemize}

\section{Summary and Recommendation }

In this document, we have investigated the depth requirements for a
large underground detector at Homestake DUSEL. The goal of this work
was to identify the best level for geotechnical studies needed to
prepare for the construction of this detector.  As part of this
evaluation, we have been advised by expert engineers on 
existing information on 
infrastructure and rock
characteristics  to narrow the choice of levels for the development of
large cavities with long lifetimes.
 
We have evaluated depth requirements for all major physics signatures
that will make up the physics program for the large underground
detector, including accelerator generated neutrinos, solar neutrinos,
atmospheric neutrinos, supernova neutrinos, and nucleon decay.  Table
\ref{tab:depth_summary} summarizes the results of these studies 
for both water Cherenkov and
liquid argon detector technologies.  None of the signatures requires a
depth greater than the 4850 level at Homestake ($\sim$4300 mwe). We therefore
recommend that geotechnical studies for the large detector be carried
out at the 4850 ft level as soon as possible. 
  This depth is sufficient to carry out an
excellent physics program, and takes the best advantage of the
infrastructure and rock conditions at the Homestake Mine.

\newpage

\section{\bf Appendix 1: Requirements for a Long Baseline Water Cherenkov Detector}  
\label{appen1} 

\underline{\bf Assumptions}  
\begin{itemize} 

\item  
Broadband neutrino beam from Fermilab spanning energy range 1-10 GeV. 
Total event rate of 20000 to 50000  events per 100 kTon of detector mass per year. 

\item  
Detector should have sufficient total fiducial mass to reach
 CP violation sensitivity for 
$\sin^2 2 \theta_{13} \approx 0.01$; the exposure 
needed for this  is  $\ge$300 kTon$\times$1 MW$\times$~5yrs.

\item Detector must have capability to search for nucleon decay beyond existing 
bounds. 

\item Detector must have capability to search for low energy 
phenomena with threshold $>$5 MeV. 
The exact threshold will be one of the scope parameters for the project. 

\item Above requirements set the total mass of the detector to be $>$300 kTon.  

\end{itemize}

\underline{\bf Preliminary Technical Requirements} 
\begin{itemize} 

\item Detector to be in $\le 3 $ cavities. 

\item Detector should be located with $>4300$ mwe shielding. 

\item Detector modules should not be more than 5 km apart to obtain 
maximum beam intensity and to reduce the spectral differences between 
detector modules. 

\item Maximum height of water (exerting pressure on photomultiplier assemblies): 
$< 60 m$. 

\item Maximum length through water (to reduce light loss through attenuation):
$< 80 m$. 

\item Maximum width of cavern to be determined by studies of cavern stability.

\item Cavern must be lined to maintain reasonable water loss.  

\item Each detector module fiducial volume be maintained: $>100 kTon$. 

\item Cavern liner should accommodate $\sim 50000$ photomultiplier tubes. 

\item Magnetic field suppression coils. 

\item Total power needed for operation: $\sim 2-4 MW$ clean power, $\sim 2 MW$ standard power. 
Exact power number to be determined by analysis of cooling requirements for the 
water and the photo-multipliers.

\item Other underground space needed for: electronics racks, water system, and control room
for each detector module. 

\item Standard internet connectivity and communication links. 

\item Detector materials must be compatible with ultra high purity water. 

\item Eventual addition of Gd salts to be considered. Material and safety considerable 
should include effects of Gd loading. 

\end{itemize}

\newpage

\section{\bf Appendix 2: Requirements for a Long Baseline 
Liquid Argon Time Projection Chamber}  

\label{appen2}

\underline{\bf Assumptions}   
\begin{itemize} 

\item  
Broadband neutrino beam from Fermilab spanning energy range 1-10 GeV. 
Total event rate of 10000 to 20000 events per 50 kTon of detector mass 
per year.

\item  
Liquid argon Detector should have sufficient total fiducial mass to reach
 CP violation sensitivity for 
$\sin^2 2 \theta_{13} \approx 0.01$; the exposure 
needed for this  is  $\ge$50 kTon$\times$1 MW$\times$~5yrs with currently 
assumed  event reconstruction  
performance. The exact performance will be known as a result of 
ongoing R\&D. 

\item Detector must have capability to search for nucleon decay in the 
channels unique to liquid argon including $p\to \bar\nu K^+$. 

\item Detector must have capability to search for low energy 
phenomena with threshold $>$5 MeV. 
The exact threshold will be one of the scope parameters for this detector
 project. 

\end{itemize} 

\underline{\bf Preliminary Technical Requirements} 
\begin{itemize} 

\item 
Detector will be built in a program of increasing size and capability. 
The initial size will be a 5 kTon module; second step will be 25 kTon detector; 
final goal is to reach $\ge$50 kTon
of fiducial mass. 

\item Liquid argon detector should be within ~5~km of the core of the FNAL neutrino 
beam. 

\item Minimum single drift volume 5~m$\times$5~m $\times$ 30~m

\item The 5~kTon module cavern 15~m$\times$15~m $\times$ 30~m

\item The 25~kTon to be built in $\le$6 modules.  

\item Minimum wire pitch 3 mm. Leads to 250k to 1 M channels for 25 kTon.  

\item Heat load from electronics 50~kW to 80~kW total.  

\item 
LAR purification system will be either internal or external. If external
additional space will be needed. 

\item 
Additional experimental support drifts will be needed to accommodate 
service equipment, data acquisition, and control rooms. 

\item Clean experimental power, HVAC, UPS power, chiller water, lighting and 
other laboratory facilities underground.  

\item Above ground N$_2$ storage facility. Above group Ar delivery
and storage facility. 

\item  LAR specific Safety infrastructure: \\
	i. Fluid tight doors/bulkheads \\
	ii. Dedicated ventilation \\
	iii. Multi level containment \\
	iv. Vent to surface \\
	v. ODH monitoring system \\
	vi. Seismic/vibration damping  \\
	vii. Floor barriers to dam/collect/direct liquid and gas 
	to ventilation shaft. \\

\item Detector cavern isolation. Detector modules will be housed 
so that the liquid
argon area can be isolated.

\end{itemize}


\begin{thebibliography}{99}
\bibitem{skcollab} 
This figure of a simulation in the Super-Kamioka detector is a courtesy of 
the SK collaboration.
\end{thebibliography}
\end{document}